\documentclass[pra,superscriptaddress,twocolumn,showpacs,preprintnumbers,nofootinbib,amsmath,amssymb,longbibliography]{revtex4-2}
\usepackage{comment}
\usepackage{amsmath}
\usepackage{amssymb}
\usepackage{graphicx}
\usepackage{float}
\usepackage{bm}
\usepackage[english]{babel}
\usepackage{dsfont}
\usepackage{epstopdf}
\usepackage{textcmds}
\usepackage{soul}
\usepackage{color}
\usepackage{braket}
\usepackage[colorlinks=true,
            linkcolor=magenta,
            urlcolor=blue,
            citecolor=blue]{hyperref}

\usepackage[normalem]{ulem}
\usepackage{orcidlink}

\begin{document}

\title{Flat band  mediated photon-photon interactions in 2D waveguide QED networks}

\author{Matija Te\v{c}er\orcidlink{0009-0009-5903-9499}}
\affiliation{Dipartimento di Fisica e Astronomia ``G. Galilei'', via Marzolo 8, I-35131 Padova, Italy}

\author{Giuseppe Calaj\'o \orcidlink{0000-0002-5749-2224}}
\affiliation{Dipartimento di Fisica e Astronomia ``G. Galilei'', via Marzolo 8, I-35131 Padova, Italy}
\affiliation{Istituto Nazionale di Fisica Nucleare (INFN), Sezione di Padova, I-35131 Padova, Italy}

\author{Marco Di Liberto\orcidlink{0000-0002-3574-7003}}
\affiliation{Dipartimento di Fisica e Astronomia ``G. Galilei'', via Marzolo 8, I-35131 Padova, Italy}
\affiliation{Padua Quantum Technologies Research Center, Universit\'a degli Studi di Padova}
\affiliation{Istituto Nazionale di Fisica Nucleare (INFN), Sezione di Padova, I-35131 Padova, Italy}

\date{\today}

\begin{abstract}

We investigate a Lieb lattice of quantum emitters coupled to a two-dimensional waveguide network and demonstrate that this system supports an energetically isolated flat band, enabling localization despite the presence of long-range photon-mediated couplings. We then explore the two-excitation dynamics in both the softcore and hardcore interaction regimes, which arise from the nonlinearity of the emitters. In the softcore regime, we observe interaction-induced photon transport within the flat band, mediated by the formation of bound photon pairs. In the hardcore regime, corresponding to the two-level atom limit, we instead find the emergence of metastable exciton-like dressed states involving both flat and dispersive bands. Our findings highlight how the interplay between the collective behavior of emitters and effective photon-photon interactions can provide a platform for studying highly correlated photonic states in flat-band systems.

\end{abstract}

\maketitle

\section{Introduction}
Systems featuring dispersionless (flat) bands provide an ideal setting for the emergence of strong particle correlations, as interactions dominate over single-particle energy scales \cite{doi:10.1142/S0217979215300078}. 
Besides the famous Landau level problem, which gives rise to fractional quantum Hall states \cite{Laughlin1983}, and the recent discovery of strongly correlated states in twisted bilayer graphene \cite{Cao2018,Cao2018b}, flat bands can also arise in a variety of lattice models \cite{Leykam01012018,danieli2024flat}.
By definition, single particles in a flat band exhibit zero group velocity and infinite effective mass. 
While in trivial flat bands, this comes from having  decoupled sites, in  flat bands embedded within multi-band systems, it results from destructive wave interference, leading to the formation of highly localized states. 
These states, known as compact localized states (CLS), are superpositions of Bloch states confined to a few lattice sites~\cite{PhysRevB.95.115135,Ramachandran2017,Rhim2019,PhysRevB.95.115309}, which can even result in the extreme situation of full localization with all-bands flat known as Aharonov-Bohm caging \cite{Vidal1998, Mukherjee2018}. 
In the multi-particles case, interactions can induce transport that occurs in pairs \cite{PhysRevLett.85.3906,PhysRevB.82.184502, Junemann2017, Tovmasyan2018, Burgher2025, PhysRevResearch.5.043259}, something reminiscent of Cooper pairing in superconductors, that leads to superfluid behavior~\cite{peotta2015superfluidity, Salerno2023, Torma2022}. 
Remarkably, the effective mass of these pairs has been shown to have a geometric origin based on the quantum metric of the flat band~\cite{Torma_PRL,Torma_PRB}. 
These fascinating effects of flat-band physics have been extensively studied and experimentally probed~\cite{Xia2016,PhysRevLett.114.245504,GuzmnSilva2014,Taie2015,PhysRevLett.118.175301,PhysRevB.93.075126,Centaa2023} in models with finite-range hopping.
In contrast, long-range hopping has been identified as a mean to introduce non-trivial topology to flat bands (non-zero Chern number)\cite{bergholtz2013topological,PhysRevB.90.115132}. 
However, it is generally understood that incorporating fine-tuned long-range terms for topological bands breaks the single-particle localization due to the obstruction theorem and the non-zero Chern number, thus inducing a finite, possibly small, band dispersion, making the described effects less robust \cite{Sun2011}.

\begin{figure}[b!]
    \includegraphics[width=0.99\linewidth]{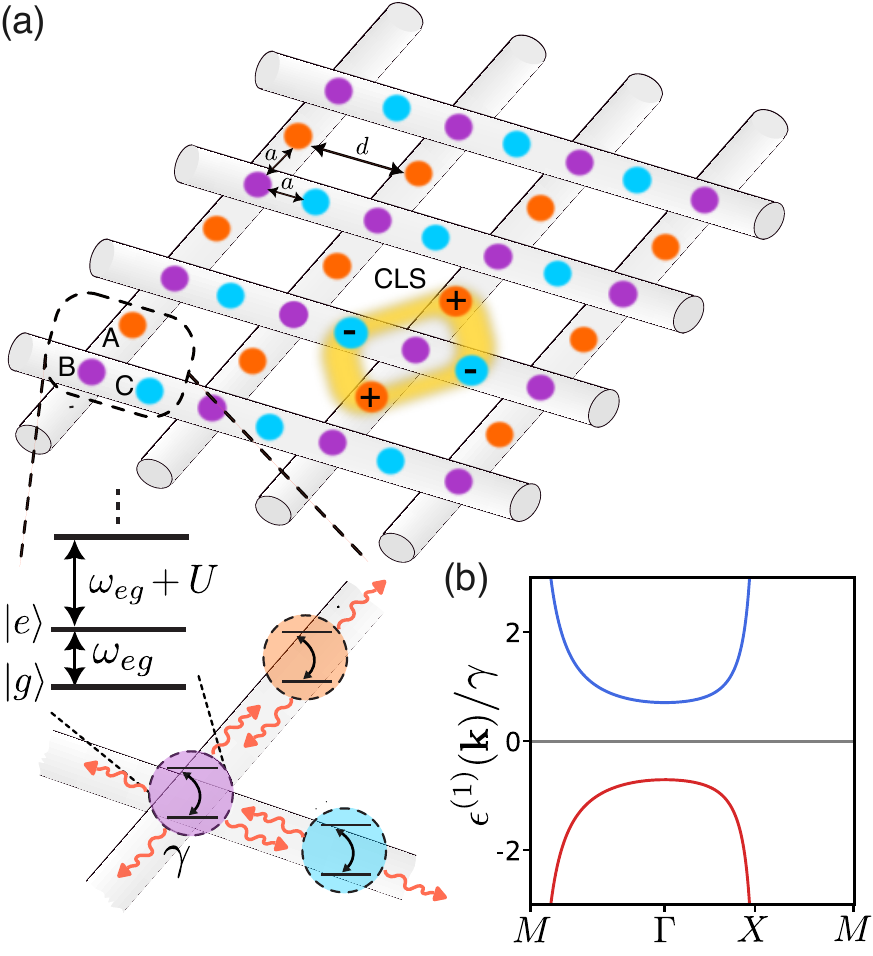}
    \caption{(a) A Lieb lattice of \textcolor{red}{$N$} quantum emitters with lattice constant $d$ and intracell distance $a$ is coupled to a square grid of one-dimensional waveguides. 
The unit cell of the lattice consists of a corner site and two edge sites coupled to two waveguides and an individual waveguide, respectively. 
The quantum emitters are modeled as nonlinear resonators with on-site repulsive interaction $U$.
(b) The single-excitation dispersion within the first Brillouin zone: two dispersion branches are separated by an isolated flat band at zero energy. 
Here, we set $k_0d = \pi$ and $a = \frac{d}{2}$.}
    \label{fig:1}
\end{figure}

Infinitely long-range photon-mediated interactions arise naturally among multiple emitters coupled to a one-dimensional photonic channel. 
This scenario, referred to as waveguide QED~\cite{RevModPhys.95.015002,RevModPhys.89.021001,Ciccarello:24}, has attracted increasing attention due to its feasibility for experimental implementation either at optical~\cite{lodahl2015interfacing,hood2016atom,corzo2019waveguide,tiranov2023collective,prasad2020correlating} or microwave frequencies \cite{astafiev2010resonance,brehm2021waveguide,mirhosseini2019cavity,kannan2023demand,scigliuzzo2022controlling,shah2024stabilizing,jouanny2025high}.
Extensive studies have been conducted to tailor atom-atom interactions through photonic bath engineering. 
In particular, when the atomic frequency falls within a photonic bandgap, it becomes possible to induce coherent atom-atom interactions with tunable ranges~\cite{gonzalez2024light,Douglas2015,Tudela2015sub,RevModPhys.90.031002}. 
Further control over these interactions can be achieved by coupling the atoms to topological photonic lattices~\cite{PhysRevLett.126.103603,PRXQuantum.4.030306,PhysRevResearch.5.023031,PhysRevB.107.054301} or photonic flat bands~\cite{di2024dipole,o2025circuit,wang2024cavity}. 
A fundamentally different approach to engineering coherent and dissipative interaction properties involves leveraging the collective emission of ordered atomic arrays coupled to propagating photons.
In this scenario, collective phenomena can significantly alter the excitation lifetime within the array, leading to either superradiant~\cite{sierra2022dicke,Cardenas,Arno_super} or subradiant states~\cite{asenjo2017exponential,albrecht2019subradiant,dimer_shermet,tiranov2023collective,Molmer,needham2019subradiance,kornovan2019extremely,ostermann2019super,poshakinskiy2021dimerization}. 
Such collective interactions also influence the coherent properties of the emitters, giving rise to polariton-like dispersion branches~\cite{albrecht2019subradiant,calajo2022emergence,Schrinski_polariton}, which determine the propagation speed of excitations within the array.
The engineering of these dispersions has been studied at the single-excitation level in two dimensions~\cite{PhysRevA.86.031602,
bettles2016enhanced,shahmoon2017cooperative,manzoni2018optimization,PhysRevLett.117.243601,PhysRevA.108.030101}, uncovering the presence of topological~\cite{perczel2017topological,perczel2017photonic,perczel2020topological} and flat bands~\cite{perczel2020topological,PhysRevA.103.033702,PhysRevResearch.4.043039}. 
However, in the latter case, no demonstration of compact localized state formation or investigation of multi-excitation physics was conducted.
Meanwhile, the coupling of quantum emitters with confined light  is also known to induce effective photon-photon interactions, leading to highly correlated light observable as photon anti-bunching~\cite{shen2007strongly,shen2007stronglyL,Schrinski_polariton} and bunching~\cite{zheng2011cavity,mahmoodian2018strongly,prasad2020correlating,mahmoodian2020dynamics,le2022dynamical}. 
These phenomena, involving single or multiple emitters, have been extensively explored in one-dimensional channels~~\cite{tomm2023photon,zhang2020subradiant,poddubny2020quasiflat,bakkensen2021photonic,calajo2022emergence,PhysRevA.108.023707}, as well as in arrays of emitters coupled to waveguide networks~\cite{marques2021bound,schrinski2023photon}, or to two-dimensional waveguides~\cite{PhysRevLett.132.163602}.

In this work, we combine two key features, array dispersion engineering and effective photon-photon interactions, by exploring a Lieb lattice geometry of emitters coupled to a square grid of one-dimensional waveguides. 
We demonstrate that long-range photon-mediated interactions between emitters can give rise to an energetically isolated non-trivial flat band supporting well-defined compact localized states (CLS).
We then investigate the dynamics of two initially excited nearby CLS in the regime of 
softcore photon interactions revealing interaction-induced photon transport mediated by the flat band. 
In this regime, propagation occurs through the formation of bound photon pairs with finite group velocity and effective mass, which escape the lattice through the array edges.
In the opposite regime of hardcore emitters, coinciding with the two-level atoms limit, we observe the formation of metastable exciton-like dressed states involving the flat and dispersive bands. 
These results lay the groundwork for studying highly correlated states involving many interacting photons in flat-band systems.

The manuscript is organized as follows.  In Sec.~\ref{sec:model}, we introduce the model consisting of a Lieb lattice of quantum nonlinear emitters coupled to a square grid of one-dimensional waveguides, described within the Born-Markov approximation. In Sec.~\ref{sec:flatband}, we derive the single-particle spectrum, which consists of two dispersive bands separated by an isolated flat band. For this flat band, we demonstrate the existence of compact localized states and compute the quantum geometric tensor, discussing its key properties. 
In Sec.~\ref{sec:interactions}, we analyze the two-excitation properties, treating the softcore and hardcore nonlinear regimes separately. In the first case, we observe interaction-induced photon transport within the flat band, enabled by the formation of bound photon pairs. In the second case, we discuss the emergence of metastable, exciton-like dressed states that involve hybridization between flat and dispersive bands.
Finally, in Sec.~\ref{sec:conclusions} we draw our conclusions.

\section{Model}
\label{sec:model}

We consider an array of $N$ quantum emitters (atoms)  coupled to a two-dimensional waveguide network.
We describe the quantum emitters as bosons
represented by the field operator $\hat{b}_i$ ($\hat{b}_i^\dagger$)  annihilating (creating) a photon at position $i$ and satisfying  $[\hat{b}_i, \hat{b}_j^\dagger] = \delta_{ij}$. 
To make energetically unfavorable for an emitter to host multiple excitations, we include an onsite photon-photon repulsion of strength $U$, that describes a photonic nonlinearity~\cite{RevModPhys.85.299}. 
In the limit of infinite interaction strength $U$, the emitters effectively behave as two-level systems (qubits) with ground and excited states $|g\rangle,|e\rangle$ with corresponding transition frequency $\omega_{eg}$.

Before presenting the waveguide network considered in this work, we first review the derivation of the effective waveguide QED Hamiltonian for an array of quantum emitters coupled to a one dimensional photonic channel, starting from the original atom–light interaction model.

\subsection{Waveguide QED Hamiltonian}

Let us assume the emitters to be arranged in an ordered array with lattice constant $a$, each coupled with the same strength to a one-dimensional waveguide through the $|g\rangle \leftrightarrow |e\rangle$ transition, resulting in an individual decay rate $\gamma$.
 Within the frequency range relevant to the system dynamics, set by $\gamma$, we assume a linear dispersion relation for the electromagnetic field in the waveguide, $\omega_k=c|k|$. We can thus separately treat left and right propagating photons traveling with opposite speed $\pm c$.
The Hamiltonian that describes this  system under the rotating-wave and dipole approximations reads ($\hbar=1$)~\cite{RevModPhys.95.015002}:
\begin{align}\label{eq: Full Hamiltionian}
    \begin{split}
        &\hat{H}=\hat{H}_0+\hat{H}_{\rm f}+\hat{H}_{\rm int}\,, \\
&\hat{H}_0=\sum_{i}\omega_{eg}\hat{b}_{i}^\dagger\hat{b}_{i}+\frac{U}{2}\sum_{i} \hat{b}_i^{\dagger}\hat{b}_i^{\dagger}\hat{b}_i\hat{b}_i\, , \\
&\hat{H}_{\rm f}=\sum_{\lambda = R,L} \int d\omega\, \omega\, \hat a^\dagger_{\lambda}(\omega) \hat a_{\lambda}(\omega) \,, \\
 & \hat H_{\rm int} =\sum_{\lambda = R,L} \sum_{i} \sqrt{\frac{\gamma}{2\pi}} \int d\omega\, \omega\, \hat a^\dagger_{\lambda}(\omega)\, \hat b_i\, e^{\mp i\omega x_i / c} + {\rm H.c.}
    \end{split}
\end{align}
where the bosonic operators $\hat{a}_{\lambda}(\omega)$ annihilate right- ($\lambda=R$) and left- ($\lambda=L$) propagating photons of frequency $\omega$ and wavevector $k=\pm \omega/c$, while $x_i$ denotes the position of the $i$-th emitter along the waveguide. 
This Hamiltonian describes the exchange of photons between the emitters mediated by the waveguide.

To make the analysis of this complex interacting system more tractable, we assume that the usual requirements for the Born–Markov approximation are fulfilled. This holds for the linear dispersion relation considered, which yields a smooth density of states, provided that retardation effects due to the finite speed of light $c$ can be neglected.
 The  condition may be expressed by demanding that the characteristic timescale of atom–photon dynamics, set by the spontaneous emission rate $\gamma$ of a single emitter into the bath, is much longer than the photon propagation time across the entire system, i.e., $\gamma \ll c/(Nd)$.  
Within this regime, photon-mediated interactions between emitters can be treated as instantaneous and the photonic modes can be adiabatically eliminated, leading to the  following Lindblad master equation governing the dynamics of the emitters density operator $\hat \rho(t)$~\cite{RevModPhys.95.015002,Chang2012,Carmichael1999,Breuer2007}:
\begin{equation}\label{eq:ME_linbl}
  \dot{\hat\rho}(t)=-i[\hat H_S,\hat\rho(t)]+ \sum_{ij}\Gamma_{ij}\left(2 \hat b_j\hat\rho \hat b_i^\dagger -\{\hat b_i^\dagger\hat b_j,\hat\rho\}\right)\,,
  \end{equation}
where 
\begin{equation}\label{eq:Hcoh}
    \hat{H}_{S} = \frac{\gamma}{2} \sum_{i,j} \sin{\left(k_0 |x_i - x_j|\right)} \, \hat{b}_{i}^{\dagger} \hat{b}_{j},
\end{equation}
describes the coherent collective dynamics of the emitters  and $k_0 = \omega_{eg}/c$ is the photon wavevector resonant with the $|g\rangle \leftrightarrow |e\rangle$ transition.
The second term in Eq.~\eqref{eq:ME_linbl} describes collective dissipation governed by the Hermitian dissipation matrix:
\begin{equation}\label{eq_diss_matrix}
    \Gamma_{ij} = \frac{\gamma}{2} \cos(k_0 |x_i - x_j|).
\end{equation}
To proceed, we rewrite the Lindblad master equation in Eq.~\eqref{eq:ME_linbl} in the form
\begin{equation}
    \label{eq: Master equation}
    \frac{d\hat{\rho}}{dt}=-i\Big{[}\hat{H}_{\rm eff}\hat{\rho}-\hat{\rho} \hat{H}_{\rm eff}^{\dagger} \Big{]}+2\sum_{i,j} \Gamma_{ij} \hat{b}_i\hat{\rho}\hat{b}_j^\dagger\,,
\end{equation}
where the anti-commutator part of the master equation has been included into the waveguide QED non-Hermitian Hamiltonian, defined as~\cite{RevModPhys.95.015002}:
\begin{equation}\label{eq:Heffwqed}
    \hat{H}_{\rm eff}=-i\frac{\gamma}{2}\sum_{i,j}e^{-ik_0|x_i-x_j|}\hat{b}_{i}^\dagger\hat{b}_j+\frac{U}{2}\sum_{i} \hat{b}_i^{\dagger}\hat{b}_i^{\dagger}\hat{b}_i\hat{b}_i\, ,
\end{equation}
where the last term accounts for the emitters nonlinearity.

This effective Hamiltonian fully describes photon-mediated interactions among the emitters within a fixed excitation sector, provided that external pumping fields are not included in the model. For each excitation sector the eigenvalues are of the form \begin{equation}\label{eq:eigenvalues}
E_{1D}^{(n_e)}(k) = \epsilon_{1D}^{(n_e)}(k) -i\gamma_{1D}^{(n_e)}(k)/2,
\end{equation} 
where $n_e$ is the number of excitations and 
$\epsilon^{(n_e)}_s$ and $\gamma^{(n_e)}_s$ denote the energy and collective decay rate of the $k$-th state, respectively.
In the single-excitation sector, the energy spectrum defines the 1D wQED dispersion relation~\cite{albrecht2019subradiant}
\begin{align}
\label{eq: 1D wQED Lieb HK elements}
        \epsilon_{1D}^{(1)}(k_i)&=\frac{\gamma}{4}\left [\mathrm{\cot}\left(\frac{(k_0+k_i)d}{2} \right)+\mathrm{\cot}\left(\frac{(k_0-k_i)d}{2} \right)   \right]\,,
\end{align}
while the decay rates are suppressed as $N^{-3}$ with increasing system size and vanish in the thermodynamic limit, where the emitter array has no edges~\cite{albrecht2019subradiant}. Consequently, the effective Hamiltonian~\eqref{eq:Heffwqed}, projected onto this excitation sector, becomes Hermitian in this limit. For higher excitations, this remains true, albeit with a different scaling of the decay rates~\cite{poddubny2020quasiflat,Molmer}, up to the limiting case of half-filling~\cite{poshakinskiy2021dimerization}.

\subsection{Two-dimensional waveguide QED network}

We now extend the previously developed formalism to treat an array of quantum emitters coupled to a square grid of one-dimensional, disconnected photonic waveguides, as illustrated in Fig.~\ref{fig:1}.
The emitters are arranged in a Lieb lattice~\cite{Leykam01012018}, composed of
two edge sites, A and C, each coupled via the $|g\rangle-|e\rangle$ transition exclusively to one vertical or horizontal waveguide, and one corner site, B, coupled via the same transition to two waveguides at their crossing point. 
We denote the lattice constant by $d$, while the intracell distance, assumed to be the same in both directions (A-B and B-C), is indicated by $a$. 
The one-dimensional waveguides mediate long-range interactions among the  emitters according to the network connectivity. 
Since the waveguides are independent, the master equation derived for a single waveguide~\eqref{eq: Master equation} can be straightforwardly generalized to the entire waveguide network, with each waveguide providing an additive contribution~\cite{PhysRevA.93.062104}.
The resulting  non-Hermitian Hamiltonian then reads:
\begin{align}
    \label{eq:Hamiltonian}
    \hat{H}_{\rm eff}&=-i\frac{\gamma}{2}\sum_{(i,j)\in \mathcal{N}}e^{ik_0|\mathbf{x}_i-\mathbf{x}_j|}\hat{b}_{i}^{\dagger}\hat{b}_{j} +\frac{U}{2}\sum_{i} \hat{b}_i^{\dagger}\hat{b}_i^{\dagger}\hat{b}_i\hat{b}_i\,,
\end{align}
where $\mathbf{x}_{i}$ represents the positions of the  atoms in the lattice.
The sum $(i,j)\in \mathcal{N}$ runs only over emitter pairs connected by the waveguide network $\mathcal{N}$, as shown in Fig.~\ref{fig:1}. 
Notice that, as each waveguide is treated as an independent photonic bath, the diagonal terms $i = j$, associated with individual spontaneous emission processes, are characterized by a rate $\gamma$ for the edge emitters A and C, and a rate $2\gamma$ for the corner sites B.
In the following analysis, we will consider an array in the plane with open boundary conditions and $N_c=N/3$ unit cells.

\section{Flat band and compact localized states}
\label{sec:flatband}

\subsection{Single particle dispersion}

In the thermodynamic limit of $N\rightarrow \infty$ the Hamiltonian~\eqref{eq:Hamiltonian} is right diagonalized by Bloch
waves, labelled by the wavevector $\mathbf{k}=(k_x,k_y)$,  \begin{equation}
    \label{eq: Bloch eigenstates}
    |\Psi_{n\mathbf{k}}\rangle=\frac{1}{\sqrt{N}}\sum_{\mathbf{R}}\sum_{\beta}e^{i\mathbf{k}\cdot \mathbf{R}}v^{(n)}_{\mathbf{k},\beta}|\mathbf{R},\beta\rangle,
\end{equation}
where $n=1,2,3$ is the band index, $\beta=A,B,C$ spans the atoms within a unit cell, $\mathbf{R}$ is the lattice vector running over all the unit cells and $v^{(n)}_{\mathbf{k},\beta}$ are the Bloch coefficients with the same periodicity of the lattice.
Using this ansatz we obtain the following Bloch Hamiltonian for each value of $\mathbf{k}$:
 \begin{equation}
     \label{eq: Bloch Hamiltonian of wQED network}
      \small
      {\mathcal{\hat{H}(\mathbf{k})}= \begin{pmatrix}
\epsilon_{1D}^{(1)}(k_x)+\epsilon_{1D}^{(1)}(k_y) & t\left(k_x\right)&  t\left(k_y\right) \\
t^*\left(k_x\right) & \epsilon_{1D}^{(1)}(k_x)  & 0 \\
t^*\left(k_y\right) & 0&\epsilon_{1D}^{(1)}(k_y)
\end{pmatrix}},
 \end{equation}
 where $\epsilon^{(1)}_{1D}(k_i)$, with  $i=x,y$, is the single excitation dispersion of a 1D atomic array coupled to a one-dimensional waveguide given in Eq.~\eqref{eq: 1D wQED Lieb HK elements}, while the off-diagonal terms $t(k_i)$ read:
 \begin{align}
 \begin{split}
    &t(k_i)=\epsilon_{1D}^{(1)}(k_i)\cos(k_0 a)+\frac{\gamma}{2}\sin(k_0a) \\
        &-i\frac{\gamma}{4}\sin(k_0 a)\left [\mathrm{\cot}\left(\frac{(k_0+k_i)d}{2} \right)
        -\mathrm{\cot}\left(\frac{(k_0-k_i)d}{2} \right)   \right]\,.
        \end{split}
       \end{align}

In order to achieve a flat band even in the presence of long-range interactions, we aim at finding the condition for the model to possess chiral symmetry $\mathcal{\hat{C}}$, \emph{i.e.} the Bloch Hamiltonian should transform under $\mathcal{\hat{C}}$ as $\mathcal{\hat{C}}\mathcal{\hat{H}}(\mathbf{k})\mathcal{\hat{C}} = -\mathcal{\hat{H}}(\mathbf{k})$, where $\mathcal{\hat{C}}$ is a unitary operator~\cite{Ramachandran2017}. 
This requirement is naturally satisfied in bipartite lattices. 
In our setting, to make the lattice bipartite, we must set the diagonal elements of the Bloch Hamiltonian~\eqref{eq: Bloch Hamiltonian of wQED network} to zero, namely the couplings between emitters belonging to the same sublattice have to vanish.
This can be achieved by exploiting the perfect periodicity of the plane-wave-mediated atom-atom interactions in Eq.~\eqref{eq:Hamiltonian}. 
By setting $k_0d = m\pi, m \in \mathbb{N}$, destructive interference completely suppresses coherent interactions between atoms of the same species. As a result, the emitters organize into a bipartite atomic lattice, where the majority sublattice consists of edge sites and the minority sublattice comprises corner sites.
Note that in spite of this choice the atom-atom interactions remain long-range.
With this arrangement we obtain
 $\epsilon_{1D}^{(1)}(k_i) = 0$, and
 \begin{equation}
    \label{eq: chiral 1D wQED network Lieb offdiagonal element}
    t(k_i)=\frac{\gamma}{2}\sin(k_0a)\left(1+i\tan\left(\frac{k_id}{2}\right)\right)\,.
\end{equation}
It is convenient to rewrite Hamiltonian~\eqref{eq: Bloch Hamiltonian of wQED network}  in terms of the Gell-Mann matrices 
\begin{equation}
\begin{split}
    \label{eq: write down Gell mann matrices}
&\lambda_1=\begin{pmatrix}
         0 & 1 & 0 \\1 & 0 & 0 \\0 & 0 & 0 \\
    \end{pmatrix}, \quad 
    \lambda_2=\begin{pmatrix}
         0 & -i & 0 \\i & 0 & 0 \\0 & 0 & 0 \\
    \end{pmatrix}, \\
        &\lambda_4=\begin{pmatrix}
         0 &0  & 1 \\0 & 0 & 0 \\1 & 0 & 0 \\
    \end{pmatrix}, \quad 
    \lambda_5=\begin{pmatrix}
         0 & 0 & -i \\0 & 0 & 0 \\i & 0 & 0 \\
    \end{pmatrix},
    \end{split}
\end{equation}
as:
\begin{equation}
    \label{eq: Hamiltonian expanded in Gell-Mann matrices}
    \begin{split}
            \hat{\mathcal{H}}(\mathbf{k})&=\frac{\gamma}{2}\sin(k_0 a)\Big{[}(\lambda_1+\lambda_4) \\ 
            &-\tan\left(\frac{k_x d}{2}\right)\lambda_2-\tan\left(\frac{k_y d}{2} \right)\lambda_5\Big{]}\,.
    \end{split}
\end{equation}
By diagonalizing the Bloch Hamiltonian in Eq.~\eqref{eq: Hamiltonian expanded in Gell-Mann matrices}, we obtain the single-excitation spectrum, $\epsilon^{(1)}(\mathbf{k})$, shown in Fig.~\ref{fig:1}(b). 
This spectrum consists of two polariton-like dispersive bands separated by a flat band with \mbox{$\epsilon^{(1)}(\mathbf{k}) = 0$}.
The presence of the gap at the $\Gamma$ point is due to the $(\lambda_1 + \lambda_4)$-term in the Hamiltonian $\hat{\mathcal{H}}(\mathbf{k})$. 
In the standard nearest-neighbor Lieb lattice model, this term appears, for example, when breaking inversion symmetry, while here originates from the long-range structure of the couplings. 
The two dispersive bands are separated from the flat-band by an energy gap $\Delta \epsilon = \frac{\gamma}{\sqrt{2}}|\sin(k_0a)|$. 
This gap is crucial for defining a complete basis of compact localized states (CLS), as we  discuss in the following. 
A large energy gap is also essential for observing flat-band-mediated photon-photon interactions, which will be explored in the next section. 
Therefore, throughout the paper, we maximize this gap to $\Delta \epsilon = \gamma/\sqrt{2}$ by setting $a = d/2$.
Finally, note that the dispersive bands diverge at the edge of the Brillouin zone, as typically occurs in 1D wQED systems~\cite{RevModPhys.95.015002}. 
These divergences are associated with superradiant modes at the resonant wavevector $|\mathbf{k}| = k_0$~\cite{albrecht2019subradiant,zhang2019theory,kumlin2020nonexponential}.

\subsection{Compact localized states}

To construct a compact localized state (CLS), we consider a general superposition of Bloch states in the flat band (band index $n = 2$), which is itself an eigenstate:
\begin{align}
    \label{eq: Superposition of FB eigenstates}
|\chi^{(\mathbf{R})}\rangle&=\sum_{\mathbf{k}\in BZ}\alpha_\mathbf{k} e^{-i\mathbf{k}\cdot \mathbf{R}}|\Psi_\mathbf{k}\rangle=\frac{1}{\sqrt{N}}\sum_{\mathbf{R}'}\sum_{\beta}A^{(\mathbf{R})}_{\mathbf{R}',\beta}|\mathbf{R}',\beta\rangle\,, 
\end{align}
with the coefficients 
\begin{align}
\label{eq: A_RR'beta expansion coeficient for CLS}
A^{(\mathbf{R})}_{\mathbf{R}',\beta}&=\sum_{\mathbf{k}\in BZ}\alpha_{\mathbf{k}}v_{\mathbf{k},\beta}e^{-i\mathbf{k}\cdot (\mathbf{R}-\mathbf{R}')}\,,
\end{align}
where we omitted the band index.
If there exists a set of coefficients $\alpha_{\mathbf{k}}$ such that $A^{(\mathbf{R})}_{{\mathbf{R}',\beta}}$ is nonzero only for a finite number of unit cells around $\mathbf{R}$, then $|\chi^{(\mathbf{R})}\rangle$ is called a compact localized state. This implies that each component $\alpha_\mathbf{k} v_{\mathbf{k},\beta}$ must be given by a finite sum of Bloch phases.
Sufficient conditions to fulfill this requirement  are the presence of a flat band and  finite-range hoppings in the Hamiltonian~\cite{Rhim2019, PhysRevB.95.115309, Ramachandran2017}. Importantly, while the latter condition is not met in our setting as it hosts long-range hoppings, it is still possible to find suitable coefficients $\alpha_\mathbf{k}$ such that $\alpha_\mathbf{k} v_{\mathbf{k},\beta}$ becomes a finite sum of Bloch phases (see App.~\ref{App.A}). This is possible again due to destructive interference of the plane-waves-mediated interactions in the network. 
A generic compact localized state  centered around the lattice position $\mathbf{R_0}$ is given by:
\begin{align}
\label{eq: Lieb wQED network CLS in real space}
\begin{split}
    |\chi^{\left(\mathbf{R_0}\right)}_{CLS}\rangle &=\hat c^\dagger_\mathbf{R_0}|0\rangle =\frac{1}{2} \Big{(} \hat b^\dagger_{\mathbf{R_0},A}+\hat b^\dagger_{\mathbf{R_0} - d\mathbf{\hat{y},A}}\\
    & -\hat b^\dagger_{\mathbf{R_0},C}-\hat b^\dagger_{\mathbf{R_0} - d\mathbf{\hat{x},C}}
\Big{)} |0\rangle 
    \end{split}
\end{align}
where we defined the operator $\hat{c}_{\mathbf{R}_0}^{\dagger}$, which creates a CLS centered at the lattice sitee $\mathbf{R}_0$, and
$d\mathbf{\hat{x}}$ ($d\mathbf{\hat{y}}$) indicates the unit lattice vector along $x$ ($y$). 
The state is localized within the four edge sites surrounding a corner emitter, as sketched in Fig.~\ref{fig:1}. 
By construction, this state remains perfectly dark even in finite-sized arrays, similar to the behavior of bound states in the continuum observed between two atoms separated by multiples of half the wavelength in 1D waveguide QED~\cite{longhi2007bound,PhysRevA.87.040103,PhysRevA.94.043839,PhysRevLett.126.063601}. 
Note that the structure of this CLS is markedly different from that of the nearest-neighbor Lieb model, where the CLS is localized within a plaquette. Once the form of the CLS is identified, it is possible to construct a complete (but not orthogonal) basis by translating the state by multiples of the lattice vectors until the entire flat band is covered~\cite{Rhim2019}.

\subsection{Quantum geometric tensor}

\begin{figure}[t!] 
\includegraphics[width=0.99\linewidth]{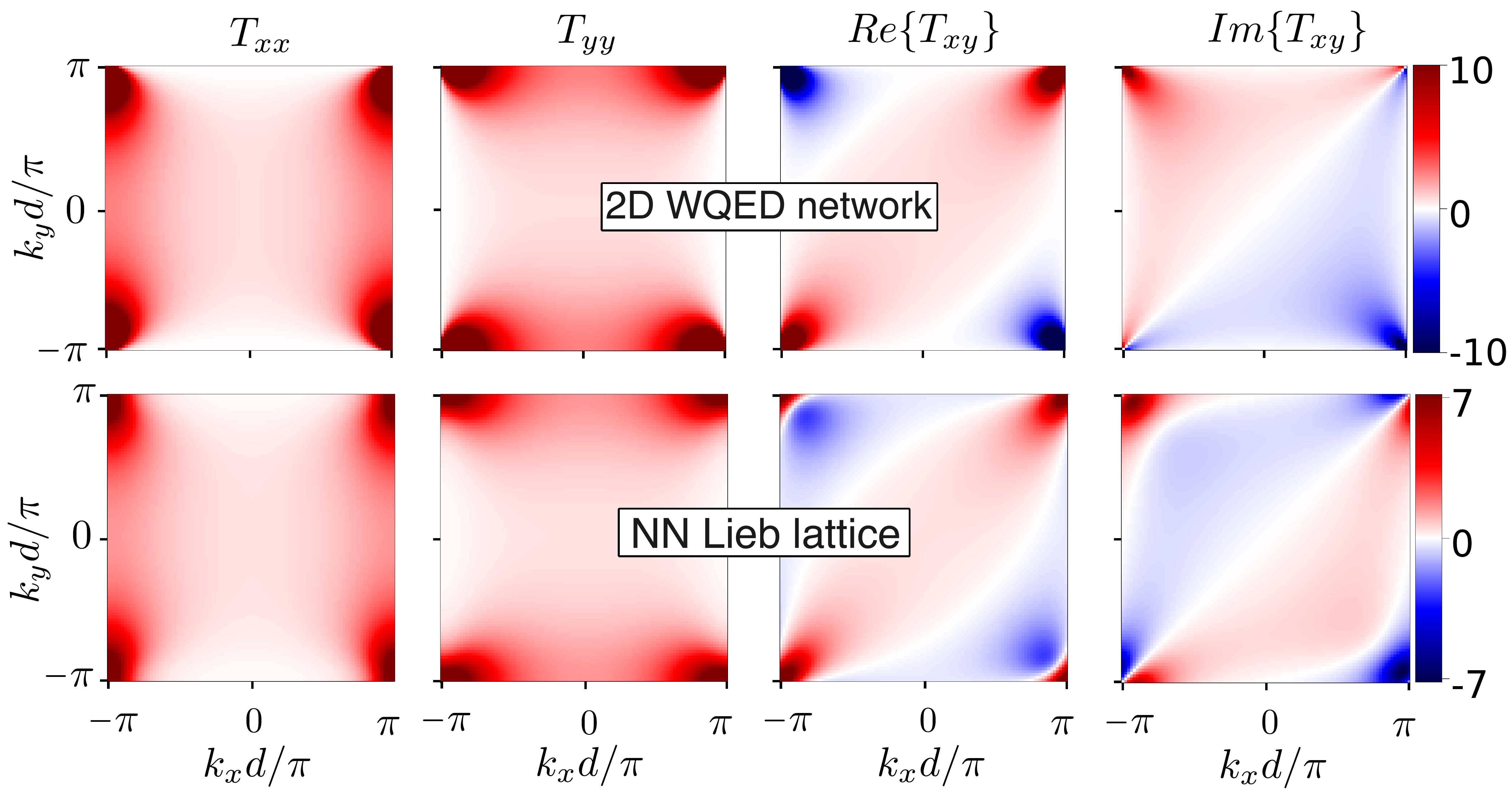}
    \caption{The components of the Quantum Geometric Tensor, as computed in Eq.~\eqref{eq: Calculating geometric tensor}, are shown as functions of the wavevector components $k_x$ and $k_y$. The first line corresponds to the result obtained for the Lieb lattice coupled to a two-dimensional waveguide network, with $k_0d=\pi$ and $a=d/2$, as discussed in the main text. To provide a direct comparison, we plotted in the second line  the result for the nearest-neighbor where the two intracell are set to $t_1=1.2$ (coupling between A and B sites) and $t_2=0.8$ (coupling between B and C sites) 
    } 
    \label{fig:QGT}
\end{figure}
To conclude our analysis of the single-particle properties of the system and to further support the nontrivial nature of the flat band, we compute the Quantum Geometric Tensor (QGT) \cite{Bleu2018,PhysRevResearch.6.L022020,PhysRevB.103.125302}, which characterizes the quantum geometry of each band, labeled by $n$:
\begin{equation}
    \label{eq: Calculating geometric tensor}
    T^{(n)}_{ij}(\mathbf{k})=\sum_{m\neq n }\frac{\langle v^{(m)}_{\mathbf{k}} |\partial_{k_i} \mathcal{H}(\mathbf{k}) |v^{(n)}_\mathbf{k}\rangle \langle v^{(n)}_{\mathbf{k}} |\partial_{k_j} \mathcal{H}(\mathbf{k}) |v^{(m)}_\mathbf{k}\rangle    }{\left(E_m-E_n\right)^2}\,,
\end{equation}
where $i,j\in \{x,y\}$. 
The imaginary part of the QGT is proportional to the Berry curvature $\Omega_{ij}(\mathbf{k})$ of a given band, which characterizes the system's topological properties. The real part instead is related to the quantum metric $g_{ij}(\mathbf{k})$, which plays a key role to establish superfluidity in flat-band systems~\cite{PhysRevLett.85.3906,peotta2015superfluidity}.

If we consider the Bloch Hamiltonian~\eqref{eq: Bloch Hamiltonian of wQED network} with chiral symmetry obtained when $k_0d = m\pi, m \in \mathbb{N}$, we obtain the following components of the  QGT tensor for the flat band (see App.~\ref{App.B}):
\begin{align}
    \label{eq: Elements of the QGT for the flatband for the general 2D chiral Hamiltonian}
   \begin{split}
       T_{xx}(k_x,k_y)&=\frac{\big{|}t_y\big{|}^2\big{|}\partial_{k_x}t_x\big{|}^2}{\Big{(}|t_x|^2+|t_y|^2\Big{)}^2}\in \mathbb{R}, \\
       T_{yy}(k_x,k_y)&=\frac{\big{|}t_x\big{|}^2\big{|}\partial_{k_y}t_y\big{|}^2}{\Big{(}|t_x|^2+|t_y|^2\Big{)}^2}\in \mathbb{R}, \\
       T_{xy}(k_x,k_y)&=-\frac{t_x^*(\partial_{k_x} t_x)\,t_y(\partial_{k_y}t_y)^*}{\Big{(}|t_x|^2+|t_y|^2\Big{)}^2}\,,
   \end{split} 
\end{align}
where the coefficients $t_i=t(k_i)$ are the ones given in Eq. \eqref{eq: chiral 1D wQED network Lieb offdiagonal element} and we omitted the flat band index. 
The flat band has non-vanishing Berry curvature, $\text{Im}(T_{xy})$, but due to time-reversal symmetry the Chern number is zero.
In contrast, the integral of the quantum metric, $\text{Re}(T_{xy})$, results in a finite value: 
\begin{equation}
     \label{eq: Geometric tensor integrals across BZ for wQED chiral network}
            \int_{BZ}\mathrm{Re}(T_{xy})\,d^2\mathbf{k}=-\frac{\pi}{2}\,.
        \end{equation}

Finally, in Fig.~\ref{fig:QGT}, we present the components of the quantum geometric tensor as a function of the quasi-momenta $\mathbf{k}$. 
While they share qualitative similarities with a nearest-neighbor Lieb lattice with different intracell hoppings 
\cite{Torma_PRB}, they exhibit distinct characteristics, such as a divergence at the edge of the Brillouin zone (arising from the divergence in the dispersion of the waveguide QED polariton branches) and a different sign distribution.

\section{Interaction induced transport}

\label{sec:interactions}

After demonstrating the emergence of an isolated, non-trivial flat band hosting compact localized states, we now focus on the two-excitation subspace.
While single particles in the flat band do not propagate due to their zero group velocity and infinite effective mass, interactions in higher excitation sectors can enable particle pairs to move by acquiring a finite effective mass. In waveguide QED platforms, effective photon-photon interactions arise from emitter saturation and are enhanced by light confinement. In this section, we show that these interactions can indeed lead to interaction-induced transport, and we separately discuss the two scenarios of softcore and hardcore photon interactions.

\begin{figure}[t!] 
\includegraphics[width=0.99\linewidth]{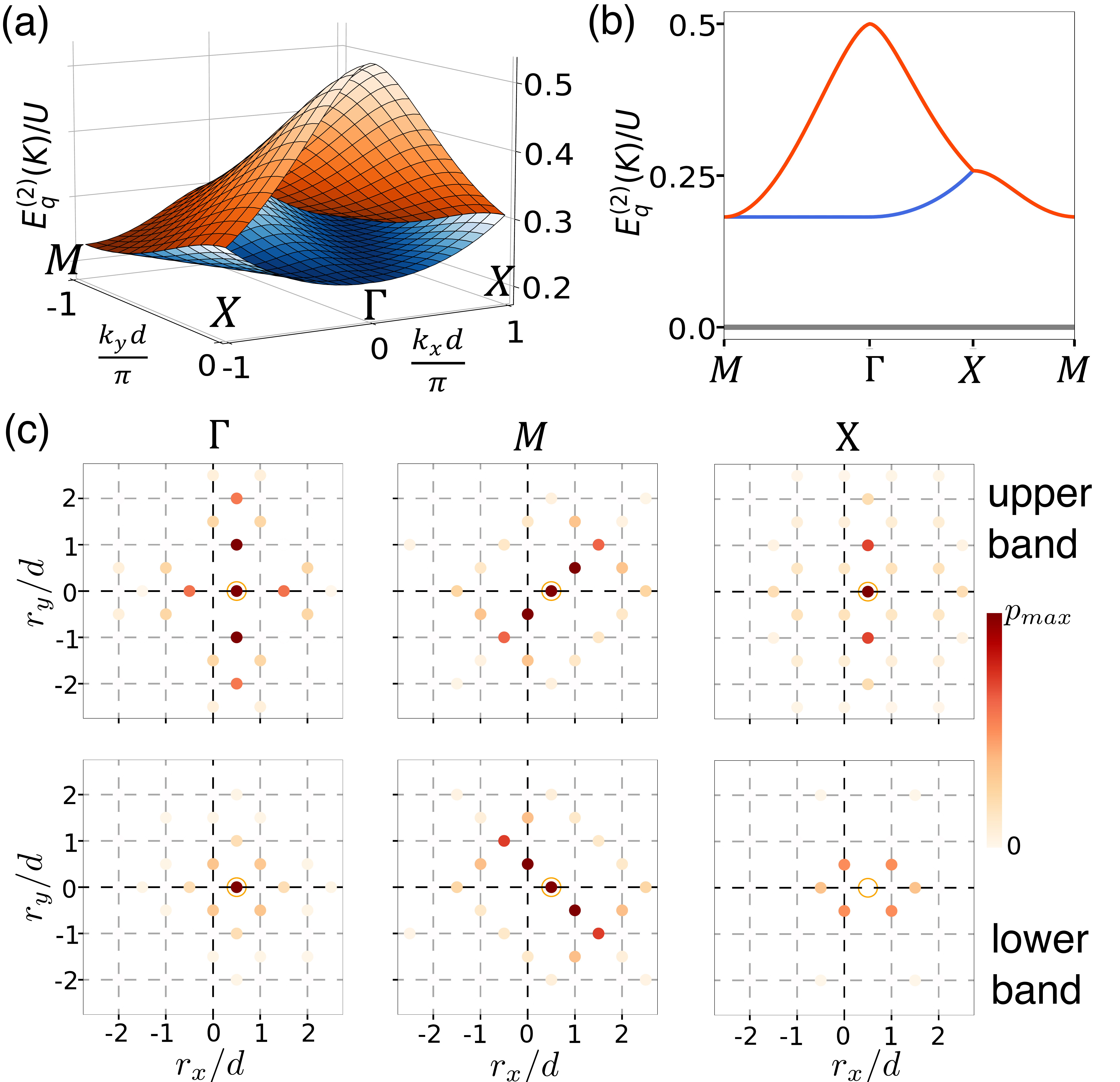}
    \caption{The two-excitation dispersion in the softcore interactions regime within the isolated-band approximation calculated by diagonalizing the matrix given in Eq.~\eqref{eq: interaction matrix} as a function of the center-of-mass momentum $\mathbf{K}$. Panel (a) displays half of the full 2D Brillouin zone ($k_y<0$) for better resolution, while panel (b) shows the dispersion along the symmetry lines of the Brillouin zone.
    (c) Relative coordinate population distribution of the bound states having the interaction-induced dispersive bands shown in panls (a)-(b) at the $M,\Gamma$ and $X$ symmetry-points. 
    The upper (lower) panel displays the upper (lower) branch of the bound states.
    The color map represents the probability distribution of finding the second excitation on different atoms in the lattice, given that the first excitation is located on the atom marked with the orange circle.
    The results were obtained for a system  of $30\times 30$ unit cells. } 
    \label{fig: two-excitation FB spectrum}
\end{figure}

\subsection{Softcore photon interactions}

\subsubsection{Two excitation spectrum}

We start by examining the softcore interaction regime, where the band separation is much larger than the emitter nonlinearity that governs effective photon-photon interactions, $U \ll \Delta \epsilon$. 
This allows us to use the isolated-band approximation~\cite{Torma_PRB}, which assumes that the interaction is sufficiently weak to prevent coupling between states from different bands, enabling us to project the two-excitation Hamiltonian onto the flat-band states alone.
Higher-excitation flat-band states can be constructed as product states of non-overlapping CLSs, which are exact zero-energy eigenstates of $\hat{H}_{\mathrm{eff}}$, independent of the value of $U$.
Specifically, the non-interacting two-excitation flat-band states can be expressed in terms of the center-of-mass (CM) coordinate, $\mathbf{R}=\frac{\mathbf{R}_1+\mathbf{R}_2}{2}$, and the relative coordinate, $\mathbf{r}=\mathbf{R}_1-\mathbf{R}_2$. We thus obtain:
\begin{align}
    \label{eq: flat band generic two-excitation state}
    \begin{split}
        &|\psi_{FB}^{(2)}(\mathbf{K},\mathbf{q})\rangle=\mathcal{N}^{(2)}\sum_{\mathbf{R,r} \atop \beta_1 \beta_2}v_{\frac{\mathbf{K}}{2}+\mathbf{q},\beta_1}v_{\frac{\mathbf{K}}{2}-\mathbf{q},\beta_2} \times \\
    &e^{i\left( \mathbf{K}\cdot \mathbf{R}+\mathbf{q}\cdot\mathbf{r}\right)} |\mathbf{R}-\frac{r}{2}\rangle |\beta_1\rangle \bigotimes |\mathbf{R}+\frac{r}{2}\rangle |\beta_2\rangle, \\
    \end{split}
\end{align}
where $\mathcal{N}^{(2)}$ is the states' norm while
$\mathbf{K}=\mathbf{k_1}+\mathbf{k_2}$ and $\mathbf{q}=\frac{k_1-k_2}{2}$ are, respectively, the center of mass momentum and relative
momentum.
To account for the effect of the interaction, we note that the interaction term in Eq.~\eqref{eq:Hamiltonian} is translationally invariant, and therefore the center-of-mass  momentum $\mathbf{K}$ is a good quantum number.
The two-excitation Hamiltonian projected onto the flat-band states~\eqref{eq: flat band generic two-excitation state} can then be diagonalized independently for each value of $\mathbf{K}$:
\begin{align}
    \label{eq: interaction matrix}
    \begin{split}
    \hat{V}_{\mathbf{q},\mathbf{q'}}(\mathbf{K})&=\langle\psi_{FB}(\mathbf{K},\mathbf{q'})|\hat{H}_{int}|\psi_{FB}(\mathbf{K,q})\rangle \\
    &=U\sum_{\beta} v_{-\mathbf{\frac{K}{2}-q'},\beta}v_{-\mathbf{\frac{K}{2}+q'},\beta}v_{\mathbf{\frac{K}{2}+q},\beta}v_{\mathbf{\frac{K}{2}-q},\beta}.
    \end{split}
\end{align}
In Fig.~\ref{fig: two-excitation FB spectrum}, we present the dispersion  $\epsilon^{(2)}(\mathbf{K})$
obtained by the diagonalization of Eq.~\eqref{eq: interaction matrix} as a function of the center-of-mass momentum $\mathbf{K}$.
The spectrum reveals $N-2$ degenerate zero-energy states for each $\mathbf{K}$, along with two dispersive branches that merge along the entire Brillouin zone boundary.
The presence of these dispersive branches underscores the interaction-induced mobility of initially localized flat-band states.
These states correspond to bound photon-photon pairs, as they appear within the energy gap formed by the degenerate flat band. Notice that these bound states obtained via the projected Hamiltonian are actually immersed in a continuum of states of the full Hamiltonian. Such a continuum originates from  the upper and lower single-particle polariton branches where each contributes to a single free photon to the two-excitation sector. We will show below that the bound states are however well-decoupled from such a continuum of modes.

Figure~\ref{fig: two-excitation FB spectrum}(c) visualizes the relative coordinate wave function populations for these bound states at high-symmetry points $\Gamma$, $M$, and $X$.
These states exhibit strong localization in their relative coordinates, as expected for bound states.

\subsubsection{Dynamics}
To probe the predicted interaction-induced transport and assess the validity of the isolated flat-band approximation a posteriori, we analyze the system dynamics by initializing it in a state consisting of two adjacent excited compact localized states,
\begin{equation}
\label{eq: FB initial state}
|\psi^{(2)}(t=0)\rangle = \hat{c}^{\dagger}_{\mathbf{R}0} \hat{c}^{\dagger}_{\mathbf{R}_0+d\mathbf{\hat{x}}} |0\rangle.
\end{equation}
Here the operator $\hat{c}_{\mathbf{R}_0}^{\dagger}$ generates a CLS centered at lattice site $\mathbf{R}_0$ as defined in Eq.~\eqref{eq: Lieb wQED network CLS in real space} and $|0\rangle$ is the vacuum state.
In the non-interacting regime ($U=0$) the state \eqref{eq: FB initial state} is an eigenstate of the system and thus remains unchanged during the evolution. 
To study the dynamics of this state in the presence of interactions, we  consider the effective Hamiltonian~\eqref{eq:Hamiltonian} projected onto the two-excitation subspace, $\hat{H}^{(2)}_{\text{eff}}$,
and we evolve the state under the Schr{\"o}dinger equation:
\begin{equation}
\label{eq: evolution FB in 2exc subspace}
i\frac{\partial |\psi^{(2)}(t)\rangle}{\partial t}=\hat{H}^{(2)}_{\text{eff}}|\psi^{(2)}(t)\rangle.
\end{equation}
The above equation captures the dynamics within the two-excitation subspace while simultaneously accounting for excitation losses, reflected as a decrease in the wave function's norm.
\begin{figure}
\includegraphics[width=0.99\linewidth]{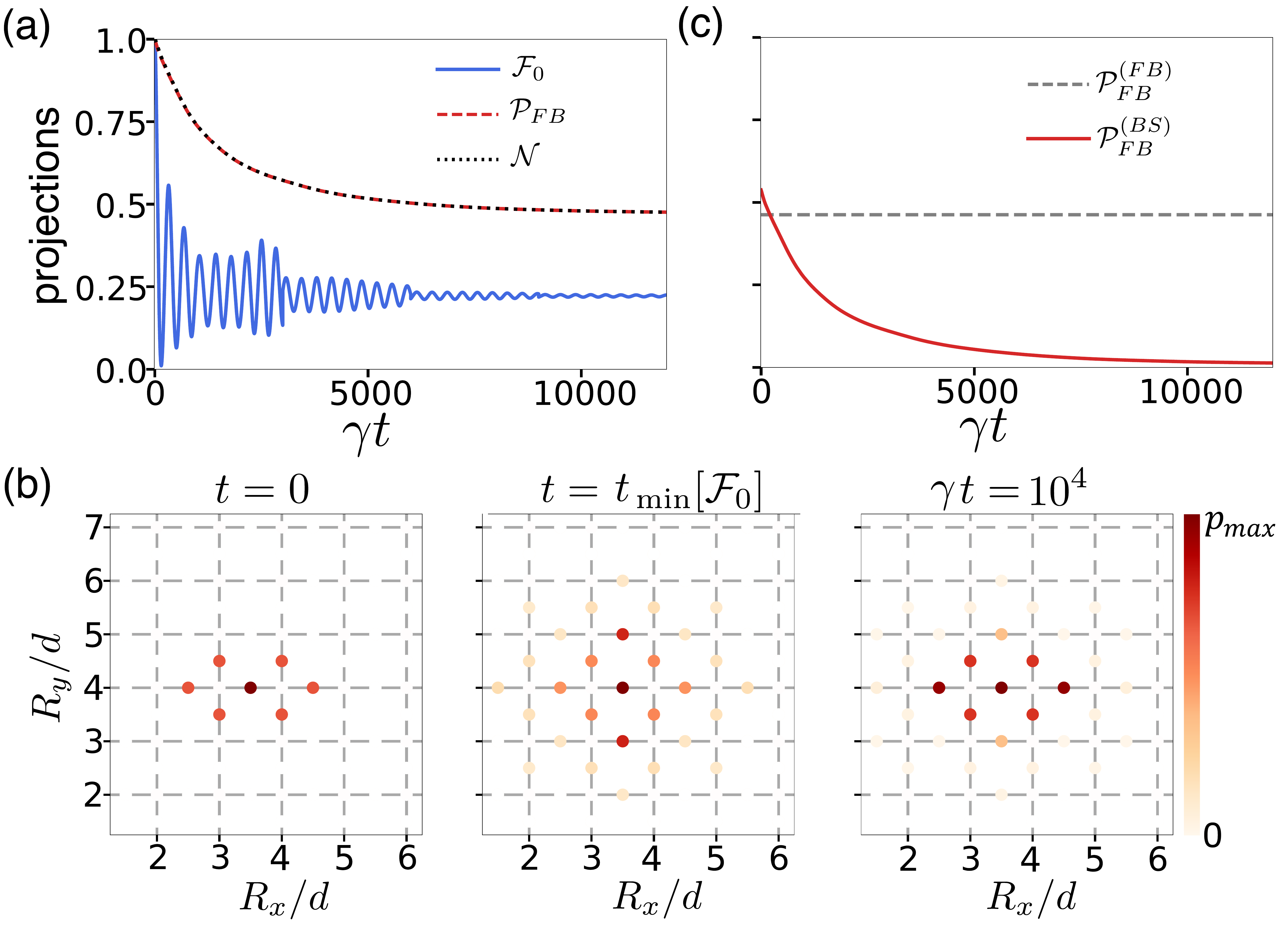}
    \caption{ Time evolution of the state defined in Eq.~\eqref{eq: FB initial state} in the softcore interactions regime with $U = 0.1 \gamma$. (a)
    Initial state fidelity $\mathcal{F}_0(t)$  (blue solid line), flat-band subspace projection $\mathcal{P}_{FB}$ (dashed red line) and  norm of the evolved state $\mathcal{N}(t)$ (black dotted line) as a function of time. (b) Snapshots of the excitation population evolution at times  $ t=0$ (left panel),  $t=t_{min}\left[\mathcal{F}_0\right]$ (center panel) and  $\gamma t=10^4$ (right panel), for two initially excited neighboring CLSs.
    The simulation was performed for  a system size  of $8\times 8$ unit cells.
    (c) Projection of the two-excitation state onto the interaction-induced dispersive bound states (red solid line) and onto the interacting non-dispersive states (gray dashed line) shown in Fig.~\ref{fig: two-excitation FB spectrum}. 
    }  
    \label{fig: FB excitation population in weak interaction limit}
\end{figure}
To quantify the deviation from the initial state, we define the initial-state fidelity as:
\begin{equation}
\label{eq: define FB initial fidelity}
\mathcal{F}_0(t) = |\langle \psi^{(2)}(t=0) | \psi^{(2)}(t) \rangle|^2\,,
\end{equation}
which is plotted in Fig.~\ref{fig: FB excitation population in weak interaction limit}(a) as a function of time.
The fidelity exhibits persistent oscillations with a frequency $\omega_0=\pi/t_{min}[\mathcal{F}_0]$, where $t_{min}[\mathcal{F}_0]$ is the time at which $\mathcal{F}_0(t)$ first reaches a minimum.  
This shows that the system evolves away from its initial configuration.
Snapshots of the excitation distribution across the lattice at different times are shown in Fig.~\ref{fig: FB excitation population in weak interaction limit}(b)  for  $U=0.1\gamma$.
The distribution is shown at three key times: the initial time, $t=0$, the intermediate time, $t=t_{min}\left[\mathcal{F}_0\right]$,  when the deviation of the evolved state from the initial state is maximal, and the late time, $t\gg (\gamma,U)$,  when the system relaxes toward a steady state.
This figure confirms that transport across the lattice occurs  in the presence of interaction.

To ensure that the observed dynamics is not due to coupling with dispersive bands, we analyze the projection of the evolving state onto the flat-band subspace defined as:
\begin{equation} 
\label{eq: projection to the flat band} \mathcal{P}_{FB}(t)=\langle \psi^{(2)}(t) |\hat{P}_{FB}|\psi^{(2)}(t)\rangle, \end{equation}
where  $\hat{P}_{FB}$ represents the projector onto the flat-band subspace of the non-interacting two-excitation Hamiltonian. The evolution of this quantity together with the  norm $\mathcal{N}(t)=|\langle \psi^{(2)}(t)|\psi^{(2)}(t)\rangle|^2$ of the two-excitation state is shown in Fig.~\ref{fig: FB excitation population in weak interaction limit}(a).
We observe that the projection  coincides with the norm, indicating that all the dynamics takes place entirely within manifold of two-particle states spanned by the single-particle flat-band shown in Fig.~\ref{fig:1}.
These results validate the isolated flat-band approximation used earlier and indicate that the observed transport originates from the previously identified dispersive bound states.
We will further confirm this below, when studying the dynamics of the system as a function of the interaction strength $U$.

Finally, we notice that after sufficiently long times the system settles into a steady state.
This behavior can be understood by analyzing the projections of the state $|\psi^{(2)}(t)\rangle$ onto two distinct subspaces: one consisting of interaction-induced dispersive bound states and another formed by the zero-energy (unbound) flat-band continuum  shown in Fig.~\ref{fig: two-excitation FB spectrum}(b). 
As depicted in Fig.~\ref{fig: FB excitation population in weak interaction limit}(c), the projection onto the interaction-induced dispersive states exhibits an exponential decay over time.
In contrast, the population within the flat-band subspace remains unchanged, since these states remain entirely dark even in the presence of interactions.
The system's evolution effectively halts when the contribution of the dispersive states decays completely. This occurs due to loss of propagating bound photon pairs at the system's edges.

In Ref.~\cite{Torma_PRB} a connection between isolated flat bands and the inverse mass tensor ($m_{ij}^{-1}$) of interacting low-energy bound states was shown to involve the quantum metric and, in particular, its integral via $m_{ij}^{-1}\sim \int \text d\mathbf{k} \, \text{Re}(T_{ij}(\mathbf k))$, which in our case we found to be finite (see Eq.~\eqref{eq: Geometric tensor integrals across BZ for wQED chiral network}). 
Whether this connection is also occurring here, and under what conditions, remains however an open question. 
In particular, we would like to highlight that Ref.~\cite{Torma_PRB} established this connection only for a single bound-state branch, despite the fact that multiple branches, each separated by an energy gap, occurred in the two-body spectrum. 
In our case, the multiple bound-state branches are energy intertwined, see Fig.~\ref{fig: two-excitation FB spectrum}(b), which implicitly introduces the notion of an emergent flavor for the bound pairs. 
A similar phenomenon was recently shown to occur at high energy for strongly-interacting nearest-neighbor bound pairs \cite{Salerno2020} based on the concept of line-graph lattice for their center of mass. 
This provided a clear geometrical explanation behind the origin of multi-band bound states branches. 
Here in the softcore interactions regime a less straightforward mechanism might be in place originating from flat band mediated correlations. 
This fact prevents a plain-vanilla employment of the reasonings discussed in  Ref.~\cite{Torma_PRB} to connect the bound states mass to the quantum metric, and possibly requiring a non-abelian/multi-branch treatment to include the bound state flavors into the treatment.

\begin{figure}[t!] 
\includegraphics[width=0.99\linewidth]{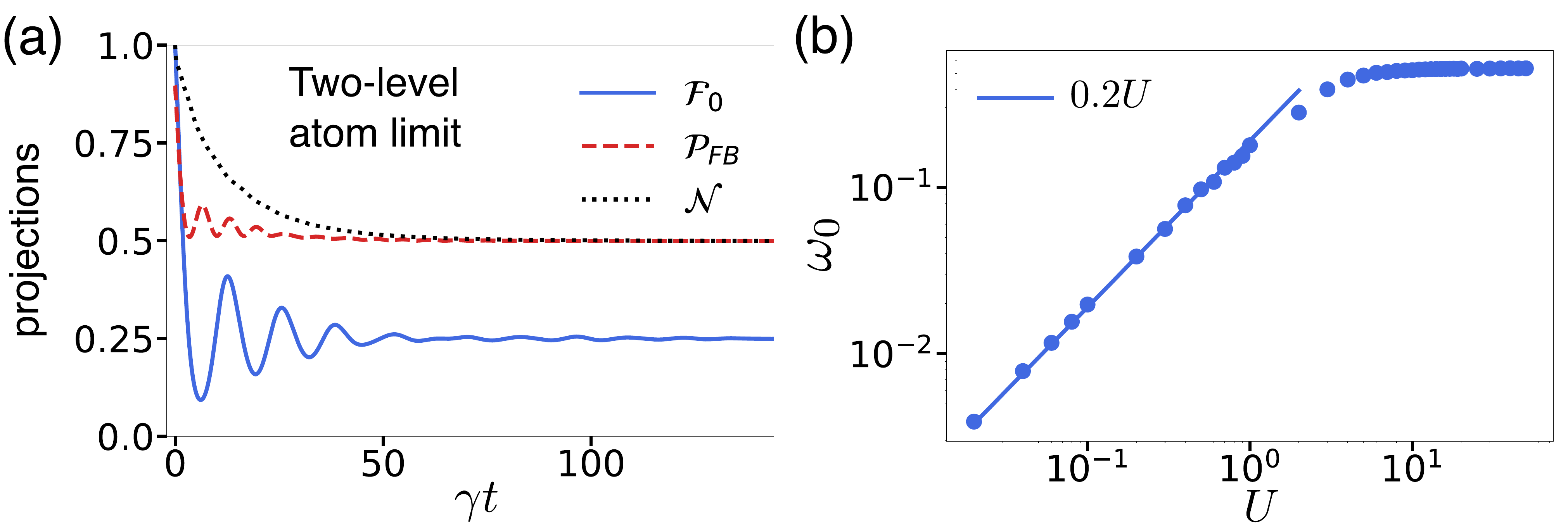}
    \caption{(a) Same as in Fig.~\ref{fig: FB excitation population in weak interaction limit}(a) in the hardcore interactions regime ($U\rightarrow\infty)$, for the initial condition defined in Eq.~\eqref{eq: initial condition FB hardcore bosons}. (b) Scaling of the fidelity oscillation frequency, $\omega_0$ for the initial condition defined in Eq.~\eqref{eq: initial condition FB hardcore bosons}, as a function of interaction strength $U$. The continuous line indicates the fit of the numerical data in the linear regime.
    }  
    \label{fig: dynamics analysis FB}
\end{figure}

\subsection{Two level atom limit}
Having examined the softcore interaction regime, we now turn our focus to the opposite limit of hardcore bosons ($U\rightarrow \infty$), which describes two level atoms coupled to a waveguide network.
In Fig.~\ref{fig: dynamics analysis FB}(a), we perform an analysis similar to that of the softcore interaction regime.
In this limit, where two-level atoms cannot be doubly excited,  the initial state must differ from that used previously. 
To maintain consistency,  we replace the bosonic creation operators in Eq.~\eqref{eq: Lieb wQED network CLS in real space} with spin operators. The considered initial state then reads:
\begin{equation}
\label{eq: initial condition FB hardcore bosons}
|\tilde{\psi}^{(2)}(t=0)\rangle=\hat{s}^{+}_{\mathbf{R}_0}\hat{s}^{+}_{\mathbf{R}_0+d\mathbf{\hat{x}}}|0\rangle,
\end{equation}
where we 
re-defined the CLS creation operator in terms of spin operators as:
\begin{align}
 \hat s^\dagger_\mathbf{R_0} =\frac{1}{2} \Big{(} \hat \sigma^+_{\mathbf{R_0},A}+\hat \sigma^+_{\mathbf{R_0} - d\mathbf{\hat{y},A}}-\hat \sigma^+_{\mathbf{R_0},C}-\hat \sigma^+_{\mathbf{R_0} - d\mathbf{\hat{x},C}}
\Big{)}.
\end{align}

The key distinction between this initial state and the one given in Eq.~\eqref{eq: FB initial state} is that, in this case, the edge atom $A$ of unit cell $\mathbf{R}_0$ cannot  be doubly occupied.
Consequently, the initial projection onto the flat-band subspace is not unity but slightly lower, ${\mathcal{P}}_{FB}(t=0) = 0.9$.

The transport behavior, as measured by $\mathcal{F}_0$, exhibits qualitative similarities to the softcore interacting boson case: an initial rapid decrease followed by the relaxation into a steady-state flat-band states.
However, a key distinction is that while transport in the softcore interactions regime depends on $U$, in this scenario, the timescale is  governed only by the decay rate $\gamma$. 
The observed mismatch between the state norm, $\mathcal{N}(t)$, and flat-band projection, $\mathcal{P}_{FB}(t)$, suggests that the unbound dispersive bands of Fig.~\ref{fig:1} play a dominant role in the transient dynamics before the system reaches a steady state. 
This process involves a coherent energy exchange between the dispersive and flat bands, as indicated by the oscillatory behavior of $\mathcal{P}_{FB}(t)$, which can be interpreted as the result of a metastable exciton-like dressed states between the two bands.
Ultimately, this coupling to the dispersive bands is what enables transport and eventually leads to excitation loss in this regime.

\subsection{Softcore-to-hardcore interaction tradeoff}

The transition between the softcore interaction regime and the hardcore boson limit is illustrated in Fig.~\ref{fig: dynamics analysis FB}(b), which shows the dependence of the fidelity oscillation frequency $\omega_0$ on the interaction strength $U$. 
For small $U$, $\omega_0$ exhibits a linear scaling with $U$, a hallmark of flat-band-mediated interactions further confirming our projection theory analysis, as discussed in Ref.~\cite{Torma_PRB}, where the group velocity and the
inverse of the effective mass scale were shown to scale linearly with the interaction strength.
When $U \sim \gamma$, $\omega_0$ smoothly transitions to the hardcore boson limit, where the oscillation frequency saturates and  no longer depends on $U$. In this regime this oscillation frequency is ruled by the flat-dispersive band-gap $\Delta \epsilon$, which sets the characteristic timescale of the dynamics.
Note that although the system's evolution in this plot has been considered for the initial state given in Eq.~\eqref{eq: initial condition FB hardcore bosons} to capture the saturation in the hardcore regime, the same behavior is observed in the softcore regime when the system is initialized with the state given in Eq.~\eqref{eq: FB initial state}.

\section{Conclusion}

\label{sec:conclusions}

In this paper, we have shown that in a waveguide QED system consisting of a 2D Lieb-like array of quantum emitters coupled to a network of 1D waveguides, infinitely long-range interactions can give rise to an energetically isolated flat band. 
We prove that this band is nontrivial, possessing nonvanishing geometric tensor components and hosting perfectly subradiant compact localized states.
We then demonstrate that for two excitations, emitter nonlinearity can induce strong effective photon-photon interactions, leading to interaction-induced transport of bound photon pairs. 
In the softcore interactions regime, propagation occurs at a rate that scales linearly with the effective interaction strength, mirroring predictions for fermionic systems~\cite{Torma_PRB}. 
This raises the intriguing question whether in the half-filling regime~\cite{PhysRevB.95.115135} this platform could host superfluid states~\cite{Torma_PRL,Ramachandran2017,Rhim2019,PhysRevB.95.115309}.
The ability to couple superconducting qubits to transmission lines in various architectures   with negligible coupling to other dissipation channels~\cite{brehm2021waveguide,mirhosseini2019cavity,kannan2023demand,scigliuzzo2022controlling,shah2024stabilizing} makes the microwave regime a promising platform for experimentally probing our findings. 
Additionally, the tunability of the emitters nonlinearity would enable the exploration of the different interacting regimes discussed in this work. 
The main limitation in this setting arises from the possible breakdown of the Markov approximation, which imposes an upper bound on the atom-field coupling,  the lattice spacing and the number of emitters~\cite{ferreira2024deterministic}. The former can be experimentally controlled by ensuring that the coupling frequency remains smaller than the bandwidth of the transmission line. The latter can be controlled by fixing the emitter spacing to $\lambda/2$, as required in our proposal. For experimental parameters similar to those in Ref.~\cite{shah2024stabilizing}, this approximation is expected to hold except for very large arrays on the order of several hundred emitters.
Alternatively, our results could be probed in the optical regime by interfacing waveguide networks with cold or artificial atoms~\cite{lodahl2015interfacing,hood2016atom,corzo2019waveguide,tiranov2023collective,prasad2020correlating}, or, in the single-excitation linear regime, by utilizing other photonic modes ~\cite{crespi2019experimental}. In this scenario, the main challenges concern trapping atoms near complex photonic structures and the presence of decay into radiative modes other than the waveguide.

Finally, an intriguing open question is about the impact of long-range couplings either in topological bands or in generating purely interacting topological bound states \cite{Salerno2020}.
 
 \section{Acknowledgments}
The authors thank Francesco Ciccarello,
Enrico Di Benedetto, Simone Montangero, Sebastiano Peotta,
Federico Roccati, Filippo Romanato and Pietro Silvi  for fruitful discussions.
This work was supported by the EU QuantERA2021 project T-NiSQ, the  Quantum Technology Flagship project PASQuanS2,   the  Quantum Computing and Simulation Center of Padova University, the INFN project Iniziativa Specifica IS-Quantum
and by the Italian Ministry of University and Research via the Excellence grant 2023-2027 ``Quantum Frontiers",  the Rita Levi-Montalcini program and the Fondazione Cassa di Risparmio di Padova e Rovigo" (Fondazione Cariparo).
The authors also acknowledge computational resources by Cloud Veneto.

\appendix

\section{Compact localized states derivation}\label{App.A}
In this appendix, we derive the compact localized states in waveguide QED network defined in~\eqref{eq: Lieb wQED network CLS in real space}. 
The Bloch eigenstates of the flat-band are given by:
\begin{equation}
    \label{eq: bloch state FB}
    v^{(0)}_\mathbf{k}=\frac{1}{\sqrt{|t_x|^2+|t_y|^2}}\begin{pmatrix}
        0\\-t_y\\t_x
    \end{pmatrix},
\end{equation}
which have  support only on A and C sites.
By rewriting the off-diagonal terms of the Bloch Hamiltonian~\eqref{eq: chiral 1D wQED network Lieb offdiagonal element}  as:
\begin{equation}
    \label{eq: rewrite 1+itan}
    \begin{split}
    1+i\tan\left(\frac{k_i d}{2}\right)&=\frac{\cos\left(\frac{k_i d}{2}\right)+i\sin\left(\frac{k_i d}{2}\right)}{\cos{\left(\frac{k_i d}{2}\right)}}\\
            &=2\frac{e^{i\frac{k_i d}{2}}}{e^{i\frac{k_i d}{2}}+e^{i\frac{k_i d}{2}}}=\frac{2}{1+e^{-ik_i d}}\,,
    \end{split}
\end{equation}
 the flat band Bloch eigenstates~\eqref{eq: bloch state FB} read:
\begin{equation}
    \label{eq: rewrite bloch FB eigenstate}
    v^{(0)}_\mathbf{k}=\frac{1}{\sqrt{|t_x|^2+|t_y|^2}} \begin{pmatrix}
0 \\
-\frac{1}{1+e^{-ik_yd}} \\
\frac{1}{1+e^{-ik_xd}} 
 \end{pmatrix}.
\end{equation}
To construct a compact localized state, we need to determine the coefficients $\alpha_{\mathbf{k}}$ such that $\alpha_\mathbf{k} v_{\mathbf{k},\beta}^{(0)}$ can be expressed as a finite sum of Bloch phases.
This condition is satisfied by selecting:
\begin{equation}
    \alpha_{\mathbf{k}}=\left(1+e^{-ik_yd}\right) \left(1+e^{-ik_xd} \right)\sqrt{|t_x|^2+|t_y|^2}.
\end{equation}
By this choice, $\alpha_{\mathbf{k}}v^{(0)}_\mathbf{k}$ is given by the sum of Bloch phases:
\begin{equation}
    \label{eq: alphak vk for 1D qQED Lieb network}
    \alpha_{\mathbf{k}}v^{(0)}_\mathbf{k}=\begin{pmatrix}
0 \\
-\left(1+e^{-ik_xd}\right)  \\
1+e^{-ik_yd}
 \end{pmatrix}.
\end{equation}

By performing the discrete Fourier transform eq.~\eqref{eq: A_RR'beta expansion coeficient for CLS}, we obtain the compact localized state in real space located around $\mathbf{R^{(0)}}$ 
\begin{align}
    \label{eq: Obtainin CLS Lieb wQED network}
    \begin{split}
A^{\left(\mathbf{R^{(0)}}\right)}_{\mathbf{R},A}&=-\sum_{k_x} e^{-ik_x\left(R_x-R^{(0)}_x\right) }\left(1+e^{-ik_xd}\right)\sum_{-k_y}e^{ik_y\left(R_y-R^{(0)}_y\right) }\\
&=\delta_{R_y,R_y^{(0)}}\left(\delta_{R_x,R_x^{(0)}}+\delta_{R_x,R_x^{(0)}-d} \right), \\
    A^{\left(\mathbf{R^{(0)}}\right)}_{\mathbf{R},C}&=\sum_{k_y} e^{-ik_y\left(R_y-R^{(0)}_y\right) }\left(1+e^{-ik_yd}\right)\sum_{k_x}e^{-ik_x\left(R_x-R^{(0)}_x\right) }\\
&=\delta_{R_x,R_x^{(0)}}\left(\delta_{R_y,R_y^{(0)}}+\delta_{R_y,R_y^{(0)}-d} \right).
    \end{split}
\end{align}
Finally, substituting the above result into Eq.~\eqref{eq: Superposition of FB eigenstates}, we obtain a compact localized state  in real space, confined to three unit cells surrounding the cell $\mathbf{R}_0$:
\begin{align}
\begin{split}
|\chi^{\left(\mathbf{R_0}\right)}_{CLS}\rangle &= \frac{1}{2} \Big{(} |\mathbf{R_0},A\rangle + |\mathbf{R_0} - d\mathbf{\hat{y}},A\rangle -\\&|\mathbf{R_0},C\rangle - |\mathbf{R_0} - d\mathbf{\hat{x}},C\rangle\Big{)} 
    \end{split}
\end{align}

\section{Quantum geometric tensor calculation}\label{App.B}

Let us calculate the QGT for a 2D bipartite (chiral) Hamiltonian  of the general form:
\begin{equation}
    \label{eq: General 2D chiral Hamiltonian in k space}
    \mathcal{H}(\mathbf{k})= \begin{pmatrix}
0 & t_x &t_y \\
t_x^* & 0  & 0 \\
t_y^*& 0&0
\end{pmatrix}.
\end{equation}
The eigenenergies of the system for the three bands are given by:
\begin{align}
    \label{eq: eigenenergies of the general 2D chiral Hamiltonian in k space}
    \begin{split}
        E_0&=0, \\
        E_{\pm}(k_x,k_y)&=\pm E=\pm\sqrt{|t_x|^2+|t_y|^2},
    \end{split}
\end{align}
while the corresponding eigenvectors given by:
\begin{equation}
    \label{eq: eigenvectors of the general 2D chiral Hamiltonian in k space}
v^{(0)}_\mathbf{k}=\mathcal{N}^{(0)}_\mathbf{k}\begin{pmatrix}
        0\\-t_y\\t_x
    \end{pmatrix}, \quad v^{(\pm)}_\mathbf{k}=\mathcal{N}^{(\pm)}_\mathbf{k}\begin{pmatrix}
        \pm E\\t_x^*\\t_y^*
    \end{pmatrix}\,,
\end{equation}
where $\mathcal{N}^{(0)}_{\mathbf{k}}=\left(|t_y|^2+|t_x|^2\right)^{-\frac{1}{2}}$ and $\mathcal{N}^{(\pm)}_\mathbf{k}=\left(|t_y|^2+|t_x|^2+E^2\right)^{-\frac{1}{2}}$ are the eigenstate norms.
By using Eq.~\eqref{eq: Calculating geometric tensor}, we obtain the QGT flat band components 
given in Eq.~\eqref{eq: Elements of the QGT for the flatband for the general 2D chiral Hamiltonian}.
Plugging in the expression  for the coefficients $t_i=t(k_i)$ given in Eq.~\eqref{eq: chiral 1D wQED network Lieb offdiagonal element} into \eqref{eq: Elements of the QGT for the flatband for the general 2D chiral Hamiltonian} we finally obtain:

\begin{align}
    \label{eq: Elements of the QGT for the flatband for the wQED 2D network chiral Hamiltonian}
   \begin{split}
       T_{xx}(k_x,k_y)&=\frac{d^2}{4}\frac{\cos^2\left(\frac{k_yd}{2}\right)}{\left( \cos^2\left(\frac{k_xd}{2}\right)+\cos^2\left(\frac{k_yd}{2} \right) \right)^2},\\
       T_{yy}(k_x,k_y)&=\frac{d^2}{4}\frac{\cos^2\left(\frac{k_xd}{2}\right)}{\left( \cos^2\left(\frac{k_xd}{2}\right)+\cos^2\left(\frac{k_yd}{2} \right) \right)^2},\\
       T_{xy}(k_x,k_y)&=-\frac{d^2}{4}\frac{\cos\left(\frac{k_xd}{2}\right)\cos\left(\frac{k_yd}{2}\right)}{\left( \cos^2\left(\frac{k_xd}{2}\right)+\cos^2\left(\frac{k_yd}{2} \right) \right)^2}\times\\
       &\left(\cos\left(\frac{(k_y-k_x)d}{2}\right) +i\sin\left(\frac{(k_y-k_x)d}{2} \right) \right).
   \end{split} 
\end{align}

\bibliography{refs}

\begin{thebibliography}{113}%
\makeatletter
\providecommand \@ifxundefined [1]{%
 \@ifx{#1\undefined}
}%
\providecommand \@ifnum [1]{%
 \ifnum #1\expandafter \@firstoftwo
 \else \expandafter \@secondoftwo
 \fi
}%
\providecommand \@ifx [1]{%
 \ifx #1\expandafter \@firstoftwo
 \else \expandafter \@secondoftwo
 \fi
}%
\providecommand \natexlab [1]{#1}%
\providecommand \enquote  [1]{``#1''}%
\providecommand \bibnamefont  [1]{#1}%
\providecommand \bibfnamefont [1]{#1}%
\providecommand \citenamefont [1]{#1}%
\providecommand \href@noop [0]{\@secondoftwo}%
\providecommand \href [0]{\begingroup \@sanitize@url \@href}%
\providecommand \@href[1]{\@@startlink{#1}\@@href}%
\providecommand \@@href[1]{\endgroup#1\@@endlink}%
\providecommand \@sanitize@url [0]{\catcode `\\12\catcode `\$12\catcode
  `\&12\catcode `\#12\catcode `\^12\catcode `\_12\catcode `\%12\relax}%
\providecommand \@@startlink[1]{}%
\providecommand \@@endlink[0]{}%
\providecommand \url  [0]{\begingroup\@sanitize@url \@url }%
\providecommand \@url [1]{\endgroup\@href {#1}{\urlprefix }}%
\providecommand \urlprefix  [0]{URL }%
\providecommand \Eprint [0]{\href }%
\providecommand \doibase [0]{https://doi.org/}%
\providecommand \selectlanguage [0]{\@gobble}%
\providecommand \bibinfo  [0]{\@secondoftwo}%
\providecommand \bibfield  [0]{\@secondoftwo}%
\providecommand \translation [1]{[#1]}%
\providecommand \BibitemOpen [0]{}%
\providecommand \bibitemStop [0]{}%
\providecommand \bibitemNoStop [0]{.\EOS\space}%
\providecommand \EOS [0]{\spacefactor3000\relax}%
\providecommand \BibitemShut  [1]{\csname bibitem#1\endcsname}%
\let\auto@bib@innerbib\@empty
\bibitem [{\citenamefont {Derzhko}\ \emph {et~al.}(2015)\citenamefont
  {Derzhko}, \citenamefont {Richter},\ and\ \citenamefont
  {Maksymenko}}]{doi:10.1142/S0217979215300078}%
  \BibitemOpen
  \bibfield  {author} {\bibinfo {author} {\bibfnamefont {O.}~\bibnamefont
  {Derzhko}}, \bibinfo {author} {\bibfnamefont {J.}~\bibnamefont {Richter}},\
  and\ \bibinfo {author} {\bibfnamefont {M.}~\bibnamefont {Maksymenko}},\
  }\bibfield  {title} {\bibinfo {title} {Strongly correlated flat-band systems:
  The route from heisenberg spins to hubbard electrons},\ }\href
  {https://doi.org/10.1142/S0217979215300078} {\bibfield  {journal} {\bibinfo
  {journal} {International Journal of Modern Physics B}\ }\textbf {\bibinfo
  {volume} {29}},\ \bibinfo {pages} {1530007} (\bibinfo {year} {2015})},\
  \Eprint {https://arxiv.org/abs/https://doi.org/10.1142/S0217979215300078}
  {https://doi.org/10.1142/S0217979215300078} \BibitemShut {NoStop}%
\bibitem [{\citenamefont {Laughlin}(1983)}]{Laughlin1983}%
  \BibitemOpen
  \bibfield  {author} {\bibinfo {author} {\bibfnamefont {R.~B.}\ \bibnamefont
  {Laughlin}},\ }\bibfield  {title} {\bibinfo {title} {Anomalous quantum hall
  effect: An incompressible quantum fluid with fractionally charged
  excitations},\ }\href {https://doi.org/10.1103/PhysRevLett.50.1395}
  {\bibfield  {journal} {\bibinfo  {journal} {Phys. Rev. Lett.}\ }\textbf
  {\bibinfo {volume} {50}},\ \bibinfo {pages} {1395} (\bibinfo {year}
  {1983})}\BibitemShut {NoStop}%
\bibitem [{\citenamefont {Cao}\ \emph {et~al.}(2018{\natexlab{a}})\citenamefont
  {Cao}, \citenamefont {Fatemi}, \citenamefont {Fang}, \citenamefont
  {Watanabe}, \citenamefont {Taniguchi}, \citenamefont {Kaxiras},\ and\
  \citenamefont {Jarillo-Herrero}}]{Cao2018}%
  \BibitemOpen
  \bibfield  {author} {\bibinfo {author} {\bibfnamefont {Y.}~\bibnamefont
  {Cao}}, \bibinfo {author} {\bibfnamefont {V.}~\bibnamefont {Fatemi}},
  \bibinfo {author} {\bibfnamefont {S.}~\bibnamefont {Fang}}, \bibinfo {author}
  {\bibfnamefont {K.}~\bibnamefont {Watanabe}}, \bibinfo {author}
  {\bibfnamefont {T.}~\bibnamefont {Taniguchi}}, \bibinfo {author}
  {\bibfnamefont {E.}~\bibnamefont {Kaxiras}},\ and\ \bibinfo {author}
  {\bibfnamefont {P.}~\bibnamefont {Jarillo-Herrero}},\ }\bibfield  {title}
  {\bibinfo {title} {Unconventional superconductivity in magic-angle graphene
  superlattices},\ }\href {https://doi.org/10.1038/nature26160} {\bibfield
  {journal} {\bibinfo  {journal} {Nature}\ }\textbf {\bibinfo {volume} {556}},\
  \bibinfo {pages} {43} (\bibinfo {year} {2018}{\natexlab{a}})}\BibitemShut
  {NoStop}%
\bibitem [{\citenamefont {Cao}\ \emph {et~al.}(2018{\natexlab{b}})\citenamefont
  {Cao}, \citenamefont {Fatemi}, \citenamefont {Demir}, \citenamefont {Fang},
  \citenamefont {Tomarken}, \citenamefont {Luo}, \citenamefont
  {Sanchez-Yamagishi}, \citenamefont {Watanabe}, \citenamefont {Taniguchi},
  \citenamefont {Kaxiras}, \citenamefont {Ashoori},\ and\ \citenamefont
  {Jarillo-Herrero}}]{Cao2018b}%
  \BibitemOpen
  \bibfield  {author} {\bibinfo {author} {\bibfnamefont {Y.}~\bibnamefont
  {Cao}}, \bibinfo {author} {\bibfnamefont {V.}~\bibnamefont {Fatemi}},
  \bibinfo {author} {\bibfnamefont {A.}~\bibnamefont {Demir}}, \bibinfo
  {author} {\bibfnamefont {S.}~\bibnamefont {Fang}}, \bibinfo {author}
  {\bibfnamefont {S.~L.}\ \bibnamefont {Tomarken}}, \bibinfo {author}
  {\bibfnamefont {J.~Y.}\ \bibnamefont {Luo}}, \bibinfo {author} {\bibfnamefont
  {J.~D.}\ \bibnamefont {Sanchez-Yamagishi}}, \bibinfo {author} {\bibfnamefont
  {K.}~\bibnamefont {Watanabe}}, \bibinfo {author} {\bibfnamefont
  {T.}~\bibnamefont {Taniguchi}}, \bibinfo {author} {\bibfnamefont
  {E.}~\bibnamefont {Kaxiras}}, \bibinfo {author} {\bibfnamefont {R.~C.}\
  \bibnamefont {Ashoori}},\ and\ \bibinfo {author} {\bibfnamefont
  {P.}~\bibnamefont {Jarillo-Herrero}},\ }\bibfield  {title} {\bibinfo {title}
  {Correlated insulator behaviour at half-filling in magic-angle graphene
  superlattices},\ }\href {https://doi.org/10.1038/nature26154} {\bibfield
  {journal} {\bibinfo  {journal} {Nature}\ }\textbf {\bibinfo {volume} {556}},\
  \bibinfo {pages} {80} (\bibinfo {year} {2018}{\natexlab{b}})}\BibitemShut
  {NoStop}%
\bibitem [{\citenamefont {Daniel~Leykam}\ and\ \citenamefont
  {Flach}(2018)}]{Leykam01012018}%
  \BibitemOpen
  \bibfield  {author} {\bibinfo {author} {\bibfnamefont {A.~A.}\ \bibnamefont
  {Daniel~Leykam}}\ and\ \bibinfo {author} {\bibfnamefont {S.}~\bibnamefont
  {Flach}},\ }\bibfield  {title} {\bibinfo {title} {Artificial flat band
  systems: from lattice models to experiments},\ }\href
  {https://doi.org/10.1080/23746149.2018.1473052} {\bibfield  {journal}
  {\bibinfo  {journal} {Advances in Physics: X}\ }\textbf {\bibinfo {volume}
  {3}},\ \bibinfo {pages} {1473052} (\bibinfo {year} {2018})},\ \Eprint
  {https://arxiv.org/abs/https://doi.org/10.1080/23746149.2018.1473052}
  {https://doi.org/10.1080/23746149.2018.1473052} \BibitemShut {NoStop}%
\bibitem [{\citenamefont {Danieli}\ \emph {et~al.}(2024)\citenamefont
  {Danieli}, \citenamefont {Andreanov}, \citenamefont {Leykam},\ and\
  \citenamefont {Flach}}]{danieli2024flat}%
  \BibitemOpen
  \bibfield  {author} {\bibinfo {author} {\bibfnamefont {C.}~\bibnamefont
  {Danieli}}, \bibinfo {author} {\bibfnamefont {A.}~\bibnamefont {Andreanov}},
  \bibinfo {author} {\bibfnamefont {D.}~\bibnamefont {Leykam}},\ and\ \bibinfo
  {author} {\bibfnamefont {S.}~\bibnamefont {Flach}},\ }\bibfield  {title}
  {\bibinfo {title} {Flat band fine-tuning and its photonic applications},\
  }\href {https://doi.org/doi.org/10.1515/nanoph-2024-0135} {\bibfield
  {journal} {\bibinfo  {journal} {Nanophotonics}\ }\textbf {\bibinfo {volume}
  {13}},\ \bibinfo {pages} {3925} (\bibinfo {year} {2024})}\BibitemShut
  {NoStop}%
\bibitem [{\citenamefont {Maimaiti}\ \emph {et~al.}(2017)\citenamefont
  {Maimaiti}, \citenamefont {Andreanov}, \citenamefont {Park}, \citenamefont
  {Gendelman},\ and\ \citenamefont {Flach}}]{PhysRevB.95.115135}%
  \BibitemOpen
  \bibfield  {author} {\bibinfo {author} {\bibfnamefont {W.}~\bibnamefont
  {Maimaiti}}, \bibinfo {author} {\bibfnamefont {A.}~\bibnamefont {Andreanov}},
  \bibinfo {author} {\bibfnamefont {H.~C.}\ \bibnamefont {Park}}, \bibinfo
  {author} {\bibfnamefont {O.}~\bibnamefont {Gendelman}},\ and\ \bibinfo
  {author} {\bibfnamefont {S.}~\bibnamefont {Flach}},\ }\bibfield  {title}
  {\bibinfo {title} {Compact localized states and flat-band generators in one
  dimension},\ }\href {https://doi.org/10.1103/PhysRevB.95.115135} {\bibfield
  {journal} {\bibinfo  {journal} {Phys. Rev. B}\ }\textbf {\bibinfo {volume}
  {95}},\ \bibinfo {pages} {115135} (\bibinfo {year} {2017})}\BibitemShut
  {NoStop}%
\bibitem [{\citenamefont {Ramachandran}\ \emph {et~al.}(2017)\citenamefont
  {Ramachandran}, \citenamefont {Andreanov},\ and\ \citenamefont
  {Flach}}]{Ramachandran2017}%
  \BibitemOpen
  \bibfield  {author} {\bibinfo {author} {\bibfnamefont {A.}~\bibnamefont
  {Ramachandran}}, \bibinfo {author} {\bibfnamefont {A.}~\bibnamefont
  {Andreanov}},\ and\ \bibinfo {author} {\bibfnamefont {S.}~\bibnamefont
  {Flach}},\ }\bibfield  {title} {\bibinfo {title} {Chiral flat bands:
  Existence, engineering, and stability},\ }\bibfield  {journal} {\bibinfo
  {journal} {Physical Review B}\ }\textbf {\bibinfo {volume} {96}},\ \href
  {https://doi.org/10.1103/physrevb.96.161104} {10.1103/physrevb.96.161104}
  (\bibinfo {year} {2017})\BibitemShut {NoStop}%
\bibitem [{\citenamefont {Rhim}\ and\ \citenamefont {Yang}(2019)}]{Rhim2019}%
  \BibitemOpen
  \bibfield  {author} {\bibinfo {author} {\bibfnamefont {J.-W.}\ \bibnamefont
  {Rhim}}\ and\ \bibinfo {author} {\bibfnamefont {B.-J.}\ \bibnamefont
  {Yang}},\ }\bibfield  {title} {\bibinfo {title} {Classification of flat bands
  according to the band-crossing singularity of bloch wave functions},\
  }\bibfield  {journal} {\bibinfo  {journal} {Physical Review B}\ }\textbf
  {\bibinfo {volume} {99}},\ \href {https://doi.org/10.1103/physrevb.99.045107}
  {10.1103/physrevb.99.045107} (\bibinfo {year} {2019})\BibitemShut {NoStop}%
\bibitem [{\citenamefont {Read}(2017)}]{PhysRevB.95.115309}%
  \BibitemOpen
  \bibfield  {author} {\bibinfo {author} {\bibfnamefont {N.}~\bibnamefont
  {Read}},\ }\bibfield  {title} {\bibinfo {title} {Compactly supported wannier
  functions and algebraic $k$-theory},\ }\href
  {https://doi.org/10.1103/PhysRevB.95.115309} {\bibfield  {journal} {\bibinfo
  {journal} {Phys. Rev. B}\ }\textbf {\bibinfo {volume} {95}},\ \bibinfo
  {pages} {115309} (\bibinfo {year} {2017})}\BibitemShut {NoStop}%
\bibitem [{\citenamefont {Vidal}\ \emph {et~al.}(1998)\citenamefont {Vidal},
  \citenamefont {Mosseri},\ and\ \citenamefont {Dou\ifmmode~\mbox{\c{c}}\else
  \c{c}\fi{}ot}}]{Vidal1998}%
  \BibitemOpen
  \bibfield  {author} {\bibinfo {author} {\bibfnamefont {J.}~\bibnamefont
  {Vidal}}, \bibinfo {author} {\bibfnamefont {R.}~\bibnamefont {Mosseri}},\
  and\ \bibinfo {author} {\bibfnamefont {B.}~\bibnamefont
  {Dou\ifmmode~\mbox{\c{c}}\else \c{c}\fi{}ot}},\ }\bibfield  {title} {\bibinfo
  {title} {Aharonov-bohm cages in two-dimensional structures},\ }\href
  {https://doi.org/10.1103/PhysRevLett.81.5888} {\bibfield  {journal} {\bibinfo
   {journal} {Phys. Rev. Lett.}\ }\textbf {\bibinfo {volume} {81}},\ \bibinfo
  {pages} {5888} (\bibinfo {year} {1998})}\BibitemShut {NoStop}%
\bibitem [{\citenamefont {Mukherjee}\ \emph {et~al.}(2018)\citenamefont
  {Mukherjee}, \citenamefont {Di~Liberto}, \citenamefont {\"Ohberg},
  \citenamefont {Thomson},\ and\ \citenamefont {Goldman}}]{Mukherjee2018}%
  \BibitemOpen
  \bibfield  {author} {\bibinfo {author} {\bibfnamefont {S.}~\bibnamefont
  {Mukherjee}}, \bibinfo {author} {\bibfnamefont {M.}~\bibnamefont
  {Di~Liberto}}, \bibinfo {author} {\bibfnamefont {P.}~\bibnamefont
  {\"Ohberg}}, \bibinfo {author} {\bibfnamefont {R.~R.}\ \bibnamefont
  {Thomson}},\ and\ \bibinfo {author} {\bibfnamefont {N.}~\bibnamefont
  {Goldman}},\ }\bibfield  {title} {\bibinfo {title} {Experimental observation
  of aharonov-bohm cages in photonic lattices},\ }\href
  {https://doi.org/10.1103/PhysRevLett.121.075502} {\bibfield  {journal}
  {\bibinfo  {journal} {Phys. Rev. Lett.}\ }\textbf {\bibinfo {volume} {121}},\
  \bibinfo {pages} {075502} (\bibinfo {year} {2018})}\BibitemShut {NoStop}%
\bibitem [{\citenamefont {Vidal}\ \emph {et~al.}(2000)\citenamefont {Vidal},
  \citenamefont {Dou\ifmmode~\mbox{\c{c}}\else \c{c}\fi{}ot}, \citenamefont
  {Mosseri},\ and\ \citenamefont {Butaud}}]{PhysRevLett.85.3906}%
  \BibitemOpen
  \bibfield  {author} {\bibinfo {author} {\bibfnamefont {J.}~\bibnamefont
  {Vidal}}, \bibinfo {author} {\bibfnamefont {B.}~\bibnamefont
  {Dou\ifmmode~\mbox{\c{c}}\else \c{c}\fi{}ot}}, \bibinfo {author}
  {\bibfnamefont {R.}~\bibnamefont {Mosseri}},\ and\ \bibinfo {author}
  {\bibfnamefont {P.}~\bibnamefont {Butaud}},\ }\bibfield  {title} {\bibinfo
  {title} {Interaction induced delocalization for two particles in a periodic
  potential},\ }\href {https://doi.org/10.1103/PhysRevLett.85.3906} {\bibfield
  {journal} {\bibinfo  {journal} {Phys. Rev. Lett.}\ }\textbf {\bibinfo
  {volume} {85}},\ \bibinfo {pages} {3906} (\bibinfo {year}
  {2000})}\BibitemShut {NoStop}%
\bibitem [{\citenamefont {Huber}\ and\ \citenamefont
  {Altman}(2010)}]{PhysRevB.82.184502}%
  \BibitemOpen
  \bibfield  {author} {\bibinfo {author} {\bibfnamefont {S.~D.}\ \bibnamefont
  {Huber}}\ and\ \bibinfo {author} {\bibfnamefont {E.}~\bibnamefont {Altman}},\
  }\bibfield  {title} {\bibinfo {title} {Bose condensation in flat bands},\
  }\href {https://doi.org/10.1103/PhysRevB.82.184502} {\bibfield  {journal}
  {\bibinfo  {journal} {Phys. Rev. B}\ }\textbf {\bibinfo {volume} {82}},\
  \bibinfo {pages} {184502} (\bibinfo {year} {2010})}\BibitemShut {NoStop}%
\bibitem [{\citenamefont {J\"unemann}\ \emph {et~al.}(2017)\citenamefont
  {J\"unemann}, \citenamefont {Piga}, \citenamefont {Ran}, \citenamefont
  {Lewenstein}, \citenamefont {Rizzi},\ and\ \citenamefont
  {Bermudez}}]{Junemann2017}%
  \BibitemOpen
  \bibfield  {author} {\bibinfo {author} {\bibfnamefont {J.}~\bibnamefont
  {J\"unemann}}, \bibinfo {author} {\bibfnamefont {A.}~\bibnamefont {Piga}},
  \bibinfo {author} {\bibfnamefont {S.-J.}\ \bibnamefont {Ran}}, \bibinfo
  {author} {\bibfnamefont {M.}~\bibnamefont {Lewenstein}}, \bibinfo {author}
  {\bibfnamefont {M.}~\bibnamefont {Rizzi}},\ and\ \bibinfo {author}
  {\bibfnamefont {A.}~\bibnamefont {Bermudez}},\ }\bibfield  {title} {\bibinfo
  {title} {Exploring interacting topological insulators with ultracold atoms:
  The synthetic creutz-hubbard model},\ }\href
  {https://doi.org/10.1103/PhysRevX.7.031057} {\bibfield  {journal} {\bibinfo
  {journal} {Phys. Rev. X}\ }\textbf {\bibinfo {volume} {7}},\ \bibinfo {pages}
  {031057} (\bibinfo {year} {2017})}\BibitemShut {NoStop}%
\bibitem [{\citenamefont {Tovmasyan}\ \emph {et~al.}(2018)\citenamefont
  {Tovmasyan}, \citenamefont {Peotta}, \citenamefont {Liang}, \citenamefont
  {T\"orm\"a},\ and\ \citenamefont {Huber}}]{Tovmasyan2018}%
  \BibitemOpen
  \bibfield  {author} {\bibinfo {author} {\bibfnamefont {M.}~\bibnamefont
  {Tovmasyan}}, \bibinfo {author} {\bibfnamefont {S.}~\bibnamefont {Peotta}},
  \bibinfo {author} {\bibfnamefont {L.}~\bibnamefont {Liang}}, \bibinfo
  {author} {\bibfnamefont {P.}~\bibnamefont {T\"orm\"a}},\ and\ \bibinfo
  {author} {\bibfnamefont {S.~D.}\ \bibnamefont {Huber}},\ }\bibfield  {title}
  {\bibinfo {title} {Preformed pairs in flat bloch bands},\ }\href
  {https://doi.org/10.1103/PhysRevB.98.134513} {\bibfield  {journal} {\bibinfo
  {journal} {Phys. Rev. B}\ }\textbf {\bibinfo {volume} {98}},\ \bibinfo
  {pages} {134513} (\bibinfo {year} {2018})}\BibitemShut {NoStop}%
\bibitem [{\citenamefont {Burgher}\ \emph {et~al.}(2025)\citenamefont
  {Burgher}, \citenamefont {Di~Liberto}, \citenamefont {Goldman},\ and\
  \citenamefont {Amelio}}]{Burgher2025}%
  \BibitemOpen
  \bibfield  {author} {\bibinfo {author} {\bibfnamefont {M.}~\bibnamefont
  {Burgher}}, \bibinfo {author} {\bibfnamefont {M.}~\bibnamefont {Di~Liberto}},
  \bibinfo {author} {\bibfnamefont {N.}~\bibnamefont {Goldman}},\ and\ \bibinfo
  {author} {\bibfnamefont {I.}~\bibnamefont {Amelio}},\ }\bibfield  {title}
  {\bibinfo {title} {Fate of chiral order and impurity self-pinning in flat
  bands with local symmetry},\ }\href
  {https://doi.org/10.1103/PhysRevB.111.064514} {\bibfield  {journal} {\bibinfo
   {journal} {Phys. Rev. B}\ }\textbf {\bibinfo {volume} {111}},\ \bibinfo
  {pages} {064514} (\bibinfo {year} {2025})}\BibitemShut {NoStop}%
\bibitem [{\citenamefont {Pyykk\"onen}\ \emph {et~al.}(2023)\citenamefont
  {Pyykk\"onen}, \citenamefont {Salerno}, \citenamefont {K\"ah\"ar\"a},\ and\
  \citenamefont {T\"orm\"a}}]{PhysRevResearch.5.043259}%
  \BibitemOpen
  \bibfield  {author} {\bibinfo {author} {\bibfnamefont {V.~A.~J.}\
  \bibnamefont {Pyykk\"onen}}, \bibinfo {author} {\bibfnamefont
  {G.}~\bibnamefont {Salerno}}, \bibinfo {author} {\bibfnamefont
  {J.}~\bibnamefont {K\"ah\"ar\"a}},\ and\ \bibinfo {author} {\bibfnamefont
  {P.}~\bibnamefont {T\"orm\"a}},\ }\bibfield  {title} {\bibinfo {title}
  {All-optical switching at the two-photon limit with interference-localized
  states},\ }\href {https://doi.org/10.1103/PhysRevResearch.5.043259}
  {\bibfield  {journal} {\bibinfo  {journal} {Phys. Rev. Res.}\ }\textbf
  {\bibinfo {volume} {5}},\ \bibinfo {pages} {043259} (\bibinfo {year}
  {2023})}\BibitemShut {NoStop}%
\bibitem [{\citenamefont {Peotta}\ and\ \citenamefont
  {T{\"o}rm{\"a}}(2015)}]{peotta2015superfluidity}%
  \BibitemOpen
  \bibfield  {author} {\bibinfo {author} {\bibfnamefont {S.}~\bibnamefont
  {Peotta}}\ and\ \bibinfo {author} {\bibfnamefont {P.}~\bibnamefont
  {T{\"o}rm{\"a}}},\ }\bibfield  {title} {\bibinfo {title} {Superfluidity in
  topologically nontrivial flat bands},\ }\href
  {https://www.nature.com/articles/ncomms9944.pdf} {\bibfield  {journal}
  {\bibinfo  {journal} {Nature communications}\ }\textbf {\bibinfo {volume}
  {6}},\ \bibinfo {pages} {8944} (\bibinfo {year} {2015})}\BibitemShut
  {NoStop}%
\bibitem [{\citenamefont {Salerno}\ \emph {et~al.}(2023)\citenamefont
  {Salerno}, \citenamefont {Ozawa},\ and\ \citenamefont
  {T\"orm\"a}}]{Salerno2023}%
  \BibitemOpen
  \bibfield  {author} {\bibinfo {author} {\bibfnamefont {G.}~\bibnamefont
  {Salerno}}, \bibinfo {author} {\bibfnamefont {T.}~\bibnamefont {Ozawa}},\
  and\ \bibinfo {author} {\bibfnamefont {P.}~\bibnamefont {T\"orm\"a}},\
  }\bibfield  {title} {\bibinfo {title} {Drude weight and the many-body quantum
  metric in one-dimensional bose systems},\ }\href
  {https://doi.org/10.1103/PhysRevB.108.L140503} {\bibfield  {journal}
  {\bibinfo  {journal} {Phys. Rev. B}\ }\textbf {\bibinfo {volume} {108}},\
  \bibinfo {pages} {L140503} (\bibinfo {year} {2023})}\BibitemShut {NoStop}%
\bibitem [{\citenamefont {T{\"o}rm{\"a}}\ \emph {et~al.}(2022)\citenamefont
  {T{\"o}rm{\"a}}, \citenamefont {Peotta},\ and\ \citenamefont
  {Bernevig}}]{Torma2022}%
  \BibitemOpen
  \bibfield  {author} {\bibinfo {author} {\bibfnamefont {P.}~\bibnamefont
  {T{\"o}rm{\"a}}}, \bibinfo {author} {\bibfnamefont {S.}~\bibnamefont
  {Peotta}},\ and\ \bibinfo {author} {\bibfnamefont {B.~A.}\ \bibnamefont
  {Bernevig}},\ }\bibfield  {title} {\bibinfo {title} {Superconductivity,
  superfluidity and quantum geometry in twisted multilayer systems},\ }\href
  {https://doi.org/10.1038/s42254-022-00466-y} {\bibfield  {journal} {\bibinfo
  {journal} {Nature Reviews Physics}\ }\textbf {\bibinfo {volume} {4}},\
  \bibinfo {pages} {528} (\bibinfo {year} {2022})}\BibitemShut {NoStop}%
\bibitem [{\citenamefont {Julku}\ \emph {et~al.}(2016)\citenamefont {Julku},
  \citenamefont {Peotta}, \citenamefont {Vanhala}, \citenamefont {Kim},\ and\
  \citenamefont {T\"orm\"a}}]{Torma_PRL}%
  \BibitemOpen
  \bibfield  {author} {\bibinfo {author} {\bibfnamefont {A.}~\bibnamefont
  {Julku}}, \bibinfo {author} {\bibfnamefont {S.}~\bibnamefont {Peotta}},
  \bibinfo {author} {\bibfnamefont {T.~I.}\ \bibnamefont {Vanhala}}, \bibinfo
  {author} {\bibfnamefont {D.-H.}\ \bibnamefont {Kim}},\ and\ \bibinfo {author}
  {\bibfnamefont {P.}~\bibnamefont {T\"orm\"a}},\ }\bibfield  {title} {\bibinfo
  {title} {Geometric origin of superfluidity in the lieb-lattice flat band},\
  }\href {https://doi.org/10.1103/PhysRevLett.117.045303} {\bibfield  {journal}
  {\bibinfo  {journal} {Phys. Rev. Lett.}\ }\textbf {\bibinfo {volume} {117}},\
  \bibinfo {pages} {045303} (\bibinfo {year} {2016})}\BibitemShut {NoStop}%
\bibitem [{\citenamefont {T\"orm\"a}\ \emph {et~al.}(2018)\citenamefont
  {T\"orm\"a}, \citenamefont {Liang},\ and\ \citenamefont
  {Peotta}}]{Torma_PRB}%
  \BibitemOpen
  \bibfield  {author} {\bibinfo {author} {\bibfnamefont {P.}~\bibnamefont
  {T\"orm\"a}}, \bibinfo {author} {\bibfnamefont {L.}~\bibnamefont {Liang}},\
  and\ \bibinfo {author} {\bibfnamefont {S.}~\bibnamefont {Peotta}},\
  }\bibfield  {title} {\bibinfo {title} {Quantum metric and effective mass of a
  two-body bound state in a flat band},\ }\href
  {https://doi.org/10.1103/PhysRevB.98.220511} {\bibfield  {journal} {\bibinfo
  {journal} {Phys. Rev. B}\ }\textbf {\bibinfo {volume} {98}},\ \bibinfo
  {pages} {220511} (\bibinfo {year} {2018})}\BibitemShut {NoStop}%
\bibitem [{\citenamefont {Xia}\ \emph {et~al.}(2016)\citenamefont {Xia},
  \citenamefont {Hu}, \citenamefont {Song}, \citenamefont {Zong}, \citenamefont
  {Tang},\ and\ \citenamefont {Chen}}]{Xia2016}%
  \BibitemOpen
  \bibfield  {author} {\bibinfo {author} {\bibfnamefont {S.}~\bibnamefont
  {Xia}}, \bibinfo {author} {\bibfnamefont {Y.}~\bibnamefont {Hu}}, \bibinfo
  {author} {\bibfnamefont {D.}~\bibnamefont {Song}}, \bibinfo {author}
  {\bibfnamefont {Y.}~\bibnamefont {Zong}}, \bibinfo {author} {\bibfnamefont
  {L.}~\bibnamefont {Tang}},\ and\ \bibinfo {author} {\bibfnamefont
  {Z.}~\bibnamefont {Chen}},\ }\bibfield  {title} {\bibinfo {title}
  {Demonstration of flat-band image transmission in optically induced lieb
  photonic lattices},\ }\href {https://doi.org/10.1364/ol.41.001435} {\bibfield
   {journal} {\bibinfo  {journal} {Optics Letters}\ }\textbf {\bibinfo {volume}
  {41}},\ \bibinfo {pages} {1435} (\bibinfo {year} {2016})}\BibitemShut
  {NoStop}%
\bibitem [{\citenamefont {Mukherjee}\ \emph {et~al.}(2015)\citenamefont
  {Mukherjee}, \citenamefont {Spracklen}, \citenamefont {Choudhury},
  \citenamefont {Goldman}, \citenamefont {\"Ohberg}, \citenamefont
  {Andersson},\ and\ \citenamefont {Thomson}}]{PhysRevLett.114.245504}%
  \BibitemOpen
  \bibfield  {author} {\bibinfo {author} {\bibfnamefont {S.}~\bibnamefont
  {Mukherjee}}, \bibinfo {author} {\bibfnamefont {A.}~\bibnamefont
  {Spracklen}}, \bibinfo {author} {\bibfnamefont {D.}~\bibnamefont
  {Choudhury}}, \bibinfo {author} {\bibfnamefont {N.}~\bibnamefont {Goldman}},
  \bibinfo {author} {\bibfnamefont {P.}~\bibnamefont {\"Ohberg}}, \bibinfo
  {author} {\bibfnamefont {E.}~\bibnamefont {Andersson}},\ and\ \bibinfo
  {author} {\bibfnamefont {R.~R.}\ \bibnamefont {Thomson}},\ }\bibfield
  {title} {\bibinfo {title} {Observation of a localized flat-band state in a
  photonic lieb lattice},\ }\href
  {https://doi.org/10.1103/PhysRevLett.114.245504} {\bibfield  {journal}
  {\bibinfo  {journal} {Phys. Rev. Lett.}\ }\textbf {\bibinfo {volume} {114}},\
  \bibinfo {pages} {245504} (\bibinfo {year} {2015})}\BibitemShut {NoStop}%
\bibitem [{\citenamefont {Guzmán-Silva}\ \emph {et~al.}(2014)\citenamefont
  {Guzmán-Silva}, \citenamefont {Mejía-Cortés}, \citenamefont {Bandres},
  \citenamefont {Rechtsman}, \citenamefont {Weimann}, \citenamefont {Nolte},
  \citenamefont {Segev}, \citenamefont {Szameit},\ and\ \citenamefont
  {Vicencio}}]{GuzmnSilva2014}%
  \BibitemOpen
  \bibfield  {author} {\bibinfo {author} {\bibfnamefont {D.}~\bibnamefont
  {Guzmán-Silva}}, \bibinfo {author} {\bibfnamefont {C.}~\bibnamefont
  {Mejía-Cortés}}, \bibinfo {author} {\bibfnamefont {M.~A.}\ \bibnamefont
  {Bandres}}, \bibinfo {author} {\bibfnamefont {M.~C.}\ \bibnamefont
  {Rechtsman}}, \bibinfo {author} {\bibfnamefont {S.}~\bibnamefont {Weimann}},
  \bibinfo {author} {\bibfnamefont {S.}~\bibnamefont {Nolte}}, \bibinfo
  {author} {\bibfnamefont {M.}~\bibnamefont {Segev}}, \bibinfo {author}
  {\bibfnamefont {A.}~\bibnamefont {Szameit}},\ and\ \bibinfo {author}
  {\bibfnamefont {R.~A.}\ \bibnamefont {Vicencio}},\ }\bibfield  {title}
  {\bibinfo {title} {Experimental observation of bulk and edge transport in
  photonic lieb lattices},\ }\href
  {https://doi.org/10.1088/1367-2630/16/6/063061} {\bibfield  {journal}
  {\bibinfo  {journal} {New Journal of Physics}\ }\textbf {\bibinfo {volume}
  {16}},\ \bibinfo {pages} {063061} (\bibinfo {year} {2014})}\BibitemShut
  {NoStop}%
\bibitem [{\citenamefont {Taie}\ \emph {et~al.}(2015)\citenamefont {Taie},
  \citenamefont {Ozawa}, \citenamefont {Ichinose}, \citenamefont {Nishio},
  \citenamefont {Nakajima},\ and\ \citenamefont {Takahashi}}]{Taie2015}%
  \BibitemOpen
  \bibfield  {author} {\bibinfo {author} {\bibfnamefont {S.}~\bibnamefont
  {Taie}}, \bibinfo {author} {\bibfnamefont {H.}~\bibnamefont {Ozawa}},
  \bibinfo {author} {\bibfnamefont {T.}~\bibnamefont {Ichinose}}, \bibinfo
  {author} {\bibfnamefont {T.}~\bibnamefont {Nishio}}, \bibinfo {author}
  {\bibfnamefont {S.}~\bibnamefont {Nakajima}},\ and\ \bibinfo {author}
  {\bibfnamefont {Y.}~\bibnamefont {Takahashi}},\ }\bibfield  {title} {\bibinfo
  {title} {Coherent driving and freezing of bosonic matter wave in an optical
  lieb lattice},\ }\bibfield  {journal} {\bibinfo  {journal} {Science
  Advances}\ }\textbf {\bibinfo {volume} {1}},\ \href
  {https://doi.org/10.1126/sciadv.1500854} {10.1126/sciadv.1500854} (\bibinfo
  {year} {2015})\BibitemShut {NoStop}%
\bibitem [{\citenamefont {Ozawa}\ \emph {et~al.}(2017)\citenamefont {Ozawa},
  \citenamefont {Taie}, \citenamefont {Ichinose},\ and\ \citenamefont
  {Takahashi}}]{PhysRevLett.118.175301}%
  \BibitemOpen
  \bibfield  {author} {\bibinfo {author} {\bibfnamefont {H.}~\bibnamefont
  {Ozawa}}, \bibinfo {author} {\bibfnamefont {S.}~\bibnamefont {Taie}},
  \bibinfo {author} {\bibfnamefont {T.}~\bibnamefont {Ichinose}},\ and\
  \bibinfo {author} {\bibfnamefont {Y.}~\bibnamefont {Takahashi}},\ }\bibfield
  {title} {\bibinfo {title} {Interaction-driven shift and distortion of a flat
  band in an optical lieb lattice},\ }\href
  {https://doi.org/10.1103/PhysRevLett.118.175301} {\bibfield  {journal}
  {\bibinfo  {journal} {Phys. Rev. Lett.}\ }\textbf {\bibinfo {volume} {118}},\
  \bibinfo {pages} {175301} (\bibinfo {year} {2017})}\BibitemShut {NoStop}%
\bibitem [{\citenamefont {Kajiwara}\ \emph {et~al.}(2016)\citenamefont
  {Kajiwara}, \citenamefont {Urade}, \citenamefont {Nakata}, \citenamefont
  {Nakanishi},\ and\ \citenamefont {Kitano}}]{PhysRevB.93.075126}%
  \BibitemOpen
  \bibfield  {author} {\bibinfo {author} {\bibfnamefont {S.}~\bibnamefont
  {Kajiwara}}, \bibinfo {author} {\bibfnamefont {Y.}~\bibnamefont {Urade}},
  \bibinfo {author} {\bibfnamefont {Y.}~\bibnamefont {Nakata}}, \bibinfo
  {author} {\bibfnamefont {T.}~\bibnamefont {Nakanishi}},\ and\ \bibinfo
  {author} {\bibfnamefont {M.}~\bibnamefont {Kitano}},\ }\bibfield  {title}
  {\bibinfo {title} {Observation of a nonradiative flat band for spoof surface
  plasmons in a metallic lieb lattice},\ }\href
  {https://doi.org/10.1103/PhysRevB.93.075126} {\bibfield  {journal} {\bibinfo
  {journal} {Phys. Rev. B}\ }\textbf {\bibinfo {volume} {93}},\ \bibinfo
  {pages} {075126} (\bibinfo {year} {2016})}\BibitemShut {NoStop}%
\bibitem [{\citenamefont {Centała}\ and\ \citenamefont
  {Kłos}(2023)}]{Centaa2023}%
  \BibitemOpen
  \bibfield  {author} {\bibinfo {author} {\bibfnamefont {G.}~\bibnamefont
  {Centała}}\ and\ \bibinfo {author} {\bibfnamefont {J.~W.}\ \bibnamefont
  {Kłos}},\ }\bibfield  {title} {\bibinfo {title} {Compact localized states in
  magnonic lieb lattices},\ }\bibfield  {journal} {\bibinfo  {journal}
  {Scientific Reports}\ }\textbf {\bibinfo {volume} {13}},\ \href
  {https://doi.org/10.1038/s41598-023-39816-w} {10.1038/s41598-023-39816-w}
  (\bibinfo {year} {2023})\BibitemShut {NoStop}%
\bibitem [{\citenamefont {Bergholtz}\ and\ \citenamefont
  {Liu}(2013)}]{bergholtz2013topological}%
  \BibitemOpen
  \bibfield  {author} {\bibinfo {author} {\bibfnamefont {E.~J.}\ \bibnamefont
  {Bergholtz}}\ and\ \bibinfo {author} {\bibfnamefont {Z.}~\bibnamefont
  {Liu}},\ }\bibfield  {title} {\bibinfo {title} {Topological flat band models
  and fractional chern insulators},\ }\href
  {https://doi.org/10.1142/s021797921330017x} {\bibfield  {journal} {\bibinfo
  {journal} {International Journal of Modern Physics B}\ }\textbf {\bibinfo
  {volume} {27}},\ \bibinfo {pages} {1330017} (\bibinfo {year}
  {2013})}\BibitemShut {NoStop}%
\bibitem [{\citenamefont {Scaffidi}\ and\ \citenamefont
  {Simon}(2014)}]{PhysRevB.90.115132}%
  \BibitemOpen
  \bibfield  {author} {\bibinfo {author} {\bibfnamefont {T.}~\bibnamefont
  {Scaffidi}}\ and\ \bibinfo {author} {\bibfnamefont {S.~H.}\ \bibnamefont
  {Simon}},\ }\bibfield  {title} {\bibinfo {title} {Exact solutions of
  fractional chern insulators: Interacting particles in the hofstadter model at
  finite size},\ }\href {https://doi.org/10.1103/PhysRevB.90.115132} {\bibfield
   {journal} {\bibinfo  {journal} {Phys. Rev. B}\ }\textbf {\bibinfo {volume}
  {90}},\ \bibinfo {pages} {115132} (\bibinfo {year} {2014})}\BibitemShut
  {NoStop}%
\bibitem [{\citenamefont {Sun}\ \emph {et~al.}(2011)\citenamefont {Sun},
  \citenamefont {Gu}, \citenamefont {Katsura},\ and\ \citenamefont
  {Das~Sarma}}]{Sun2011}%
  \BibitemOpen
  \bibfield  {author} {\bibinfo {author} {\bibfnamefont {K.}~\bibnamefont
  {Sun}}, \bibinfo {author} {\bibfnamefont {Z.}~\bibnamefont {Gu}}, \bibinfo
  {author} {\bibfnamefont {H.}~\bibnamefont {Katsura}},\ and\ \bibinfo {author}
  {\bibfnamefont {S.}~\bibnamefont {Das~Sarma}},\ }\bibfield  {title} {\bibinfo
  {title} {Nearly flatbands with nontrivial topology},\ }\href
  {https://doi.org/10.1103/PhysRevLett.106.236803} {\bibfield  {journal}
  {\bibinfo  {journal} {Phys. Rev. Lett.}\ }\textbf {\bibinfo {volume} {106}},\
  \bibinfo {pages} {236803} (\bibinfo {year} {2011})}\BibitemShut {NoStop}%
\bibitem [{\citenamefont {Sheremet}\ \emph {et~al.}(2023)\citenamefont
  {Sheremet}, \citenamefont {Petrov}, \citenamefont {Iorsh}, \citenamefont
  {Poshakinskiy},\ and\ \citenamefont {Poddubny}}]{RevModPhys.95.015002}%
  \BibitemOpen
  \bibfield  {author} {\bibinfo {author} {\bibfnamefont {A.~S.}\ \bibnamefont
  {Sheremet}}, \bibinfo {author} {\bibfnamefont {M.~I.}\ \bibnamefont
  {Petrov}}, \bibinfo {author} {\bibfnamefont {I.~V.}\ \bibnamefont {Iorsh}},
  \bibinfo {author} {\bibfnamefont {A.~V.}\ \bibnamefont {Poshakinskiy}},\ and\
  \bibinfo {author} {\bibfnamefont {A.~N.}\ \bibnamefont {Poddubny}},\
  }\bibfield  {title} {\bibinfo {title} {Waveguide quantum electrodynamics:
  Collective radiance and photon-photon correlations},\ }\href
  {https://doi.org/10.1103/RevModPhys.95.015002} {\bibfield  {journal}
  {\bibinfo  {journal} {Rev. Mod. Phys.}\ }\textbf {\bibinfo {volume} {95}},\
  \bibinfo {pages} {015002} (\bibinfo {year} {2023})}\BibitemShut {NoStop}%
\bibitem [{\citenamefont {Roy}\ \emph {et~al.}(2017)\citenamefont {Roy},
  \citenamefont {Wilson},\ and\ \citenamefont
  {Firstenberg}}]{RevModPhys.89.021001}%
  \BibitemOpen
  \bibfield  {author} {\bibinfo {author} {\bibfnamefont {D.}~\bibnamefont
  {Roy}}, \bibinfo {author} {\bibfnamefont {C.~M.}\ \bibnamefont {Wilson}},\
  and\ \bibinfo {author} {\bibfnamefont {O.}~\bibnamefont {Firstenberg}},\
  }\bibfield  {title} {\bibinfo {title} {Colloquium: Strongly interacting
  photons in one-dimensional continuum},\ }\href
  {https://doi.org/10.1103/RevModPhys.89.021001} {\bibfield  {journal}
  {\bibinfo  {journal} {Rev. Mod. Phys.}\ }\textbf {\bibinfo {volume} {89}},\
  \bibinfo {pages} {021001} (\bibinfo {year} {2017})}\BibitemShut {NoStop}%
\bibitem [{\citenamefont {Ciccarello}\ \emph {et~al.}(2024)\citenamefont
  {Ciccarello}, \citenamefont {Lodahl},\ and\ \citenamefont
  {Schneble}}]{Ciccarello:24}%
  \BibitemOpen
  \bibfield  {author} {\bibinfo {author} {\bibfnamefont {F.}~\bibnamefont
  {Ciccarello}}, \bibinfo {author} {\bibfnamefont {P.}~\bibnamefont {Lodahl}},\
  and\ \bibinfo {author} {\bibfnamefont {D.}~\bibnamefont {Schneble}},\
  }\bibfield  {title} {\bibinfo {title} {Waveguide quantum electrodynamics},\
  }\href {https://doi.org/10.1364/OPN.35.1.000034} {\bibfield  {journal}
  {\bibinfo  {journal} {Opt. Photon. News}\ }\textbf {\bibinfo {volume} {35}},\
  \bibinfo {pages} {34} (\bibinfo {year} {2024})}\BibitemShut {NoStop}%
\bibitem [{\citenamefont {Lodahl}\ \emph {et~al.}(2015)\citenamefont {Lodahl},
  \citenamefont {Mahmoodian},\ and\ \citenamefont
  {Stobbe}}]{lodahl2015interfacing}%
  \BibitemOpen
  \bibfield  {author} {\bibinfo {author} {\bibfnamefont {P.}~\bibnamefont
  {Lodahl}}, \bibinfo {author} {\bibfnamefont {S.}~\bibnamefont {Mahmoodian}},\
  and\ \bibinfo {author} {\bibfnamefont {S.}~\bibnamefont {Stobbe}},\
  }\bibfield  {title} {\bibinfo {title} {Interfacing single photons and single
  quantum dots with photonic nanostructures},\ }\href
  {https://doi.org/10.1103/RevModPhys.87.347} {\bibfield  {journal} {\bibinfo
  {journal} {Rev. Mod. Phys.}\ }\textbf {\bibinfo {volume} {87}},\ \bibinfo
  {pages} {347} (\bibinfo {year} {2015})}\BibitemShut {NoStop}%
\bibitem [{\citenamefont {Hood}\ \emph {et~al.}(2016)\citenamefont {Hood},
  \citenamefont {Goban}, \citenamefont {Asenjo-Garcia}, \citenamefont {Lu},
  \citenamefont {Yu}, \citenamefont {Chang},\ and\ \citenamefont
  {Kimble}}]{hood2016atom}%
  \BibitemOpen
  \bibfield  {author} {\bibinfo {author} {\bibfnamefont {J.~D.}\ \bibnamefont
  {Hood}}, \bibinfo {author} {\bibfnamefont {A.}~\bibnamefont {Goban}},
  \bibinfo {author} {\bibfnamefont {A.}~\bibnamefont {Asenjo-Garcia}}, \bibinfo
  {author} {\bibfnamefont {M.}~\bibnamefont {Lu}}, \bibinfo {author}
  {\bibfnamefont {S.-P.}\ \bibnamefont {Yu}}, \bibinfo {author} {\bibfnamefont
  {D.~E.}\ \bibnamefont {Chang}},\ and\ \bibinfo {author} {\bibfnamefont
  {H.~J.}\ \bibnamefont {Kimble}},\ }\bibfield  {title} {\bibinfo {title}
  {Atom–atom interactions around the band edge of a photonic crystal
  waveguide},\ }\href {https://doi.org/10.1073/pnas.1603788113} {\bibfield
  {journal} {\bibinfo  {journal} {Proceedings of the National Academy of
  Sciences}\ }\textbf {\bibinfo {volume} {113}},\ \bibinfo {pages}
  {10507–10512} (\bibinfo {year} {2016})}\BibitemShut {NoStop}%
\bibitem [{\citenamefont {Corzo}\ \emph {et~al.}(2019)\citenamefont {Corzo},
  \citenamefont {Raskop}, \citenamefont {Chandra}, \citenamefont {Sheremet},
  \citenamefont {Gouraud},\ and\ \citenamefont {Laurat}}]{corzo2019waveguide}%
  \BibitemOpen
  \bibfield  {author} {\bibinfo {author} {\bibfnamefont {N.~V.}\ \bibnamefont
  {Corzo}}, \bibinfo {author} {\bibfnamefont {J.}~\bibnamefont {Raskop}},
  \bibinfo {author} {\bibfnamefont {A.}~\bibnamefont {Chandra}}, \bibinfo
  {author} {\bibfnamefont {A.~S.}\ \bibnamefont {Sheremet}}, \bibinfo {author}
  {\bibfnamefont {B.}~\bibnamefont {Gouraud}},\ and\ \bibinfo {author}
  {\bibfnamefont {J.}~\bibnamefont {Laurat}},\ }\bibfield  {title} {\bibinfo
  {title} {Waveguide-coupled single collective excitation of atomic arrays},\
  }\href {https://doi.org/10.1038/s41586-019-0902-3} {\bibfield  {journal}
  {\bibinfo  {journal} {Nature}\ }\textbf {\bibinfo {volume} {566}},\ \bibinfo
  {pages} {359–362} (\bibinfo {year} {2019})}\BibitemShut {NoStop}%
\bibitem [{\citenamefont {Tiranov}\ \emph {et~al.}(2023)\citenamefont
  {Tiranov}, \citenamefont {Angelopoulou}, \citenamefont {van Diepen},
  \citenamefont {Schrinski}, \citenamefont {Sandberg}, \citenamefont {Wang},
  \citenamefont {Midolo}, \citenamefont {Scholz}, \citenamefont {Wieck},
  \citenamefont {Ludwig}, \citenamefont {Sørensen},\ and\ \citenamefont
  {Lodahl}}]{tiranov2023collective}%
  \BibitemOpen
  \bibfield  {author} {\bibinfo {author} {\bibfnamefont {A.}~\bibnamefont
  {Tiranov}}, \bibinfo {author} {\bibfnamefont {V.}~\bibnamefont
  {Angelopoulou}}, \bibinfo {author} {\bibfnamefont {C.~J.}\ \bibnamefont {van
  Diepen}}, \bibinfo {author} {\bibfnamefont {B.}~\bibnamefont {Schrinski}},
  \bibinfo {author} {\bibfnamefont {O.~A.~D.}\ \bibnamefont {Sandberg}},
  \bibinfo {author} {\bibfnamefont {Y.}~\bibnamefont {Wang}}, \bibinfo {author}
  {\bibfnamefont {L.}~\bibnamefont {Midolo}}, \bibinfo {author} {\bibfnamefont
  {S.}~\bibnamefont {Scholz}}, \bibinfo {author} {\bibfnamefont {A.~D.}\
  \bibnamefont {Wieck}}, \bibinfo {author} {\bibfnamefont {A.}~\bibnamefont
  {Ludwig}}, \bibinfo {author} {\bibfnamefont {A.~S.}\ \bibnamefont
  {Sørensen}},\ and\ \bibinfo {author} {\bibfnamefont {P.}~\bibnamefont
  {Lodahl}},\ }\bibfield  {title} {\bibinfo {title} {Collective super- and
  subradiant dynamics between distant optical quantum emitters},\ }\href
  {https://doi.org/10.1126/science.ade9324} {\bibfield  {journal} {\bibinfo
  {journal} {Science}\ }\textbf {\bibinfo {volume} {379}},\ \bibinfo {pages}
  {389–393} (\bibinfo {year} {2023})}\BibitemShut {NoStop}%
\bibitem [{\citenamefont {Prasad}\ \emph {et~al.}(2020)\citenamefont {Prasad},
  \citenamefont {Hinney}, \citenamefont {Mahmoodian}, \citenamefont {Hammerer},
  \citenamefont {Rind}, \citenamefont {Schneeweiss}, \citenamefont {Sørensen},
  \citenamefont {Volz},\ and\ \citenamefont
  {Rauschenbeutel}}]{prasad2020correlating}%
  \BibitemOpen
  \bibfield  {author} {\bibinfo {author} {\bibfnamefont {A.~S.}\ \bibnamefont
  {Prasad}}, \bibinfo {author} {\bibfnamefont {J.}~\bibnamefont {Hinney}},
  \bibinfo {author} {\bibfnamefont {S.}~\bibnamefont {Mahmoodian}}, \bibinfo
  {author} {\bibfnamefont {K.}~\bibnamefont {Hammerer}}, \bibinfo {author}
  {\bibfnamefont {S.}~\bibnamefont {Rind}}, \bibinfo {author} {\bibfnamefont
  {P.}~\bibnamefont {Schneeweiss}}, \bibinfo {author} {\bibfnamefont {A.~S.}\
  \bibnamefont {Sørensen}}, \bibinfo {author} {\bibfnamefont {J.}~\bibnamefont
  {Volz}},\ and\ \bibinfo {author} {\bibfnamefont {A.}~\bibnamefont
  {Rauschenbeutel}},\ }\bibfield  {title} {\bibinfo {title} {Correlating
  photons using the collective nonlinear response of atoms weakly coupled to an
  optical mode},\ }\href {https://doi.org/10.1038/s41566-020-0692-z} {\bibfield
   {journal} {\bibinfo  {journal} {Nature Photonics}\ }\textbf {\bibinfo
  {volume} {14}},\ \bibinfo {pages} {719–722} (\bibinfo {year}
  {2020})}\BibitemShut {NoStop}%
\bibitem [{\citenamefont {Astafiev}\ \emph {et~al.}(2010)\citenamefont
  {Astafiev}, \citenamefont {Zagoskin}, \citenamefont {Abdumalikov},
  \citenamefont {Pashkin}, \citenamefont {Yamamoto}, \citenamefont {Inomata},
  \citenamefont {Nakamura},\ and\ \citenamefont
  {Tsai}}]{astafiev2010resonance}%
  \BibitemOpen
  \bibfield  {author} {\bibinfo {author} {\bibfnamefont {O.}~\bibnamefont
  {Astafiev}}, \bibinfo {author} {\bibfnamefont {A.~M.}\ \bibnamefont
  {Zagoskin}}, \bibinfo {author} {\bibfnamefont {A.~A.}\ \bibnamefont
  {Abdumalikov}}, \bibinfo {author} {\bibfnamefont {Y.~A.}\ \bibnamefont
  {Pashkin}}, \bibinfo {author} {\bibfnamefont {T.}~\bibnamefont {Yamamoto}},
  \bibinfo {author} {\bibfnamefont {K.}~\bibnamefont {Inomata}}, \bibinfo
  {author} {\bibfnamefont {Y.}~\bibnamefont {Nakamura}},\ and\ \bibinfo
  {author} {\bibfnamefont {J.~S.}\ \bibnamefont {Tsai}},\ }\bibfield  {title}
  {\bibinfo {title} {Resonance fluorescence of a single artificial atom},\
  }\href {https://doi.org/10.1126/science.1181918} {\bibfield  {journal}
  {\bibinfo  {journal} {Science}\ }\textbf {\bibinfo {volume} {327}},\ \bibinfo
  {pages} {840–843} (\bibinfo {year} {2010})}\BibitemShut {NoStop}%
\bibitem [{\citenamefont {Brehm}\ \emph {et~al.}(2021)\citenamefont {Brehm},
  \citenamefont {Poddubny}, \citenamefont {Stehli}, \citenamefont {Wolz},
  \citenamefont {Rotzinger},\ and\ \citenamefont
  {Ustinov}}]{brehm2021waveguide}%
  \BibitemOpen
  \bibfield  {author} {\bibinfo {author} {\bibfnamefont {J.~D.}\ \bibnamefont
  {Brehm}}, \bibinfo {author} {\bibfnamefont {A.~N.}\ \bibnamefont {Poddubny}},
  \bibinfo {author} {\bibfnamefont {A.}~\bibnamefont {Stehli}}, \bibinfo
  {author} {\bibfnamefont {T.}~\bibnamefont {Wolz}}, \bibinfo {author}
  {\bibfnamefont {H.}~\bibnamefont {Rotzinger}},\ and\ \bibinfo {author}
  {\bibfnamefont {A.~V.}\ \bibnamefont {Ustinov}},\ }\bibfield  {title}
  {\bibinfo {title} {Waveguide bandgap engineering with an array of
  superconducting qubits},\ }\href
  {http://dx.doi.org/10.1038/s41535-021-00310-z} {\bibfield  {journal}
  {\bibinfo  {journal} {npj Quantum Materials}\ }\textbf {\bibinfo {volume}
  {6}} (\bibinfo {year} {2021})}\BibitemShut {NoStop}%
\bibitem [{\citenamefont {Mirhosseini}\ \emph {et~al.}(2019)\citenamefont
  {Mirhosseini}, \citenamefont {Kim}, \citenamefont {Zhang}, \citenamefont
  {Sipahigil}, \citenamefont {Dieterle}, \citenamefont {Keller}, \citenamefont
  {Asenjo-Garcia}, \citenamefont {Chang},\ and\ \citenamefont
  {Painter}}]{mirhosseini2019cavity}%
  \BibitemOpen
  \bibfield  {author} {\bibinfo {author} {\bibfnamefont {M.}~\bibnamefont
  {Mirhosseini}}, \bibinfo {author} {\bibfnamefont {E.}~\bibnamefont {Kim}},
  \bibinfo {author} {\bibfnamefont {X.}~\bibnamefont {Zhang}}, \bibinfo
  {author} {\bibfnamefont {A.}~\bibnamefont {Sipahigil}}, \bibinfo {author}
  {\bibfnamefont {P.~B.}\ \bibnamefont {Dieterle}}, \bibinfo {author}
  {\bibfnamefont {A.~J.}\ \bibnamefont {Keller}}, \bibinfo {author}
  {\bibfnamefont {A.}~\bibnamefont {Asenjo-Garcia}}, \bibinfo {author}
  {\bibfnamefont {D.~E.}\ \bibnamefont {Chang}},\ and\ \bibinfo {author}
  {\bibfnamefont {O.}~\bibnamefont {Painter}},\ }\bibfield  {title} {\bibinfo
  {title} {Cavity quantum electrodynamics with atom-like mirrors},\ }\href
  {https://doi.org/10.1038/s41586-019-1196-1} {\bibfield  {journal} {\bibinfo
  {journal} {Nature}\ }\textbf {\bibinfo {volume} {569}},\ \bibinfo {pages}
  {692–697} (\bibinfo {year} {2019})}\BibitemShut {NoStop}%
\bibitem [{\citenamefont {Kannan}\ \emph {et~al.}(2023)\citenamefont {Kannan},
  \citenamefont {Almanakly}, \citenamefont {Sung}, \citenamefont {Di~Paolo},
  \citenamefont {Rower}, \citenamefont {Braumüller}, \citenamefont {Melville},
  \citenamefont {Niedzielski}, \citenamefont {Karamlou}, \citenamefont
  {Serniak}, \citenamefont {Vepsäläinen}, \citenamefont {Schwartz},
  \citenamefont {Yoder}, \citenamefont {Winik}, \citenamefont {Wang},
  \citenamefont {Orlando}, \citenamefont {Gustavsson}, \citenamefont {Grover},\
  and\ \citenamefont {Oliver}}]{kannan2023demand}%
  \BibitemOpen
  \bibfield  {author} {\bibinfo {author} {\bibfnamefont {B.}~\bibnamefont
  {Kannan}}, \bibinfo {author} {\bibfnamefont {A.}~\bibnamefont {Almanakly}},
  \bibinfo {author} {\bibfnamefont {Y.}~\bibnamefont {Sung}}, \bibinfo {author}
  {\bibfnamefont {A.}~\bibnamefont {Di~Paolo}}, \bibinfo {author}
  {\bibfnamefont {D.~A.}\ \bibnamefont {Rower}}, \bibinfo {author}
  {\bibfnamefont {J.}~\bibnamefont {Braumüller}}, \bibinfo {author}
  {\bibfnamefont {A.}~\bibnamefont {Melville}}, \bibinfo {author}
  {\bibfnamefont {B.~M.}\ \bibnamefont {Niedzielski}}, \bibinfo {author}
  {\bibfnamefont {A.}~\bibnamefont {Karamlou}}, \bibinfo {author}
  {\bibfnamefont {K.}~\bibnamefont {Serniak}}, \bibinfo {author} {\bibfnamefont
  {A.}~\bibnamefont {Vepsäläinen}}, \bibinfo {author} {\bibfnamefont {M.~E.}\
  \bibnamefont {Schwartz}}, \bibinfo {author} {\bibfnamefont {J.~L.}\
  \bibnamefont {Yoder}}, \bibinfo {author} {\bibfnamefont {R.}~\bibnamefont
  {Winik}}, \bibinfo {author} {\bibfnamefont {J.~I.-J.}\ \bibnamefont {Wang}},
  \bibinfo {author} {\bibfnamefont {T.~P.}\ \bibnamefont {Orlando}}, \bibinfo
  {author} {\bibfnamefont {S.}~\bibnamefont {Gustavsson}}, \bibinfo {author}
  {\bibfnamefont {J.~A.}\ \bibnamefont {Grover}},\ and\ \bibinfo {author}
  {\bibfnamefont {W.~D.}\ \bibnamefont {Oliver}},\ }\bibfield  {title}
  {\bibinfo {title} {On-demand directional microwave photon emission using
  waveguide quantum electrodynamics},\ }\href
  {https://doi.org/10.1038/s41567-022-01869-5} {\bibfield  {journal} {\bibinfo
  {journal} {Nature Physics}\ }\textbf {\bibinfo {volume} {19}},\ \bibinfo
  {pages} {394–400} (\bibinfo {year} {2023})}\BibitemShut {NoStop}%
\bibitem [{\citenamefont {Scigliuzzo}\ \emph {et~al.}(2022)\citenamefont
  {Scigliuzzo}, \citenamefont {Calaj\`o}, \citenamefont {Ciccarello},
  \citenamefont {Perez~Lozano}, \citenamefont {Bengtsson}, \citenamefont
  {Scarlino}, \citenamefont {Wallraff}, \citenamefont {Chang}, \citenamefont
  {Delsing},\ and\ \citenamefont {Gasparinetti}}]{scigliuzzo2022controlling}%
  \BibitemOpen
  \bibfield  {author} {\bibinfo {author} {\bibfnamefont {M.}~\bibnamefont
  {Scigliuzzo}}, \bibinfo {author} {\bibfnamefont {G.}~\bibnamefont
  {Calaj\`o}}, \bibinfo {author} {\bibfnamefont {F.}~\bibnamefont
  {Ciccarello}}, \bibinfo {author} {\bibfnamefont {D.}~\bibnamefont
  {Perez~Lozano}}, \bibinfo {author} {\bibfnamefont {A.}~\bibnamefont
  {Bengtsson}}, \bibinfo {author} {\bibfnamefont {P.}~\bibnamefont {Scarlino}},
  \bibinfo {author} {\bibfnamefont {A.}~\bibnamefont {Wallraff}}, \bibinfo
  {author} {\bibfnamefont {D.}~\bibnamefont {Chang}}, \bibinfo {author}
  {\bibfnamefont {P.}~\bibnamefont {Delsing}},\ and\ \bibinfo {author}
  {\bibfnamefont {S.}~\bibnamefont {Gasparinetti}},\ }\bibfield  {title}
  {\bibinfo {title} {Controlling atom-photon bound states in an array of
  josephson-junction resonators},\ }\href
  {https://doi.org/10.1103/PhysRevX.12.031036} {\bibfield  {journal} {\bibinfo
  {journal} {Phys. Rev. X}\ }\textbf {\bibinfo {volume} {12}},\ \bibinfo
  {pages} {031036} (\bibinfo {year} {2022})}\BibitemShut {NoStop}%
\bibitem [{\citenamefont {Shah}\ \emph {et~al.}(2024)\citenamefont {Shah},
  \citenamefont {Yang}, \citenamefont {Joshi},\ and\ \citenamefont
  {Mirhosseini}}]{shah2024stabilizing}%
  \BibitemOpen
  \bibfield  {author} {\bibinfo {author} {\bibfnamefont {P.~S.}\ \bibnamefont
  {Shah}}, \bibinfo {author} {\bibfnamefont {F.}~\bibnamefont {Yang}}, \bibinfo
  {author} {\bibfnamefont {C.}~\bibnamefont {Joshi}},\ and\ \bibinfo {author}
  {\bibfnamefont {M.}~\bibnamefont {Mirhosseini}},\ }\bibfield  {title}
  {\bibinfo {title} {Stabilizing remote entanglement via waveguide
  dissipation},\ }\href {https://doi.org/10.1103/PRXQuantum.5.030346}
  {\bibfield  {journal} {\bibinfo  {journal} {PRX Quantum}\ }\textbf {\bibinfo
  {volume} {5}},\ \bibinfo {pages} {030346} (\bibinfo {year}
  {2024})}\BibitemShut {NoStop}%
\bibitem [{\citenamefont {Jouanny}\ \emph {et~al.}(2025)\citenamefont
  {Jouanny}, \citenamefont {Frasca}, \citenamefont {Weibel}, \citenamefont
  {Peyruchat}, \citenamefont {Scigliuzzo}, \citenamefont {Oppliger},
  \citenamefont {De~Palma}, \citenamefont {Sbroggi{\`o}}, \citenamefont
  {Beaulieu}, \citenamefont {Zilberberg} \emph {et~al.}}]{jouanny2025high}%
  \BibitemOpen
  \bibfield  {author} {\bibinfo {author} {\bibfnamefont {V.}~\bibnamefont
  {Jouanny}}, \bibinfo {author} {\bibfnamefont {S.}~\bibnamefont {Frasca}},
  \bibinfo {author} {\bibfnamefont {V.~J.}\ \bibnamefont {Weibel}}, \bibinfo
  {author} {\bibfnamefont {L.}~\bibnamefont {Peyruchat}}, \bibinfo {author}
  {\bibfnamefont {M.}~\bibnamefont {Scigliuzzo}}, \bibinfo {author}
  {\bibfnamefont {F.}~\bibnamefont {Oppliger}}, \bibinfo {author}
  {\bibfnamefont {F.}~\bibnamefont {De~Palma}}, \bibinfo {author}
  {\bibfnamefont {D.}~\bibnamefont {Sbroggi{\`o}}}, \bibinfo {author}
  {\bibfnamefont {G.}~\bibnamefont {Beaulieu}}, \bibinfo {author}
  {\bibfnamefont {O.}~\bibnamefont {Zilberberg}}, \emph {et~al.},\ }\bibfield
  {title} {\bibinfo {title} {High kinetic inductance cavity arrays for compact
  band engineering and topology-based disorder meters},\ }\href
  {https://www.nature.com/articles/s41467-025-58595-8#citeas} {\bibfield
  {journal} {\bibinfo  {journal} {Nature Communications}\ }\textbf {\bibinfo
  {volume} {16}},\ \bibinfo {pages} {3396} (\bibinfo {year}
  {2025})}\BibitemShut {NoStop}%
\bibitem [{\citenamefont {Gonz{\'a}lez-Tudela}\ \emph
  {et~al.}(2024)\citenamefont {Gonz{\'a}lez-Tudela}, \citenamefont {Reiserer},
  \citenamefont {Garc{\'\i}a-Ripoll},\ and\ \citenamefont
  {Garc{\'\i}a-Vidal}}]{gonzalez2024light}%
  \BibitemOpen
  \bibfield  {author} {\bibinfo {author} {\bibfnamefont {A.}~\bibnamefont
  {Gonz{\'a}lez-Tudela}}, \bibinfo {author} {\bibfnamefont {A.}~\bibnamefont
  {Reiserer}}, \bibinfo {author} {\bibfnamefont {J.~J.}\ \bibnamefont
  {Garc{\'\i}a-Ripoll}},\ and\ \bibinfo {author} {\bibfnamefont {F.~J.}\
  \bibnamefont {Garc{\'\i}a-Vidal}},\ }\bibfield  {title} {\bibinfo {title}
  {Light--matter interactions in quantum nanophotonic devices},\ }\href
  {https://www.nature.com/articles/s42254-023-00681-1} {\bibfield  {journal}
  {\bibinfo  {journal} {Nature Reviews Physics}\ }\textbf {\bibinfo {volume}
  {6}},\ \bibinfo {pages} {166} (\bibinfo {year} {2024})}\BibitemShut {NoStop}%
\bibitem [{\citenamefont {Douglas}\ \emph {et~al.}(2015)\citenamefont
  {Douglas}, \citenamefont {Habibian}, \citenamefont {Hung}, \citenamefont
  {Gorshkov}, \citenamefont {Kimble},\ and\ \citenamefont
  {Chang}}]{Douglas2015}%
  \BibitemOpen
  \bibfield  {author} {\bibinfo {author} {\bibfnamefont {J.~S.}\ \bibnamefont
  {Douglas}}, \bibinfo {author} {\bibfnamefont {H.}~\bibnamefont {Habibian}},
  \bibinfo {author} {\bibfnamefont {C.-L.}\ \bibnamefont {Hung}}, \bibinfo
  {author} {\bibfnamefont {A.~V.}\ \bibnamefont {Gorshkov}}, \bibinfo {author}
  {\bibfnamefont {H.~J.}\ \bibnamefont {Kimble}},\ and\ \bibinfo {author}
  {\bibfnamefont {D.~E.}\ \bibnamefont {Chang}},\ }\bibfield  {title} {\bibinfo
  {title} {Quantum many-body models with cold atoms coupled to photonic
  crystals},\ }\href {https://doi.org/10.1038/nphoton.2015.57} {\bibfield
  {journal} {\bibinfo  {journal} {Nature Photonics}\ }\textbf {\bibinfo
  {volume} {9}},\ \bibinfo {pages} {326–331} (\bibinfo {year}
  {2015})}\BibitemShut {NoStop}%
\bibitem [{\citenamefont {Gonz{\'a}lez-Tudela}\ \emph
  {et~al.}(2015)\citenamefont {Gonz{\'a}lez-Tudela}, \citenamefont {Hung},
  \citenamefont {Chang}, \citenamefont {Cirac},\ and\ \citenamefont
  {Kimble}}]{Tudela2015sub}%
  \BibitemOpen
  \bibfield  {author} {\bibinfo {author} {\bibfnamefont {A.}~\bibnamefont
  {Gonz{\'a}lez-Tudela}}, \bibinfo {author} {\bibfnamefont {C.-L.}\
  \bibnamefont {Hung}}, \bibinfo {author} {\bibfnamefont {D.~E.}\ \bibnamefont
  {Chang}}, \bibinfo {author} {\bibfnamefont {J.~I.}\ \bibnamefont {Cirac}},\
  and\ \bibinfo {author} {\bibfnamefont {H.}~\bibnamefont {Kimble}},\
  }\bibfield  {title} {\bibinfo {title} {Subwavelength vacuum lattices and
  atom--atom interactions in two-dimensional photonic crystals},\ }\href
  {https://doi.org/https://doi.org/10.1038/nphoton.2015.54} {\bibfield
  {journal} {\bibinfo  {journal} {Nature Photonics}\ }\textbf {\bibinfo
  {volume} {9}},\ \bibinfo {pages} {320} (\bibinfo {year} {2015})}\BibitemShut
  {NoStop}%
\bibitem [{\citenamefont {Chang}\ \emph {et~al.}(2018)\citenamefont {Chang},
  \citenamefont {Douglas}, \citenamefont {Gonz\'alez-Tudela}, \citenamefont
  {Hung},\ and\ \citenamefont {Kimble}}]{RevModPhys.90.031002}%
  \BibitemOpen
  \bibfield  {author} {\bibinfo {author} {\bibfnamefont {D.~E.}\ \bibnamefont
  {Chang}}, \bibinfo {author} {\bibfnamefont {J.~S.}\ \bibnamefont {Douglas}},
  \bibinfo {author} {\bibfnamefont {A.}~\bibnamefont {Gonz\'alez-Tudela}},
  \bibinfo {author} {\bibfnamefont {C.-L.}\ \bibnamefont {Hung}},\ and\
  \bibinfo {author} {\bibfnamefont {H.~J.}\ \bibnamefont {Kimble}},\ }\bibfield
   {title} {\bibinfo {title} {Colloquium: Quantum matter built from nanoscopic
  lattices of atoms and photons},\ }\href
  {https://doi.org/10.1103/RevModPhys.90.031002} {\bibfield  {journal}
  {\bibinfo  {journal} {Rev. Mod. Phys.}\ }\textbf {\bibinfo {volume} {90}},\
  \bibinfo {pages} {031002} (\bibinfo {year} {2018})}\BibitemShut {NoStop}%
\bibitem [{\citenamefont {De~Bernardis}\ \emph {et~al.}(2021)\citenamefont
  {De~Bernardis}, \citenamefont {Cian}, \citenamefont {Carusotto},
  \citenamefont {Hafezi},\ and\ \citenamefont {Rabl}}]{PhysRevLett.126.103603}%
  \BibitemOpen
  \bibfield  {author} {\bibinfo {author} {\bibfnamefont {D.}~\bibnamefont
  {De~Bernardis}}, \bibinfo {author} {\bibfnamefont {Z.-P.}\ \bibnamefont
  {Cian}}, \bibinfo {author} {\bibfnamefont {I.}~\bibnamefont {Carusotto}},
  \bibinfo {author} {\bibfnamefont {M.}~\bibnamefont {Hafezi}},\ and\ \bibinfo
  {author} {\bibfnamefont {P.}~\bibnamefont {Rabl}},\ }\bibfield  {title}
  {\bibinfo {title} {Light-matter interactions in synthetic magnetic fields:
  Landau-photon polaritons},\ }\href
  {https://doi.org/10.1103/PhysRevLett.126.103603} {\bibfield  {journal}
  {\bibinfo  {journal} {Phys. Rev. Lett.}\ }\textbf {\bibinfo {volume} {126}},\
  \bibinfo {pages} {103603} (\bibinfo {year} {2021})}\BibitemShut {NoStop}%
\bibitem [{\citenamefont {De~Bernardis}\ \emph {et~al.}(2023)\citenamefont
  {De~Bernardis}, \citenamefont {Piccioli}, \citenamefont {Rabl},\ and\
  \citenamefont {Carusotto}}]{PRXQuantum.4.030306}%
  \BibitemOpen
  \bibfield  {author} {\bibinfo {author} {\bibfnamefont {D.}~\bibnamefont
  {De~Bernardis}}, \bibinfo {author} {\bibfnamefont {F.~S.}\ \bibnamefont
  {Piccioli}}, \bibinfo {author} {\bibfnamefont {P.}~\bibnamefont {Rabl}},\
  and\ \bibinfo {author} {\bibfnamefont {I.}~\bibnamefont {Carusotto}},\
  }\bibfield  {title} {\bibinfo {title} {Chiral quantum optics in the bulk of
  photonic quantum hall systems},\ }\href
  {https://doi.org/10.1103/PRXQuantum.4.030306} {\bibfield  {journal} {\bibinfo
   {journal} {PRX Quantum}\ }\textbf {\bibinfo {volume} {4}},\ \bibinfo {pages}
  {030306} (\bibinfo {year} {2023})}\BibitemShut {NoStop}%
\bibitem [{\citenamefont {Vega}\ \emph {et~al.}(2023)\citenamefont {Vega},
  \citenamefont {Porras},\ and\ \citenamefont
  {Gonz\'alez-Tudela}}]{PhysRevResearch.5.023031}%
  \BibitemOpen
  \bibfield  {author} {\bibinfo {author} {\bibfnamefont {C.}~\bibnamefont
  {Vega}}, \bibinfo {author} {\bibfnamefont {D.}~\bibnamefont {Porras}},\ and\
  \bibinfo {author} {\bibfnamefont {A.}~\bibnamefont {Gonz\'alez-Tudela}},\
  }\bibfield  {title} {\bibinfo {title} {Topological multimode waveguide qed},\
  }\href {https://doi.org/10.1103/PhysRevResearch.5.023031} {\bibfield
  {journal} {\bibinfo  {journal} {Phys. Rev. Res.}\ }\textbf {\bibinfo {volume}
  {5}},\ \bibinfo {pages} {023031} (\bibinfo {year} {2023})}\BibitemShut
  {NoStop}%
\bibitem [{\citenamefont {Bello}\ and\ \citenamefont
  {Cirac}(2023)}]{PhysRevB.107.054301}%
  \BibitemOpen
  \bibfield  {author} {\bibinfo {author} {\bibfnamefont {M.}~\bibnamefont
  {Bello}}\ and\ \bibinfo {author} {\bibfnamefont {J.~I.}\ \bibnamefont
  {Cirac}},\ }\bibfield  {title} {\bibinfo {title} {Topological effects in
  two-dimensional quantum emitter systems},\ }\href
  {https://doi.org/10.1103/PhysRevB.107.054301} {\bibfield  {journal} {\bibinfo
   {journal} {Phys. Rev. B}\ }\textbf {\bibinfo {volume} {107}},\ \bibinfo
  {pages} {054301} (\bibinfo {year} {2023})}\BibitemShut {NoStop}%
\bibitem [{\citenamefont {Di~Benedetto}\ \emph {et~al.}(2025)\citenamefont
  {Di~Benedetto}, \citenamefont {Gonzalez-Tudela},\ and\ \citenamefont
  {Ciccarello}}]{di2024dipole}%
  \BibitemOpen
  \bibfield  {author} {\bibinfo {author} {\bibfnamefont {E.}~\bibnamefont
  {Di~Benedetto}}, \bibinfo {author} {\bibfnamefont {A.}~\bibnamefont
  {Gonzalez-Tudela}},\ and\ \bibinfo {author} {\bibfnamefont {F.}~\bibnamefont
  {Ciccarello}},\ }\bibfield  {title} {\bibinfo {title} {Dipole-dipole
  interactions mediated by a photonic flat band},\ }\href
  {https://doi.org/10.22331/q-2025-03-25-1671} {\bibfield  {journal} {\bibinfo
  {journal} {{Quantum}}\ }\textbf {\bibinfo {volume} {9}},\ \bibinfo {pages}
  {1671} (\bibinfo {year} {2025})}\BibitemShut {NoStop}%
\bibitem [{\citenamefont {O'Brien}\ \emph {et~al.}(2025)\citenamefont
  {O'Brien}, \citenamefont {Amouzegar}, \citenamefont {Lee}, \citenamefont
  {Ritter},\ and\ \citenamefont {Koll{\'a}r}}]{o2025circuit}%
  \BibitemOpen
  \bibfield  {author} {\bibinfo {author} {\bibfnamefont {K.}~\bibnamefont
  {O'Brien}}, \bibinfo {author} {\bibfnamefont {M.}~\bibnamefont {Amouzegar}},
  \bibinfo {author} {\bibfnamefont {W.~C.}\ \bibnamefont {Lee}}, \bibinfo
  {author} {\bibfnamefont {M.}~\bibnamefont {Ritter}},\ and\ \bibinfo {author}
  {\bibfnamefont {A.~J.}\ \bibnamefont {Koll{\'a}r}},\ }\bibfield  {title}
  {\bibinfo {title} {A circuit-qed lattice system with flexible connectivity
  and gapped flat bands for photon-mediated spin models},\ }\href
  {https://arxiv.org/pdf/2505.05559} {\bibfield  {journal} {\bibinfo  {journal}
  {arXiv preprint arXiv:2505.05559}\ } (\bibinfo {year} {2025})}\BibitemShut
  {NoStop}%
\bibitem [{\citenamefont {Wang}\ \emph {et~al.}(2025)\citenamefont {Wang},
  \citenamefont {Ye}, \citenamefont {Yan}, \citenamefont {Qiao}, \citenamefont
  {Liu}, \citenamefont {Ye}, \citenamefont {Chen}, \citenamefont {Cheng},
  \citenamefont {Li}, \citenamefont {Zhang}, \citenamefont {Huang},
  \citenamefont {Meng}, \citenamefont {Zou}, \citenamefont {Zhan},
  \citenamefont {Zhao}, \citenamefont {Hu}, \citenamefont {Tee}, \citenamefont
  {Sha}, \citenamefont {Huang}, \citenamefont {Liu}, \citenamefont {Jin},
  \citenamefont {Ying},\ and\ \citenamefont {Liu}}]{wang2024cavity}%
  \BibitemOpen
  \bibfield  {author} {\bibinfo {author} {\bibfnamefont {Y.-T.}\ \bibnamefont
  {Wang}}, \bibinfo {author} {\bibfnamefont {Q.-H.}\ \bibnamefont {Ye}},
  \bibinfo {author} {\bibfnamefont {J.-Y.}\ \bibnamefont {Yan}}, \bibinfo
  {author} {\bibfnamefont {Y.}~\bibnamefont {Qiao}}, \bibinfo {author}
  {\bibfnamefont {Y.-X.}\ \bibnamefont {Liu}}, \bibinfo {author} {\bibfnamefont
  {Y.-Z.}\ \bibnamefont {Ye}}, \bibinfo {author} {\bibfnamefont
  {C.}~\bibnamefont {Chen}}, \bibinfo {author} {\bibfnamefont {X.-T.}\
  \bibnamefont {Cheng}}, \bibinfo {author} {\bibfnamefont {C.-H.}\ \bibnamefont
  {Li}}, \bibinfo {author} {\bibfnamefont {Z.-J.}\ \bibnamefont {Zhang}},
  \bibinfo {author} {\bibfnamefont {C.-N.}\ \bibnamefont {Huang}}, \bibinfo
  {author} {\bibfnamefont {Y.}~\bibnamefont {Meng}}, \bibinfo {author}
  {\bibfnamefont {K.}~\bibnamefont {Zou}}, \bibinfo {author} {\bibfnamefont
  {W.-K.}\ \bibnamefont {Zhan}}, \bibinfo {author} {\bibfnamefont
  {C.}~\bibnamefont {Zhao}}, \bibinfo {author} {\bibfnamefont {X.}~\bibnamefont
  {Hu}}, \bibinfo {author} {\bibfnamefont {C.~A. T.~H.}\ \bibnamefont {Tee}},
  \bibinfo {author} {\bibfnamefont {W.~E.~I.}\ \bibnamefont {Sha}}, \bibinfo
  {author} {\bibfnamefont {Z.}~\bibnamefont {Huang}}, \bibinfo {author}
  {\bibfnamefont {H.}~\bibnamefont {Liu}}, \bibinfo {author} {\bibfnamefont
  {C.-Y.}\ \bibnamefont {Jin}}, \bibinfo {author} {\bibfnamefont
  {L.}~\bibnamefont {Ying}},\ and\ \bibinfo {author} {\bibfnamefont
  {F.}~\bibnamefont {Liu}},\ }\bibfield  {title} {\bibinfo {title} {Moiré
  cavity quantum electrodynamics},\ }\href
  {https://doi.org/10.1126/sciadv.adv8115} {\bibfield  {journal} {\bibinfo
  {journal} {Science Advances}\ }\textbf {\bibinfo {volume} {11}},\ \bibinfo
  {pages} {eadv8115} (\bibinfo {year} {2025})}\BibitemShut {NoStop}%
\bibitem [{\citenamefont {Sierra}\ \emph {et~al.}(2022)\citenamefont {Sierra},
  \citenamefont {Masson},\ and\ \citenamefont
  {Asenjo-Garcia}}]{sierra2022dicke}%
  \BibitemOpen
  \bibfield  {author} {\bibinfo {author} {\bibfnamefont {E.}~\bibnamefont
  {Sierra}}, \bibinfo {author} {\bibfnamefont {S.~J.}\ \bibnamefont {Masson}},\
  and\ \bibinfo {author} {\bibfnamefont {A.}~\bibnamefont {Asenjo-Garcia}},\
  }\bibfield  {title} {\bibinfo {title} {Dicke superradiance in ordered
  lattices: Dimensionality matters},\ }\href
  {https://doi.org/10.1103/PhysRevResearch.4.023207} {\bibfield  {journal}
  {\bibinfo  {journal} {Phys. Rev. Res.}\ }\textbf {\bibinfo {volume} {4}},\
  \bibinfo {pages} {023207} (\bibinfo {year} {2022})}\BibitemShut {NoStop}%
\bibitem [{\citenamefont {Cardenas-Lopez}\ \emph {et~al.}(2023)\citenamefont
  {Cardenas-Lopez}, \citenamefont {Masson}, \citenamefont {Zager},\ and\
  \citenamefont {Asenjo-Garcia}}]{Cardenas}%
  \BibitemOpen
  \bibfield  {author} {\bibinfo {author} {\bibfnamefont {S.}~\bibnamefont
  {Cardenas-Lopez}}, \bibinfo {author} {\bibfnamefont {S.~J.}\ \bibnamefont
  {Masson}}, \bibinfo {author} {\bibfnamefont {Z.}~\bibnamefont {Zager}},\ and\
  \bibinfo {author} {\bibfnamefont {A.}~\bibnamefont {Asenjo-Garcia}},\
  }\bibfield  {title} {\bibinfo {title} {Many-body superradiance and dynamical
  mirror symmetry breaking in waveguide {QED}},\ }\href
  {https://doi.org/10.1103/PhysRevLett.131.033605} {\bibfield  {journal}
  {\bibinfo  {journal} {Phys. Rev. Lett.}\ }\textbf {\bibinfo {volume} {131}},\
  \bibinfo {pages} {033605} (\bibinfo {year} {2023})}\BibitemShut {NoStop}%
\bibitem [{\citenamefont {Liedl}\ \emph {et~al.}(2023)\citenamefont {Liedl},
  \citenamefont {Pucher}, \citenamefont {Tebbenjohanns}, \citenamefont
  {Schneeweiss},\ and\ \citenamefont {Rauschenbeutel}}]{Arno_super}%
  \BibitemOpen
  \bibfield  {author} {\bibinfo {author} {\bibfnamefont {C.}~\bibnamefont
  {Liedl}}, \bibinfo {author} {\bibfnamefont {S.}~\bibnamefont {Pucher}},
  \bibinfo {author} {\bibfnamefont {F.}~\bibnamefont {Tebbenjohanns}}, \bibinfo
  {author} {\bibfnamefont {P.}~\bibnamefont {Schneeweiss}},\ and\ \bibinfo
  {author} {\bibfnamefont {A.}~\bibnamefont {Rauschenbeutel}},\ }\bibfield
  {title} {\bibinfo {title} {Collective radiation of a cascaded quantum system:
  From timed {Dicke} states to inverted ensembles},\ }\href
  {https://doi.org/10.1103/PhysRevLett.130.163602} {\bibfield  {journal}
  {\bibinfo  {journal} {Phys. Rev. Lett.}\ }\textbf {\bibinfo {volume} {130}},\
  \bibinfo {pages} {163602} (\bibinfo {year} {2023})}\BibitemShut {NoStop}%
\bibitem [{\citenamefont {Asenjo-Garcia}\ \emph {et~al.}(2017)\citenamefont
  {Asenjo-Garcia}, \citenamefont {Moreno-Cardoner}, \citenamefont {Albrecht},
  \citenamefont {Kimble},\ and\ \citenamefont {Chang}}]{asenjo2017exponential}%
  \BibitemOpen
  \bibfield  {author} {\bibinfo {author} {\bibfnamefont {A.}~\bibnamefont
  {Asenjo-Garcia}}, \bibinfo {author} {\bibfnamefont {M.}~\bibnamefont
  {Moreno-Cardoner}}, \bibinfo {author} {\bibfnamefont {A.}~\bibnamefont
  {Albrecht}}, \bibinfo {author} {\bibfnamefont {H.~J.}\ \bibnamefont
  {Kimble}},\ and\ \bibinfo {author} {\bibfnamefont {D.~E.}\ \bibnamefont
  {Chang}},\ }\bibfield  {title} {\bibinfo {title} {Exponential improvement in
  photon storage fidelities using subradiance and ``selective radiance'' in
  atomic arrays},\ }\href {https://doi.org/10.1103/PhysRevX.7.031024}
  {\bibfield  {journal} {\bibinfo  {journal} {Phys. Rev. X}\ }\textbf {\bibinfo
  {volume} {7}},\ \bibinfo {pages} {031024} (\bibinfo {year}
  {2017})}\BibitemShut {NoStop}%
\bibitem [{\citenamefont {Albrecht}\ \emph {et~al.}(2019)\citenamefont
  {Albrecht}, \citenamefont {Henriet}, \citenamefont {Asenjo-Garcia},
  \citenamefont {Dieterle}, \citenamefont {Painter},\ and\ \citenamefont
  {Chang}}]{albrecht2019subradiant}%
  \BibitemOpen
  \bibfield  {author} {\bibinfo {author} {\bibfnamefont {A.}~\bibnamefont
  {Albrecht}}, \bibinfo {author} {\bibfnamefont {L.}~\bibnamefont {Henriet}},
  \bibinfo {author} {\bibfnamefont {A.}~\bibnamefont {Asenjo-Garcia}}, \bibinfo
  {author} {\bibfnamefont {P.~B.}\ \bibnamefont {Dieterle}}, \bibinfo {author}
  {\bibfnamefont {O.}~\bibnamefont {Painter}},\ and\ \bibinfo {author}
  {\bibfnamefont {D.~E.}\ \bibnamefont {Chang}},\ }\bibfield  {title} {\bibinfo
  {title} {Subradiant states of quantum bits coupled to a one-dimensional
  waveguide},\ }\href {https://doi.org/10.1088/1367-2630/ab0134} {\bibfield
  {journal} {\bibinfo  {journal} {New Journal of Physics}\ }\textbf {\bibinfo
  {volume} {21}},\ \bibinfo {pages} {025003} (\bibinfo {year}
  {2019})}\BibitemShut {NoStop}%
\bibitem [{\citenamefont {Zhang}\ and\ \citenamefont {Mølmer}(2019)}]{Molmer}%
  \BibitemOpen
  \bibfield  {author} {\bibinfo {author} {\bibfnamefont {Y.-X.}\ \bibnamefont
  {Zhang}}\ and\ \bibinfo {author} {\bibfnamefont {K.}~\bibnamefont
  {Mølmer}},\ }\bibfield  {title} {\bibinfo {title} {Theory of subradiant
  states of a one-dimensional two-level atom chain},\ }\href
  {https://doi.org/10.1103/PhysRevLett.122.203605} {\bibfield  {journal}
  {\bibinfo  {journal} {Phys. Rev. Lett.}\ }\textbf {\bibinfo {volume} {122}},\
  \bibinfo {pages} {203605} (\bibinfo {year} {2019})}\BibitemShut {NoStop}%
\bibitem [{\citenamefont {Needham}\ \emph {et~al.}(2019)\citenamefont
  {Needham}, \citenamefont {Lesanovsky},\ and\ \citenamefont
  {Olmos}}]{needham2019subradiance}%
  \BibitemOpen
  \bibfield  {author} {\bibinfo {author} {\bibfnamefont {J.~A.}\ \bibnamefont
  {Needham}}, \bibinfo {author} {\bibfnamefont {I.}~\bibnamefont
  {Lesanovsky}},\ and\ \bibinfo {author} {\bibfnamefont {B.}~\bibnamefont
  {Olmos}},\ }\bibfield  {title} {\bibinfo {title} {Subradiance-protected
  excitation transport},\ }\href {https://doi.org/10.1088/1367-2630/ab31e8}
  {\bibfield  {journal} {\bibinfo  {journal} {New Journal of Physics}\ }\textbf
  {\bibinfo {volume} {21}},\ \bibinfo {pages} {073061} (\bibinfo {year}
  {2019})}\BibitemShut {NoStop}%
\bibitem [{\citenamefont {Kornovan}\ \emph {et~al.}(2019)\citenamefont
  {Kornovan}, \citenamefont {Corzo}, \citenamefont {Laurat},\ and\
  \citenamefont {Sheremet}}]{kornovan2019extremely}%
  \BibitemOpen
  \bibfield  {author} {\bibinfo {author} {\bibfnamefont {D.~F.}\ \bibnamefont
  {Kornovan}}, \bibinfo {author} {\bibfnamefont {N.~V.}\ \bibnamefont {Corzo}},
  \bibinfo {author} {\bibfnamefont {J.}~\bibnamefont {Laurat}},\ and\ \bibinfo
  {author} {\bibfnamefont {A.~S.}\ \bibnamefont {Sheremet}},\ }\bibfield
  {title} {\bibinfo {title} {Extremely subradiant states in a periodic
  one-dimensional atomic array},\ }\href
  {https://doi.org/10.1103/PhysRevA.100.063832} {\bibfield  {journal} {\bibinfo
   {journal} {Phys. Rev. A}\ }\textbf {\bibinfo {volume} {100}},\ \bibinfo
  {pages} {063832} (\bibinfo {year} {2019})}\BibitemShut {NoStop}%
\bibitem [{\citenamefont {Ostermann}\ \emph {et~al.}(2019)\citenamefont
  {Ostermann}, \citenamefont {Meignant}, \citenamefont {Genes},\ and\
  \citenamefont {Ritsch}}]{ostermann2019super}%
  \BibitemOpen
  \bibfield  {author} {\bibinfo {author} {\bibfnamefont {L.}~\bibnamefont
  {Ostermann}}, \bibinfo {author} {\bibfnamefont {C.}~\bibnamefont {Meignant}},
  \bibinfo {author} {\bibfnamefont {C.}~\bibnamefont {Genes}},\ and\ \bibinfo
  {author} {\bibfnamefont {H.}~\bibnamefont {Ritsch}},\ }\bibfield  {title}
  {\bibinfo {title} {Super- and subradiance of clock atoms in multimode optical
  waveguides},\ }\href {https://doi.org/10.1088/1367-2630/ab05fb} {\bibfield
  {journal} {\bibinfo  {journal} {New Journal of Physics}\ }\textbf {\bibinfo
  {volume} {21}},\ \bibinfo {pages} {025004} (\bibinfo {year}
  {2019})}\BibitemShut {NoStop}%
\bibitem [{\citenamefont {Poshakinskiy}\ and\ \citenamefont
  {Poddubny}(2021)}]{poshakinskiy2021dimerization}%
  \BibitemOpen
  \bibfield  {author} {\bibinfo {author} {\bibfnamefont {A.~V.}\ \bibnamefont
  {Poshakinskiy}}\ and\ \bibinfo {author} {\bibfnamefont {A.~N.}\ \bibnamefont
  {Poddubny}},\ }\bibfield  {title} {\bibinfo {title} {Dimerization of
  many-body subradiant states in waveguide quantum electrodynamics},\ }\href
  {https://doi.org/10.1103/PhysRevLett.127.173601} {\bibfield  {journal}
  {\bibinfo  {journal} {Phys. Rev. Lett.}\ }\textbf {\bibinfo {volume} {127}},\
  \bibinfo {pages} {173601} (\bibinfo {year} {2021})}\BibitemShut {NoStop}%
\bibitem [{\citenamefont {Calaj\'o}\ and\ \citenamefont
  {Chang}(2022)}]{calajo2022emergence}%
  \BibitemOpen
  \bibfield  {author} {\bibinfo {author} {\bibfnamefont {G.}~\bibnamefont
  {Calaj\'o}}\ and\ \bibinfo {author} {\bibfnamefont {D.~E.}\ \bibnamefont
  {Chang}},\ }\bibfield  {title} {\bibinfo {title} {Emergence of solitons from
  many-body photon bound states in quantum nonlinear media},\ }\href
  {https://doi.org/10.1103/PhysRevResearch.4.023026} {\bibfield  {journal}
  {\bibinfo  {journal} {Phys. Rev. Res.}\ }\textbf {\bibinfo {volume} {4}},\
  \bibinfo {pages} {023026} (\bibinfo {year} {2022})}\BibitemShut {NoStop}%
\bibitem [{\citenamefont {Schrinski}\ and\ \citenamefont
  {Sørensen}(2022)}]{Schrinski_polariton}%
  \BibitemOpen
  \bibfield  {author} {\bibinfo {author} {\bibfnamefont {B.}~\bibnamefont
  {Schrinski}}\ and\ \bibinfo {author} {\bibfnamefont {A.~S.}\ \bibnamefont
  {Sørensen}},\ }\bibfield  {title} {\bibinfo {title} {Polariton dynamics in
  one-dimensional arrays of atoms coupled to waveguides},\ }\href
  {https://doi.org/10.1088/1367-2630/acaa4f} {\bibfield  {journal} {\bibinfo
  {journal} {New Journal of Physics}\ }\textbf {\bibinfo {volume} {24}},\
  \bibinfo {pages} {123023} (\bibinfo {year} {2022})}\BibitemShut {NoStop}%
\bibitem [{\citenamefont {Jenkins}\ and\ \citenamefont
  {Ruostekoski}(2012)}]{PhysRevA.86.031602}%
  \BibitemOpen
  \bibfield  {author} {\bibinfo {author} {\bibfnamefont {S.~D.}\ \bibnamefont
  {Jenkins}}\ and\ \bibinfo {author} {\bibfnamefont {J.}~\bibnamefont
  {Ruostekoski}},\ }\bibfield  {title} {\bibinfo {title} {Controlled
  manipulation of light by cooperative response of atoms in an optical
  lattice},\ }\href {https://doi.org/10.1103/PhysRevA.86.031602} {\bibfield
  {journal} {\bibinfo  {journal} {Phys. Rev. A}\ }\textbf {\bibinfo {volume}
  {86}},\ \bibinfo {pages} {031602} (\bibinfo {year} {2012})}\BibitemShut
  {NoStop}%
\bibitem [{\citenamefont {Bettles}\ \emph {et~al.}(2016)\citenamefont
  {Bettles}, \citenamefont {Gardiner},\ and\ \citenamefont
  {Adams}}]{bettles2016enhanced}%
  \BibitemOpen
  \bibfield  {author} {\bibinfo {author} {\bibfnamefont {R.~J.}\ \bibnamefont
  {Bettles}}, \bibinfo {author} {\bibfnamefont {S.~A.}\ \bibnamefont
  {Gardiner}},\ and\ \bibinfo {author} {\bibfnamefont {C.~S.}\ \bibnamefont
  {Adams}},\ }\bibfield  {title} {\bibinfo {title} {Enhanced optical cross
  section via collective coupling of atomic dipoles in a 2d array},\ }\href
  {https://doi.org/10.1103/PhysRevLett.116.103602} {\bibfield  {journal}
  {\bibinfo  {journal} {Phys. Rev. Lett.}\ }\textbf {\bibinfo {volume} {116}},\
  \bibinfo {pages} {103602} (\bibinfo {year} {2016})}\BibitemShut {NoStop}%
\bibitem [{\citenamefont {Shahmoon}\ \emph {et~al.}(2017)\citenamefont
  {Shahmoon}, \citenamefont {Wild}, \citenamefont {Lukin},\ and\ \citenamefont
  {Yelin}}]{shahmoon2017cooperative}%
  \BibitemOpen
  \bibfield  {author} {\bibinfo {author} {\bibfnamefont {E.}~\bibnamefont
  {Shahmoon}}, \bibinfo {author} {\bibfnamefont {D.~S.}\ \bibnamefont {Wild}},
  \bibinfo {author} {\bibfnamefont {M.~D.}\ \bibnamefont {Lukin}},\ and\
  \bibinfo {author} {\bibfnamefont {S.~F.}\ \bibnamefont {Yelin}},\ }\bibfield
  {title} {\bibinfo {title} {Cooperative resonances in light scattering from
  two-dimensional atomic arrays},\ }\href
  {https://doi.org/10.1103/PhysRevLett.118.113601} {\bibfield  {journal}
  {\bibinfo  {journal} {Phys. Rev. Lett.}\ }\textbf {\bibinfo {volume} {118}},\
  \bibinfo {pages} {113601} (\bibinfo {year} {2017})}\BibitemShut {NoStop}%
\bibitem [{\citenamefont {Manzoni}\ \emph {et~al.}(2018)\citenamefont
  {Manzoni}, \citenamefont {Moreno-Cardoner}, \citenamefont {Asenjo-Garcia},
  \citenamefont {Porto}, \citenamefont {Gorshkov},\ and\ \citenamefont
  {Chang}}]{manzoni2018optimization}%
  \BibitemOpen
  \bibfield  {author} {\bibinfo {author} {\bibfnamefont {M.~T.}\ \bibnamefont
  {Manzoni}}, \bibinfo {author} {\bibfnamefont {M.}~\bibnamefont
  {Moreno-Cardoner}}, \bibinfo {author} {\bibfnamefont {A.}~\bibnamefont
  {Asenjo-Garcia}}, \bibinfo {author} {\bibfnamefont {J.~V.}\ \bibnamefont
  {Porto}}, \bibinfo {author} {\bibfnamefont {A.~V.}\ \bibnamefont
  {Gorshkov}},\ and\ \bibinfo {author} {\bibfnamefont {D.~E.}\ \bibnamefont
  {Chang}},\ }\bibfield  {title} {\bibinfo {title} {Optimization of photon
  storage fidelity in ordered atomic arrays},\ }\href
  {https://doi.org/10.1088/1367-2630/aadb74} {\bibfield  {journal} {\bibinfo
  {journal} {New Journal of Physics}\ }\textbf {\bibinfo {volume} {20}},\
  \bibinfo {pages} {083048} (\bibinfo {year} {2018})}\BibitemShut {NoStop}%
\bibitem [{\citenamefont {Facchinetti}\ \emph {et~al.}(2016)\citenamefont
  {Facchinetti}, \citenamefont {Jenkins},\ and\ \citenamefont
  {Ruostekoski}}]{PhysRevLett.117.243601}%
  \BibitemOpen
  \bibfield  {author} {\bibinfo {author} {\bibfnamefont {G.}~\bibnamefont
  {Facchinetti}}, \bibinfo {author} {\bibfnamefont {S.~D.}\ \bibnamefont
  {Jenkins}},\ and\ \bibinfo {author} {\bibfnamefont {J.}~\bibnamefont
  {Ruostekoski}},\ }\bibfield  {title} {\bibinfo {title} {Storing light with
  subradiant correlations in arrays of atoms},\ }\href
  {https://doi.org/10.1103/PhysRevLett.117.243601} {\bibfield  {journal}
  {\bibinfo  {journal} {Phys. Rev. Lett.}\ }\textbf {\bibinfo {volume} {117}},\
  \bibinfo {pages} {243601} (\bibinfo {year} {2016})}\BibitemShut {NoStop}%
\bibitem [{\citenamefont {Ruostekoski}(2023)}]{PhysRevA.108.030101}%
  \BibitemOpen
  \bibfield  {author} {\bibinfo {author} {\bibfnamefont {J.}~\bibnamefont
  {Ruostekoski}},\ }\bibfield  {title} {\bibinfo {title} {Cooperative
  quantum-optical planar arrays of atoms},\ }\href
  {https://doi.org/10.1103/PhysRevA.108.030101} {\bibfield  {journal} {\bibinfo
   {journal} {Phys. Rev. A}\ }\textbf {\bibinfo {volume} {108}},\ \bibinfo
  {pages} {030101} (\bibinfo {year} {2023})}\BibitemShut {NoStop}%
\bibitem [{\citenamefont {Perczel}\ \emph
  {et~al.}(2017{\natexlab{a}})\citenamefont {Perczel}, \citenamefont
  {Borregaard}, \citenamefont {Chang}, \citenamefont {Pichler}, \citenamefont
  {Yelin}, \citenamefont {Zoller},\ and\ \citenamefont
  {Lukin}}]{perczel2017topological}%
  \BibitemOpen
  \bibfield  {author} {\bibinfo {author} {\bibfnamefont {J.}~\bibnamefont
  {Perczel}}, \bibinfo {author} {\bibfnamefont {J.}~\bibnamefont {Borregaard}},
  \bibinfo {author} {\bibfnamefont {D.~E.}\ \bibnamefont {Chang}}, \bibinfo
  {author} {\bibfnamefont {H.}~\bibnamefont {Pichler}}, \bibinfo {author}
  {\bibfnamefont {S.~F.}\ \bibnamefont {Yelin}}, \bibinfo {author}
  {\bibfnamefont {P.}~\bibnamefont {Zoller}},\ and\ \bibinfo {author}
  {\bibfnamefont {M.~D.}\ \bibnamefont {Lukin}},\ }\bibfield  {title} {\bibinfo
  {title} {Topological quantum optics in two-dimensional atomic arrays},\
  }\href {https://doi.org/10.1103/PhysRevLett.119.023603} {\bibfield  {journal}
  {\bibinfo  {journal} {Phys. Rev. Lett.}\ }\textbf {\bibinfo {volume} {119}},\
  \bibinfo {pages} {023603} (\bibinfo {year} {2017}{\natexlab{a}})}\BibitemShut
  {NoStop}%
\bibitem [{\citenamefont {Perczel}\ \emph
  {et~al.}(2017{\natexlab{b}})\citenamefont {Perczel}, \citenamefont
  {Borregaard}, \citenamefont {Chang}, \citenamefont {Pichler}, \citenamefont
  {Yelin}, \citenamefont {Zoller},\ and\ \citenamefont
  {Lukin}}]{perczel2017photonic}%
  \BibitemOpen
  \bibfield  {author} {\bibinfo {author} {\bibfnamefont {J.}~\bibnamefont
  {Perczel}}, \bibinfo {author} {\bibfnamefont {J.}~\bibnamefont {Borregaard}},
  \bibinfo {author} {\bibfnamefont {D.~E.}\ \bibnamefont {Chang}}, \bibinfo
  {author} {\bibfnamefont {H.}~\bibnamefont {Pichler}}, \bibinfo {author}
  {\bibfnamefont {S.~F.}\ \bibnamefont {Yelin}}, \bibinfo {author}
  {\bibfnamefont {P.}~\bibnamefont {Zoller}},\ and\ \bibinfo {author}
  {\bibfnamefont {M.~D.}\ \bibnamefont {Lukin}},\ }\bibfield  {title} {\bibinfo
  {title} {Photonic band structure of two-dimensional atomic lattices},\ }\href
  {https://doi.org/10.1103/PhysRevA.96.063801} {\bibfield  {journal} {\bibinfo
  {journal} {Phys. Rev. A}\ }\textbf {\bibinfo {volume} {96}},\ \bibinfo
  {pages} {063801} (\bibinfo {year} {2017}{\natexlab{b}})}\BibitemShut
  {NoStop}%
\bibitem [{\citenamefont {Perczel}\ \emph {et~al.}(2020)\citenamefont
  {Perczel}, \citenamefont {Borregaard}, \citenamefont {Chang}, \citenamefont
  {Yelin},\ and\ \citenamefont {Lukin}}]{perczel2020topological}%
  \BibitemOpen
  \bibfield  {author} {\bibinfo {author} {\bibfnamefont {J.}~\bibnamefont
  {Perczel}}, \bibinfo {author} {\bibfnamefont {J.}~\bibnamefont {Borregaard}},
  \bibinfo {author} {\bibfnamefont {D.~E.}\ \bibnamefont {Chang}}, \bibinfo
  {author} {\bibfnamefont {S.~F.}\ \bibnamefont {Yelin}},\ and\ \bibinfo
  {author} {\bibfnamefont {M.~D.}\ \bibnamefont {Lukin}},\ }\bibfield  {title}
  {\bibinfo {title} {Topological quantum optics using atomlike emitter arrays
  coupled to photonic crystals},\ }\href
  {https://doi.org/10.1103/PhysRevLett.124.083603} {\bibfield  {journal}
  {\bibinfo  {journal} {Phys. Rev. Lett.}\ }\textbf {\bibinfo {volume} {124}},\
  \bibinfo {pages} {083603} (\bibinfo {year} {2020})}\BibitemShut {NoStop}%
\bibitem [{\citenamefont {Marques}\ \emph
  {et~al.}(2021{\natexlab{a}})\citenamefont {Marques}, \citenamefont
  {Shelykh},\ and\ \citenamefont {Iorsh}}]{PhysRevA.103.033702}%
  \BibitemOpen
  \bibfield  {author} {\bibinfo {author} {\bibfnamefont {Y.}~\bibnamefont
  {Marques}}, \bibinfo {author} {\bibfnamefont {I.~A.}\ \bibnamefont
  {Shelykh}},\ and\ \bibinfo {author} {\bibfnamefont {I.~V.}\ \bibnamefont
  {Iorsh}},\ }\bibfield  {title} {\bibinfo {title} {Two-dimensional
  chiral-waveguide quantum electrodynamics: Long-range qubit correlations and
  flat-band dark polaritons},\ }\href
  {https://doi.org/10.1103/PhysRevA.103.033702} {\bibfield  {journal} {\bibinfo
   {journal} {Phys. Rev. A}\ }\textbf {\bibinfo {volume} {103}},\ \bibinfo
  {pages} {033702} (\bibinfo {year} {2021}{\natexlab{a}})}\BibitemShut
  {NoStop}%
\bibitem [{\citenamefont {Parmee}\ \emph {et~al.}(2022)\citenamefont {Parmee},
  \citenamefont {Ballantine},\ and\ \citenamefont
  {Ruostekoski}}]{PhysRevResearch.4.043039}%
  \BibitemOpen
  \bibfield  {author} {\bibinfo {author} {\bibfnamefont {C.~D.}\ \bibnamefont
  {Parmee}}, \bibinfo {author} {\bibfnamefont {K.~E.}\ \bibnamefont
  {Ballantine}},\ and\ \bibinfo {author} {\bibfnamefont {J.}~\bibnamefont
  {Ruostekoski}},\ }\bibfield  {title} {\bibinfo {title} {Spontaneous symmetry
  breaking in frustrated triangular atom arrays due to cooperative light
  scattering},\ }\href {https://doi.org/10.1103/PhysRevResearch.4.043039}
  {\bibfield  {journal} {\bibinfo  {journal} {Phys. Rev. Res.}\ }\textbf
  {\bibinfo {volume} {4}},\ \bibinfo {pages} {043039} (\bibinfo {year}
  {2022})}\BibitemShut {NoStop}%
\bibitem [{\citenamefont {Shen}\ and\ \citenamefont
  {Fan}(2007{\natexlab{a}})}]{shen2007strongly}%
  \BibitemOpen
  \bibfield  {author} {\bibinfo {author} {\bibfnamefont {J.-T.}\ \bibnamefont
  {Shen}}\ and\ \bibinfo {author} {\bibfnamefont {S.}~\bibnamefont {Fan}},\
  }\bibfield  {title} {\bibinfo {title} {Strongly correlated multiparticle
  transport in one dimension through a quantum impurity},\ }\href
  {https://doi.org/10.1103/PhysRevA.76.062709} {\bibfield  {journal} {\bibinfo
  {journal} {Phys. Rev. A}\ }\textbf {\bibinfo {volume} {76}},\ \bibinfo
  {pages} {062709} (\bibinfo {year} {2007}{\natexlab{a}})}\BibitemShut
  {NoStop}%
\bibitem [{\citenamefont {Shen}\ and\ \citenamefont
  {Fan}(2007{\natexlab{b}})}]{shen2007stronglyL}%
  \BibitemOpen
  \bibfield  {author} {\bibinfo {author} {\bibfnamefont {J.-T.}\ \bibnamefont
  {Shen}}\ and\ \bibinfo {author} {\bibfnamefont {S.}~\bibnamefont {Fan}},\
  }\bibfield  {title} {\bibinfo {title} {Strongly correlated two-photon
  transport in a one-dimensional waveguide coupled to a two-level system},\
  }\href {https://doi.org/10.1103/PhysRevLett.98.153003} {\bibfield  {journal}
  {\bibinfo  {journal} {Phys. Rev. Lett.}\ }\textbf {\bibinfo {volume} {98}},\
  \bibinfo {pages} {153003} (\bibinfo {year} {2007}{\natexlab{b}})}\BibitemShut
  {NoStop}%
\bibitem [{\citenamefont {Zheng}\ \emph {et~al.}(2011)\citenamefont {Zheng},
  \citenamefont {Gauthier},\ and\ \citenamefont {Baranger}}]{zheng2011cavity}%
  \BibitemOpen
  \bibfield  {author} {\bibinfo {author} {\bibfnamefont {H.}~\bibnamefont
  {Zheng}}, \bibinfo {author} {\bibfnamefont {D.~J.}\ \bibnamefont
  {Gauthier}},\ and\ \bibinfo {author} {\bibfnamefont {H.~U.}\ \bibnamefont
  {Baranger}},\ }\bibfield  {title} {\bibinfo {title} {Cavity-free photon
  blockade induced by many-body bound states},\ }\href
  {http://dx.doi.org/10.1103/PhysRevLett.107.223601} {\bibfield  {journal}
  {\bibinfo  {journal} {Physical Review Letters}\ }\textbf {\bibinfo {volume}
  {107}} (\bibinfo {year} {2011})}\BibitemShut {NoStop}%
\bibitem [{\citenamefont {Mahmoodian}\ \emph {et~al.}(2018)\citenamefont
  {Mahmoodian}, \citenamefont {\ifmmode~\check{C}\else \v{C}\fi{}epulkovskis},
  \citenamefont {Das}, \citenamefont {Lodahl}, \citenamefont {Hammerer},\ and\
  \citenamefont {S\o{}rensen}}]{mahmoodian2018strongly}%
  \BibitemOpen
  \bibfield  {author} {\bibinfo {author} {\bibfnamefont {S.}~\bibnamefont
  {Mahmoodian}}, \bibinfo {author} {\bibfnamefont {M.}~\bibnamefont
  {\ifmmode~\check{C}\else \v{C}\fi{}epulkovskis}}, \bibinfo {author}
  {\bibfnamefont {S.}~\bibnamefont {Das}}, \bibinfo {author} {\bibfnamefont
  {P.}~\bibnamefont {Lodahl}}, \bibinfo {author} {\bibfnamefont
  {K.}~\bibnamefont {Hammerer}},\ and\ \bibinfo {author} {\bibfnamefont
  {A.~S.}\ \bibnamefont {S\o{}rensen}},\ }\bibfield  {title} {\bibinfo {title}
  {Strongly correlated photon transport in waveguide quantum electrodynamics
  with weakly coupled emitters},\ }\href
  {https://doi.org/10.1103/PhysRevLett.121.143601} {\bibfield  {journal}
  {\bibinfo  {journal} {Phys. Rev. Lett.}\ }\textbf {\bibinfo {volume} {121}},\
  \bibinfo {pages} {143601} (\bibinfo {year} {2018})}\BibitemShut {NoStop}%
\bibitem [{\citenamefont {Mahmoodian}\ \emph {et~al.}(2020)\citenamefont
  {Mahmoodian}, \citenamefont {Calaj\'o}, \citenamefont {Chang}, \citenamefont
  {Hammerer},\ and\ \citenamefont {S\o{}rensen}}]{mahmoodian2020dynamics}%
  \BibitemOpen
  \bibfield  {author} {\bibinfo {author} {\bibfnamefont {S.}~\bibnamefont
  {Mahmoodian}}, \bibinfo {author} {\bibfnamefont {G.}~\bibnamefont
  {Calaj\'o}}, \bibinfo {author} {\bibfnamefont {D.~E.}\ \bibnamefont {Chang}},
  \bibinfo {author} {\bibfnamefont {K.}~\bibnamefont {Hammerer}},\ and\
  \bibinfo {author} {\bibfnamefont {A.~S.}\ \bibnamefont {S\o{}rensen}},\
  }\bibfield  {title} {\bibinfo {title} {Dynamics of many-body photon bound
  states in chiral waveguide qed},\ }\href
  {https://doi.org/10.1103/PhysRevX.10.031011} {\bibfield  {journal} {\bibinfo
  {journal} {Phys. Rev. X}\ }\textbf {\bibinfo {volume} {10}},\ \bibinfo
  {pages} {031011} (\bibinfo {year} {2020})}\BibitemShut {NoStop}%
\bibitem [{\citenamefont {Le~Jeannic}\ \emph {et~al.}(2022)\citenamefont
  {Le~Jeannic}, \citenamefont {Tiranov}, \citenamefont {Carolan}, \citenamefont
  {Ramos}, \citenamefont {Wang}, \citenamefont {Appel}, \citenamefont {Scholz},
  \citenamefont {Wieck}, \citenamefont {Ludwig}, \citenamefont {Rotenberg},
  \citenamefont {Midolo}, \citenamefont {García-Ripoll}, \citenamefont
  {Sørensen},\ and\ \citenamefont {Lodahl}}]{le2022dynamical}%
  \BibitemOpen
  \bibfield  {author} {\bibinfo {author} {\bibfnamefont {H.}~\bibnamefont
  {Le~Jeannic}}, \bibinfo {author} {\bibfnamefont {A.}~\bibnamefont {Tiranov}},
  \bibinfo {author} {\bibfnamefont {J.}~\bibnamefont {Carolan}}, \bibinfo
  {author} {\bibfnamefont {T.}~\bibnamefont {Ramos}}, \bibinfo {author}
  {\bibfnamefont {Y.}~\bibnamefont {Wang}}, \bibinfo {author} {\bibfnamefont
  {M.~H.}\ \bibnamefont {Appel}}, \bibinfo {author} {\bibfnamefont
  {S.}~\bibnamefont {Scholz}}, \bibinfo {author} {\bibfnamefont {A.~D.}\
  \bibnamefont {Wieck}}, \bibinfo {author} {\bibfnamefont {A.}~\bibnamefont
  {Ludwig}}, \bibinfo {author} {\bibfnamefont {N.}~\bibnamefont {Rotenberg}},
  \bibinfo {author} {\bibfnamefont {L.}~\bibnamefont {Midolo}}, \bibinfo
  {author} {\bibfnamefont {J.~J.}\ \bibnamefont {García-Ripoll}}, \bibinfo
  {author} {\bibfnamefont {A.~S.}\ \bibnamefont {Sørensen}},\ and\ \bibinfo
  {author} {\bibfnamefont {P.}~\bibnamefont {Lodahl}},\ }\bibfield  {title}
  {\bibinfo {title} {Dynamical photon–photon interaction mediated by a
  quantum emitter},\ }\href {https://doi.org/10.1038/s41567-022-01720-x}
  {\bibfield  {journal} {\bibinfo  {journal} {Nature Physics}\ }\textbf
  {\bibinfo {volume} {18}},\ \bibinfo {pages} {1191–1195} (\bibinfo {year}
  {2022})}\BibitemShut {NoStop}%
\bibitem [{\citenamefont {Tomm}\ \emph {et~al.}(2023)\citenamefont {Tomm},
  \citenamefont {Mahmoodian}, \citenamefont {Antoniadis}, \citenamefont
  {Schott}, \citenamefont {Valentin}, \citenamefont {Wieck}, \citenamefont
  {Ludwig}, \citenamefont {Javadi},\ and\ \citenamefont
  {Warburton}}]{tomm2023photon}%
  \BibitemOpen
  \bibfield  {author} {\bibinfo {author} {\bibfnamefont {N.}~\bibnamefont
  {Tomm}}, \bibinfo {author} {\bibfnamefont {S.}~\bibnamefont {Mahmoodian}},
  \bibinfo {author} {\bibfnamefont {N.~O.}\ \bibnamefont {Antoniadis}},
  \bibinfo {author} {\bibfnamefont {R.}~\bibnamefont {Schott}}, \bibinfo
  {author} {\bibfnamefont {S.~R.}\ \bibnamefont {Valentin}}, \bibinfo {author}
  {\bibfnamefont {A.~D.}\ \bibnamefont {Wieck}}, \bibinfo {author}
  {\bibfnamefont {A.}~\bibnamefont {Ludwig}}, \bibinfo {author} {\bibfnamefont
  {A.}~\bibnamefont {Javadi}},\ and\ \bibinfo {author} {\bibfnamefont {R.~J.}\
  \bibnamefont {Warburton}},\ }\bibfield  {title} {\bibinfo {title} {Photon
  bound state dynamics from a single artificial atom},\ }\href
  {https://doi.org/10.1038/s41567-023-01997-6} {\bibfield  {journal} {\bibinfo
  {journal} {Nature Physics}\ }\textbf {\bibinfo {volume} {19}},\ \bibinfo
  {pages} {857–862} (\bibinfo {year} {2023})}\BibitemShut {NoStop}%
\bibitem [{\citenamefont {Zhang}\ and\ \citenamefont
  {M\o{}lmer}(2020)}]{zhang2020subradiant}%
  \BibitemOpen
  \bibfield  {author} {\bibinfo {author} {\bibfnamefont {Y.-X.}\ \bibnamefont
  {Zhang}}\ and\ \bibinfo {author} {\bibfnamefont {K.}~\bibnamefont
  {M\o{}lmer}},\ }\bibfield  {title} {\bibinfo {title} {Subradiant emission
  from regular atomic arrays: Universal scaling of decay rates from the
  generalized bloch theorem},\ }\href
  {https://doi.org/10.1103/PhysRevLett.125.253601} {\bibfield  {journal}
  {\bibinfo  {journal} {Phys. Rev. Lett.}\ }\textbf {\bibinfo {volume} {125}},\
  \bibinfo {pages} {253601} (\bibinfo {year} {2020})}\BibitemShut {NoStop}%
\bibitem [{\citenamefont {Poddubny}(2020)}]{poddubny2020quasiflat}%
  \BibitemOpen
  \bibfield  {author} {\bibinfo {author} {\bibfnamefont {A.~N.}\ \bibnamefont
  {Poddubny}},\ }\bibfield  {title} {\bibinfo {title} {Quasiflat band enabling
  subradiant two-photon bound states},\ }\href
  {https://doi.org/10.1103/PhysRevA.101.043845} {\bibfield  {journal} {\bibinfo
   {journal} {Phys. Rev. A}\ }\textbf {\bibinfo {volume} {101}},\ \bibinfo
  {pages} {043845} (\bibinfo {year} {2020})}\BibitemShut {NoStop}%
\bibitem [{\citenamefont {Bakkensen}\ \emph {et~al.}(2021)\citenamefont
  {Bakkensen}, \citenamefont {Zhang}, \citenamefont {Bjerlin},\ and\
  \citenamefont {S{\o}rensen}}]{bakkensen2021photonic}%
  \BibitemOpen
  \bibfield  {author} {\bibinfo {author} {\bibfnamefont {B.}~\bibnamefont
  {Bakkensen}}, \bibinfo {author} {\bibfnamefont {Y.-X.}\ \bibnamefont
  {Zhang}}, \bibinfo {author} {\bibfnamefont {J.}~\bibnamefont {Bjerlin}},\
  and\ \bibinfo {author} {\bibfnamefont {A.~S.}\ \bibnamefont {S{\o}rensen}},\
  }\bibfield  {title} {\bibinfo {title} {Photonic bound states and scattering
  resonances in waveguide qed},\ }\href
  {https://doi.org/10.48550/arXiv.2110.06093} {\bibfield  {journal} {\bibinfo
  {journal} {arXiv preprint arXiv:2110.06093}\ } (\bibinfo {year}
  {2021})}\BibitemShut {NoStop}%
\bibitem [{\citenamefont {Poshakinskiy}\ and\ \citenamefont
  {Poddubny}(2023)}]{PhysRevA.108.023707}%
  \BibitemOpen
  \bibfield  {author} {\bibinfo {author} {\bibfnamefont {A.~V.}\ \bibnamefont
  {Poshakinskiy}}\ and\ \bibinfo {author} {\bibfnamefont {A.~N.}\ \bibnamefont
  {Poddubny}},\ }\bibfield  {title} {\bibinfo {title} {Bound state of distant
  photons in waveguide quantum electrodynamics},\ }\href
  {https://doi.org/10.1103/PhysRevA.108.023707} {\bibfield  {journal} {\bibinfo
   {journal} {Phys. Rev. A}\ }\textbf {\bibinfo {volume} {108}},\ \bibinfo
  {pages} {023707} (\bibinfo {year} {2023})}\BibitemShut {NoStop}%
\bibitem [{\citenamefont {Marques}\ \emph
  {et~al.}(2021{\natexlab{b}})\citenamefont {Marques}, \citenamefont
  {Shelykh},\ and\ \citenamefont {Iorsh}}]{marques2021bound}%
  \BibitemOpen
  \bibfield  {author} {\bibinfo {author} {\bibfnamefont {Y.}~\bibnamefont
  {Marques}}, \bibinfo {author} {\bibfnamefont {I.~A.}\ \bibnamefont
  {Shelykh}},\ and\ \bibinfo {author} {\bibfnamefont {I.~V.}\ \bibnamefont
  {Iorsh}},\ }\bibfield  {title} {\bibinfo {title} {Bound photonic pairs in 2d
  waveguide quantum electrodynamics},\ }\href
  {https://doi.org/10.1103/PhysRevLett.127.273602} {\bibfield  {journal}
  {\bibinfo  {journal} {Phys. Rev. Lett.}\ }\textbf {\bibinfo {volume} {127}},\
  \bibinfo {pages} {273602} (\bibinfo {year} {2021}{\natexlab{b}})}\BibitemShut
  {NoStop}%
\bibitem [{\citenamefont {Schrinski}\ \emph {et~al.}(2024)\citenamefont
  {Schrinski}, \citenamefont {Brimer},\ and\ \citenamefont
  {S\o{}rensen}}]{schrinski2023photon}%
  \BibitemOpen
  \bibfield  {author} {\bibinfo {author} {\bibfnamefont {B.}~\bibnamefont
  {Schrinski}}, \bibinfo {author} {\bibfnamefont {J.~A.}\ \bibnamefont
  {Brimer}},\ and\ \bibinfo {author} {\bibfnamefont {A.~S.}\ \bibnamefont
  {S\o{}rensen}},\ }\bibfield  {title} {\bibinfo {title} {Photon bound states
  in coupled waveguides},\ }\href
  {https://doi.org/10.1103/PhysRevA.110.L041701} {\bibfield  {journal}
  {\bibinfo  {journal} {Phys. Rev. A}\ }\textbf {\bibinfo {volume} {110}},\
  \bibinfo {pages} {L041701} (\bibinfo {year} {2024})}\BibitemShut {NoStop}%
\bibitem [{\citenamefont {Te\ifmmode~\check{c}\else \v{c}\fi{}er}\ \emph
  {et~al.}(2024)\citenamefont {Te\ifmmode~\check{c}\else \v{c}\fi{}er},
  \citenamefont {Di~Liberto}, \citenamefont {Silvi}, \citenamefont
  {Montangero}, \citenamefont {Romanato},\ and\ \citenamefont
  {Calaj\'o}}]{PhysRevLett.132.163602}%
  \BibitemOpen
  \bibfield  {author} {\bibinfo {author} {\bibfnamefont {M.}~\bibnamefont
  {Te\ifmmode~\check{c}\else \v{c}\fi{}er}}, \bibinfo {author} {\bibfnamefont
  {M.}~\bibnamefont {Di~Liberto}}, \bibinfo {author} {\bibfnamefont
  {P.}~\bibnamefont {Silvi}}, \bibinfo {author} {\bibfnamefont
  {S.}~\bibnamefont {Montangero}}, \bibinfo {author} {\bibfnamefont
  {F.}~\bibnamefont {Romanato}},\ and\ \bibinfo {author} {\bibfnamefont
  {G.}~\bibnamefont {Calaj\'o}},\ }\bibfield  {title} {\bibinfo {title}
  {Strongly interacting photons in 2d waveguide qed},\ }\href
  {https://doi.org/10.1103/PhysRevLett.132.163602} {\bibfield  {journal}
  {\bibinfo  {journal} {Phys. Rev. Lett.}\ }\textbf {\bibinfo {volume} {132}},\
  \bibinfo {pages} {163602} (\bibinfo {year} {2024})}\BibitemShut {NoStop}%
\bibitem [{\citenamefont {Carusotto}\ and\ \citenamefont
  {Ciuti}(2013)}]{RevModPhys.85.299}%
  \BibitemOpen
  \bibfield  {author} {\bibinfo {author} {\bibfnamefont {I.}~\bibnamefont
  {Carusotto}}\ and\ \bibinfo {author} {\bibfnamefont {C.}~\bibnamefont
  {Ciuti}},\ }\bibfield  {title} {\bibinfo {title} {Quantum fluids of light},\
  }\href {https://doi.org/10.1103/RevModPhys.85.299} {\bibfield  {journal}
  {\bibinfo  {journal} {Rev. Mod. Phys.}\ }\textbf {\bibinfo {volume} {85}},\
  \bibinfo {pages} {299} (\bibinfo {year} {2013})}\BibitemShut {NoStop}%
\bibitem [{\citenamefont {Chang}\ \emph {et~al.}(2012)\citenamefont {Chang},
  \citenamefont {Jiang}, \citenamefont {Gorshkov},\ and\ \citenamefont
  {Kimble}}]{Chang2012}%
  \BibitemOpen
  \bibfield  {author} {\bibinfo {author} {\bibfnamefont {D.~E.}\ \bibnamefont
  {Chang}}, \bibinfo {author} {\bibfnamefont {L.}~\bibnamefont {Jiang}},
  \bibinfo {author} {\bibfnamefont {A.~V.}\ \bibnamefont {Gorshkov}},\ and\
  \bibinfo {author} {\bibfnamefont {H.~J.}\ \bibnamefont {Kimble}},\ }\bibfield
   {title} {\bibinfo {title} {Cavity qed with atomic mirrors},\ }\href
  {https://doi.org/10.1088/1367-2630/14/6/063003} {\bibfield  {journal}
  {\bibinfo  {journal} {New Journal of Physics}\ }\textbf {\bibinfo {volume}
  {14}},\ \bibinfo {pages} {063003} (\bibinfo {year} {2012})}\BibitemShut
  {NoStop}%
\bibitem [{\citenamefont {Carmichael}(1999)}]{Carmichael1999}%
  \BibitemOpen
  \bibfield  {author} {\bibinfo {author} {\bibfnamefont {H.~J.}\ \bibnamefont
  {Carmichael}},\ }\href {https://doi.org/10.1007/978-3-662-03875-8} {\emph
  {\bibinfo {title} {Statistical Methods in Quantum Optics 1}}}\ (\bibinfo
  {publisher} {Springer Berlin Heidelberg},\ \bibinfo {year}
  {1999})\BibitemShut {NoStop}%
\bibitem [{\citenamefont {Breuer}\ and\ \citenamefont
  {Petruccione}(2007)}]{Breuer2007}%
  \BibitemOpen
  \bibfield  {author} {\bibinfo {author} {\bibfnamefont {H.-P.}\ \bibnamefont
  {Breuer}}\ and\ \bibinfo {author} {\bibfnamefont {F.}~\bibnamefont
  {Petruccione}},\ }\href
  {https://doi.org/10.1093/acprof:oso/9780199213900.001.0001} {\emph {\bibinfo
  {title} {The Theory of Open Quantum Systems}}}\ (\bibinfo  {publisher}
  {Oxford University {PressOxford}},\ \bibinfo {year} {2007})\BibitemShut
  {NoStop}%
\bibitem [{\citenamefont {Ramos}\ \emph {et~al.}(2016)\citenamefont {Ramos},
  \citenamefont {Vermersch}, \citenamefont {Hauke}, \citenamefont {Pichler},\
  and\ \citenamefont {Zoller}}]{PhysRevA.93.062104}%
  \BibitemOpen
  \bibfield  {author} {\bibinfo {author} {\bibfnamefont {T.}~\bibnamefont
  {Ramos}}, \bibinfo {author} {\bibfnamefont {B.}~\bibnamefont {Vermersch}},
  \bibinfo {author} {\bibfnamefont {P.}~\bibnamefont {Hauke}}, \bibinfo
  {author} {\bibfnamefont {H.}~\bibnamefont {Pichler}},\ and\ \bibinfo {author}
  {\bibfnamefont {P.}~\bibnamefont {Zoller}},\ }\bibfield  {title} {\bibinfo
  {title} {Non-markovian dynamics in chiral quantum networks with spins and
  photons},\ }\href {https://doi.org/10.1103/PhysRevA.93.062104} {\bibfield
  {journal} {\bibinfo  {journal} {Phys. Rev. A}\ }\textbf {\bibinfo {volume}
  {93}},\ \bibinfo {pages} {062104} (\bibinfo {year} {2016})}\BibitemShut
  {NoStop}%
\bibitem [{\citenamefont {Zhang}\ and\ \citenamefont
  {M\o{}lmer}(2019)}]{zhang2019theory}%
  \BibitemOpen
  \bibfield  {author} {\bibinfo {author} {\bibfnamefont {Y.-X.}\ \bibnamefont
  {Zhang}}\ and\ \bibinfo {author} {\bibfnamefont {K.}~\bibnamefont
  {M\o{}lmer}},\ }\bibfield  {title} {\bibinfo {title} {Theory of subradiant
  states of a one-dimensional two-level atom chain},\ }\href
  {https://doi.org/10.1103/PhysRevLett.122.203605} {\bibfield  {journal}
  {\bibinfo  {journal} {Phys. Rev. Lett.}\ }\textbf {\bibinfo {volume} {122}},\
  \bibinfo {pages} {203605} (\bibinfo {year} {2019})}\BibitemShut {NoStop}%
\bibitem [{\citenamefont {Kumlin}\ \emph {et~al.}(2020)\citenamefont {Kumlin},
  \citenamefont {Kleinbeck}, \citenamefont {Stiesdal}, \citenamefont {Busche},
  \citenamefont {Hofferberth},\ and\ \citenamefont
  {B\"uchler}}]{kumlin2020nonexponential}%
  \BibitemOpen
  \bibfield  {author} {\bibinfo {author} {\bibfnamefont {J.}~\bibnamefont
  {Kumlin}}, \bibinfo {author} {\bibfnamefont {K.}~\bibnamefont {Kleinbeck}},
  \bibinfo {author} {\bibfnamefont {N.}~\bibnamefont {Stiesdal}}, \bibinfo
  {author} {\bibfnamefont {H.}~\bibnamefont {Busche}}, \bibinfo {author}
  {\bibfnamefont {S.}~\bibnamefont {Hofferberth}},\ and\ \bibinfo {author}
  {\bibfnamefont {H.~P.}\ \bibnamefont {B\"uchler}},\ }\bibfield  {title}
  {\bibinfo {title} {Nonexponential decay of a collective excitation in an
  atomic ensemble coupled to a one-dimensional waveguide},\ }\href
  {https://doi.org/10.1103/PhysRevA.102.063703} {\bibfield  {journal} {\bibinfo
   {journal} {Phys. Rev. A}\ }\textbf {\bibinfo {volume} {102}},\ \bibinfo
  {pages} {063703} (\bibinfo {year} {2020})}\BibitemShut {NoStop}%
\bibitem [{\citenamefont {Longhi}(2007)}]{longhi2007bound}%
  \BibitemOpen
  \bibfield  {author} {\bibinfo {author} {\bibfnamefont {S.}~\bibnamefont
  {Longhi}},\ }\bibfield  {title} {\bibinfo {title} {Bound states in the
  continuum in a single-level fano-anderson model},\ }\href
  {https://link.springer.com/content/pdf/10.1140/epjb/e2007-00143-2.pdf}
  {\bibfield  {journal} {\bibinfo  {journal} {The European Physical Journal B}\
  }\textbf {\bibinfo {volume} {57}},\ \bibinfo {pages} {45} (\bibinfo {year}
  {2007})}\BibitemShut {NoStop}%
\bibitem [{\citenamefont {Ciccarello}\ \emph {et~al.}(2013)\citenamefont
  {Ciccarello}, \citenamefont {Palma},\ and\ \citenamefont
  {Giovannetti}}]{PhysRevA.87.040103}%
  \BibitemOpen
  \bibfield  {author} {\bibinfo {author} {\bibfnamefont {F.}~\bibnamefont
  {Ciccarello}}, \bibinfo {author} {\bibfnamefont {G.~M.}\ \bibnamefont
  {Palma}},\ and\ \bibinfo {author} {\bibfnamefont {V.}~\bibnamefont
  {Giovannetti}},\ }\bibfield  {title} {\bibinfo {title} {Collision-model-based
  approach to non-markovian quantum dynamics},\ }\href
  {https://doi.org/10.1103/PhysRevA.87.040103} {\bibfield  {journal} {\bibinfo
  {journal} {Phys. Rev. A}\ }\textbf {\bibinfo {volume} {87}},\ \bibinfo
  {pages} {040103} (\bibinfo {year} {2013})}\BibitemShut {NoStop}%
\bibitem [{\citenamefont {Facchi}\ \emph {et~al.}(2016)\citenamefont {Facchi},
  \citenamefont {Kim}, \citenamefont {Pascazio}, \citenamefont {Pepe},
  \citenamefont {Pomarico},\ and\ \citenamefont
  {Tufarelli}}]{PhysRevA.94.043839}%
  \BibitemOpen
  \bibfield  {author} {\bibinfo {author} {\bibfnamefont {P.}~\bibnamefont
  {Facchi}}, \bibinfo {author} {\bibfnamefont {M.~S.}\ \bibnamefont {Kim}},
  \bibinfo {author} {\bibfnamefont {S.}~\bibnamefont {Pascazio}}, \bibinfo
  {author} {\bibfnamefont {F.~V.}\ \bibnamefont {Pepe}}, \bibinfo {author}
  {\bibfnamefont {D.}~\bibnamefont {Pomarico}},\ and\ \bibinfo {author}
  {\bibfnamefont {T.}~\bibnamefont {Tufarelli}},\ }\bibfield  {title} {\bibinfo
  {title} {Bound states and entanglement generation in waveguide quantum
  electrodynamics},\ }\href {https://doi.org/10.1103/PhysRevA.94.043839}
  {\bibfield  {journal} {\bibinfo  {journal} {Phys. Rev. A}\ }\textbf {\bibinfo
  {volume} {94}},\ \bibinfo {pages} {043839} (\bibinfo {year}
  {2016})}\BibitemShut {NoStop}%
\bibitem [{\citenamefont {Leonforte}\ \emph {et~al.}(2021)\citenamefont
  {Leonforte}, \citenamefont {Carollo},\ and\ \citenamefont
  {Ciccarello}}]{PhysRevLett.126.063601}%
  \BibitemOpen
  \bibfield  {author} {\bibinfo {author} {\bibfnamefont {L.}~\bibnamefont
  {Leonforte}}, \bibinfo {author} {\bibfnamefont {A.}~\bibnamefont {Carollo}},\
  and\ \bibinfo {author} {\bibfnamefont {F.}~\bibnamefont {Ciccarello}},\
  }\bibfield  {title} {\bibinfo {title} {Vacancy-like dressed states in
  topological waveguide qed},\ }\href
  {https://doi.org/10.1103/PhysRevLett.126.063601} {\bibfield  {journal}
  {\bibinfo  {journal} {Phys. Rev. Lett.}\ }\textbf {\bibinfo {volume} {126}},\
  \bibinfo {pages} {063601} (\bibinfo {year} {2021})}\BibitemShut {NoStop}%
\bibitem [{\citenamefont {Bleu}\ \emph {et~al.}(2018)\citenamefont {Bleu},
  \citenamefont {Solnyshkov},\ and\ \citenamefont {Malpuech}}]{Bleu2018}%
  \BibitemOpen
  \bibfield  {author} {\bibinfo {author} {\bibfnamefont {O.}~\bibnamefont
  {Bleu}}, \bibinfo {author} {\bibfnamefont {D.~D.}\ \bibnamefont
  {Solnyshkov}},\ and\ \bibinfo {author} {\bibfnamefont {G.}~\bibnamefont
  {Malpuech}},\ }\bibfield  {title} {\bibinfo {title} {Measuring the quantum
  geometric tensor in two-dimensional photonic and exciton-polariton systems},\
  }\bibfield  {journal} {\bibinfo  {journal} {Physical Review B}\ }\textbf
  {\bibinfo {volume} {97}},\ \href {https://doi.org/10.1103/physrevb.97.195422}
  {10.1103/physrevb.97.195422} (\bibinfo {year} {2018})\BibitemShut {NoStop}%
\bibitem [{\citenamefont {Cuerda}\ \emph {et~al.}(2024)\citenamefont {Cuerda},
  \citenamefont {Taskinen}, \citenamefont {K\"allman}, \citenamefont
  {Grabitz},\ and\ \citenamefont {T\"orm\"a}}]{PhysRevResearch.6.L022020}%
  \BibitemOpen
  \bibfield  {author} {\bibinfo {author} {\bibfnamefont {J.}~\bibnamefont
  {Cuerda}}, \bibinfo {author} {\bibfnamefont {J.~M.}\ \bibnamefont
  {Taskinen}}, \bibinfo {author} {\bibfnamefont {N.}~\bibnamefont {K\"allman}},
  \bibinfo {author} {\bibfnamefont {L.}~\bibnamefont {Grabitz}},\ and\ \bibinfo
  {author} {\bibfnamefont {P.}~\bibnamefont {T\"orm\"a}},\ }\bibfield  {title}
  {\bibinfo {title} {Observation of quantum metric and non-hermitian berry
  curvature in a plasmonic lattice},\ }\href
  {https://doi.org/10.1103/PhysRevResearch.6.L022020} {\bibfield  {journal}
  {\bibinfo  {journal} {Phys. Rev. Res.}\ }\textbf {\bibinfo {volume} {6}},\
  \bibinfo {pages} {L022020} (\bibinfo {year} {2024})}\BibitemShut {NoStop}%
\bibitem [{\citenamefont {Solnyshkov}\ \emph {et~al.}(2021)\citenamefont
  {Solnyshkov}, \citenamefont {Leblanc}, \citenamefont {Bessonart},
  \citenamefont {Nalitov}, \citenamefont {Ren}, \citenamefont {Liao},
  \citenamefont {Li},\ and\ \citenamefont {Malpuech}}]{PhysRevB.103.125302}%
  \BibitemOpen
  \bibfield  {author} {\bibinfo {author} {\bibfnamefont {D.~D.}\ \bibnamefont
  {Solnyshkov}}, \bibinfo {author} {\bibfnamefont {C.}~\bibnamefont {Leblanc}},
  \bibinfo {author} {\bibfnamefont {L.}~\bibnamefont {Bessonart}}, \bibinfo
  {author} {\bibfnamefont {A.}~\bibnamefont {Nalitov}}, \bibinfo {author}
  {\bibfnamefont {J.}~\bibnamefont {Ren}}, \bibinfo {author} {\bibfnamefont
  {Q.}~\bibnamefont {Liao}}, \bibinfo {author} {\bibfnamefont {F.}~\bibnamefont
  {Li}},\ and\ \bibinfo {author} {\bibfnamefont {G.}~\bibnamefont {Malpuech}},\
  }\bibfield  {title} {\bibinfo {title} {Quantum metric and wave packets at
  exceptional points in non-hermitian systems},\ }\href
  {https://doi.org/10.1103/PhysRevB.103.125302} {\bibfield  {journal} {\bibinfo
   {journal} {Phys. Rev. B}\ }\textbf {\bibinfo {volume} {103}},\ \bibinfo
  {pages} {125302} (\bibinfo {year} {2021})}\BibitemShut {NoStop}%
\bibitem [{\citenamefont {Salerno}\ \emph {et~al.}(2020)\citenamefont
  {Salerno}, \citenamefont {Palumbo}, \citenamefont {Goldman},\ and\
  \citenamefont {Di~Liberto}}]{Salerno2020}%
  \BibitemOpen
  \bibfield  {author} {\bibinfo {author} {\bibfnamefont {G.}~\bibnamefont
  {Salerno}}, \bibinfo {author} {\bibfnamefont {G.}~\bibnamefont {Palumbo}},
  \bibinfo {author} {\bibfnamefont {N.}~\bibnamefont {Goldman}},\ and\ \bibinfo
  {author} {\bibfnamefont {M.}~\bibnamefont {Di~Liberto}},\ }\bibfield  {title}
  {\bibinfo {title} {Interaction-induced lattices for bound states: Designing
  flat bands, quantized pumps, and higher-order topological insulators for
  doublons},\ }\href {https://doi.org/10.1103/PhysRevResearch.2.013348}
  {\bibfield  {journal} {\bibinfo  {journal} {Phys. Rev. Res.}\ }\textbf
  {\bibinfo {volume} {2}},\ \bibinfo {pages} {013348} (\bibinfo {year}
  {2020})}\BibitemShut {NoStop}%
\bibitem [{\citenamefont {Ferreira}\ \emph {et~al.}(2024)\citenamefont
  {Ferreira}, \citenamefont {Kim}, \citenamefont {Butler}, \citenamefont
  {Pichler},\ and\ \citenamefont {Painter}}]{ferreira2024deterministic}%
  \BibitemOpen
  \bibfield  {author} {\bibinfo {author} {\bibfnamefont {V.~S.}\ \bibnamefont
  {Ferreira}}, \bibinfo {author} {\bibfnamefont {G.}~\bibnamefont {Kim}},
  \bibinfo {author} {\bibfnamefont {A.}~\bibnamefont {Butler}}, \bibinfo
  {author} {\bibfnamefont {H.}~\bibnamefont {Pichler}},\ and\ \bibinfo {author}
  {\bibfnamefont {O.}~\bibnamefont {Painter}},\ }\bibfield  {title} {\bibinfo
  {title} {Deterministic generation of multidimensional photonic cluster states
  with a single quantum emitter},\ }\href
  {https://doi.org/10.1038/s41567-024-02408-0} {\bibfield  {journal} {\bibinfo
  {journal} {Nature Physics}\ }\textbf {\bibinfo {volume} {20}},\ \bibinfo
  {pages} {865} (\bibinfo {year} {2024})}\BibitemShut {NoStop}%
\bibitem [{\citenamefont {Crespi}\ \emph {et~al.}(2019)\citenamefont {Crespi},
  \citenamefont {Pepe}, \citenamefont {Facchi}, \citenamefont {Sciarrino},
  \citenamefont {Mataloni}, \citenamefont {Nakazato}, \citenamefont
  {Pascazio},\ and\ \citenamefont {Osellame}}]{crespi2019experimental}%
  \BibitemOpen
  \bibfield  {author} {\bibinfo {author} {\bibfnamefont {A.}~\bibnamefont
  {Crespi}}, \bibinfo {author} {\bibfnamefont {F.~V.}\ \bibnamefont {Pepe}},
  \bibinfo {author} {\bibfnamefont {P.}~\bibnamefont {Facchi}}, \bibinfo
  {author} {\bibfnamefont {F.}~\bibnamefont {Sciarrino}}, \bibinfo {author}
  {\bibfnamefont {P.}~\bibnamefont {Mataloni}}, \bibinfo {author}
  {\bibfnamefont {H.}~\bibnamefont {Nakazato}}, \bibinfo {author}
  {\bibfnamefont {S.}~\bibnamefont {Pascazio}},\ and\ \bibinfo {author}
  {\bibfnamefont {R.}~\bibnamefont {Osellame}},\ }\bibfield  {title} {\bibinfo
  {title} {Experimental investigation of quantum decay at short, intermediate,
  and long times via integrated photonics},\ }\href
  {https://doi.org/10.1103/PhysRevLett.122.130401} {\bibfield  {journal}
  {\bibinfo  {journal} {Phys. Rev. Lett.}\ }\textbf {\bibinfo {volume} {122}},\
  \bibinfo {pages} {130401} (\bibinfo {year} {2019})}\BibitemShut {NoStop}%
\end{thebibliography}%

\end{document}